\documentclass[twocolumn]{aastex7}

\usepackage{bm}
\usepackage{enumitem}

\begin{document}
 
 
 
\title{SCALES. I. Bridging Lensed Clump -- Giant Clump Scales in Lensed Galaxies at $z=1-4$}
 
 
\correspondingauthor{Boris S. Kalita}
\email{boris.kalita@pku.edu.cn; kalita.boris.sindhu@gmail.com}
 
\author[0000-0001-9215-7053]{Boris S. Kalita}
\altaffiliation{Kavli Astrophysics Fellow and Special Boya Fellow}
\affiliation{Kavli Institute for Astronomy and Astrophysics, Peking University, Beijing 100871, People's Republic of China}
\affiliation{Kavli IPMU (WPI), UTIAS, The University of Tokyo, Kashiwa, Chiba 277-8583, Japan}
\email{boris.kalita@pku.edu.cn; kalita.boris.sindhu@gmail.com}

\author[0000-0001-7940-1816]{Adélaïde Claeyssens}
\affiliation{Univ Lyon, Univ Lyon1, ENS de Lyon, CNRS, Centre de Recherche Astrophysique de Lyon UMR5574, F-69230 Saint-Genis-Laval, France}
\email{adelaide.claeyssens@univ-lyon1.fr}

\author[0000-0001-6947-5846]{Luis C. Ho}
\affiliation{Kavli Institute for Astronomy and Astrophysics, Peking University, Beijing 100871, People's Republic of China}
\affiliation{Department of Astronomy, School of Physics, Peking University, Beijing 100871, People's Republic of China}
\email{lho.pku@gmail.com}

\author[0000-0003-0348-2917]{Miroslava Dessauges-Zavadsky}
\affiliation{Département d'Astronomie, Université de Genève, Chemin Pegasi 51, CH-1290 Versoix, Switzerland}
\email{Miroslava.Dessauges@unige.ch}

\author[0000-0003-4264-3381]{Natascha M. Förster Schreiber}
\affiliation{Max-Planck-Institut fur extraterrestrische Physik, Giessenbachstrasse, D-85748 Garching, Germany}
\email{forster@mpe.mpg.de}

\author[0000-0002-3331-9590]{Emanuele Daddi}
\affiliation{Universit\'e Paris-Saclay, Universit\'e Paris Cit\'e, CEA, CNRS, AIM, 91191, Gif-sur-Yvette, France}
\email{emanuele.daddi@cea.fr}

\author[0000-0002-8434-880X]{John D. Silverman}
\affiliation{Kavli IPMU (WPI), UTIAS, The University of Tokyo, Kashiwa, Chiba 277-8583, Japan}
\affiliation{Department of Astronomy, School of Science, The University of Tokyo, 7-3-1 Hongo, Bunkyo, Tokyo 113-0033, Japan}
\affiliation{Centre for Data-Driven Discovery, Kavli IPMU (WPI), UTIAS, The University of Tokyo, Kashiwa, Chiba 277-8583, Japan}
\affiliation{Centre for Astrophysical Sciences, Department of Physics \& Astronomy, Johns Hopkins University, Baltimore, MD 21218, USA}
\email{john.silverman@ipmu.jp}

\author[0000-0001-8812-8460]{Naval Kishor Bhadari}
\affiliation{Kavli Institute for Astronomy and Astrophysics, Peking University, Beijing 100871, People's Republic of China}
\email{naval@pku.edu.cn}

\author[0009-0003-4721-177X]{Changhao Chen}
\affiliation{Kavli Institute for Astronomy and Astrophysics, Peking University, Beijing 100871, People's Republic of China}
\affiliation{Department of Astronomy, School of Physics, Peking University, Beijing 100871, People's Republic of China}
\email{cchen_louis@stu.pku.edu.cn}

\author[0000-0001-6115-0633]{Fangzhou Jiang}
\affiliation{Kavli Institute for Astronomy and Astrophysics, Peking University, Beijing 100871, People's Republic of China}
\affiliation{Department of Astronomy, School of Physics, Peking University, Beijing 100871, People's Republic of China}
\email{fangzhou.jiang@pku.edu.cn}

\author[0000-0002-7237-3856]{Ke Wang}
\affiliation{Kavli Institute for Astronomy and Astrophysics, Peking University, Beijing 100871, People's Republic of China}
\affiliation{Department of Astronomy, School of Physics, Peking University, Beijing 100871, People's Republic of China}
\email{kwang.astro@pku.edu.cn}

\author[0000-0001-8405-2921]{Jinning Liang}
\affiliation{Kavli Institute for Astronomy and Astrophysics, Peking University, Beijing 100871, People's Republic of China}
\affiliation{Department of Astronomy, School of Physics, Peking University, Beijing 100871, People's Republic of China}
\email{jinningliangastro@gmail.com}

\author[0000-0002-4639-5285]{Alvaro Segovia Otero}
\affiliation{Department of Astronomy, Tsinghua University, Beijing 100084, People's Republic of China}
\email{asegoviao@tsinghua.edu.cn}

\author[0000-0002-4569-9009]{Jinyi Shangguan}
\affiliation{Kavli Institute for Astronomy and Astrophysics, Peking University, Beijing 100871, People's Republic of China}
\affiliation{Department of Astronomy, School of Physics, Peking University, Beijing 100871, People's Republic of China}
\email{shangguan@pku.edu.cn}

 
\begin{abstract}
 
Dense star-forming regions in $z>1$ galaxies, often termed `clumps', have been studied across two distinct regimes of telescope resolution, each motivating interpretations in tension with the other. In the SCALES series (\emph{Star-Forming Clumps As Layered Emergent Structures}), we aim to connect star-forming structures across scales, from the $\sim100\,\mathrm{pc}$ clumps in lensed studies to the $\sim\mathrm{kpc}$-scale ``giant'' clumps in unlensed surveys. In this first paper, we present a difference-of-Gaussian (DoG) decomposition that simultaneously models $\approx 50-400\,\mathrm{pc}$ (lensed) clumps and  $\sim\mathrm{kpc}$-scale component (the extended component, or EC), which would together constitute the giant clumps seen at lower resolution. Here the EC is treated as a scale-separated \emph{local background} rather than a physical object, absorbing all light that is not a compact clump. Here, we apply this method to 23 moderately lensed ($\mu \approx 2-9$) galaxies at $z_{spec} = 1-4$ to evaluate clump properties (deferring ECs and giant clumps to SCALES-II). We measure clump stellar mass function slope of $1.88 \pm 0.04$, consistent with the $\approx2$ expected for hierarchical star formation, and with that of $\sim$kpc-scale clumps in unlensed galaxies. We also show that empirical correlations of clump surface densities and clump-to-host mass ratio with redshift are exaggerated by redshift-dependent detection bias, so $\approx 50-400\,\mathrm{pc}$ clumps are largely co-similar across our range ($z = 1-4$).  Finally, combining clump mass, size and age estimates with literature gas-dispersion measurements at similar scales, we argue that $\sim100\,\mathrm{pc}$ clumps are themselves composed of star-cluster-like sub-components at much smaller scales.
\end{abstract}
 
 
\section{Introduction} \label{sec:intro}
 
Compared to the local Universe, star-formation in galaxies beyond $z \gtrsim 1$ is expected to proceed in substantially different inter-stellar medium (ISM) conditions \citep{genzel10,burkert10,carilli13, madau14,forsterschreiber20}.  The amount of gas at constant stellar mass is found to be higher by an order of magnitude, and is consumed to form stars at rates higher by a factor of $\sim2$ \citep{daddi10,tacconi10,magdis12, tacconi13,dessaugeszavadsky15,scoville16,scoville17}.  Meanwhile, the turbulent motion of this gas (both ionized and molecular) is also found to be elevated at these epochs, attributed to a range of sources that include gas accretion,  stellar feedback, and dynamical interactions \citep[][]{wisnioski15,ubler19, girard21, rizzo24,danhaive25,lee25}.  These characteristic properties have been invoked to naturally explain the ``clumpy" morphology of $z \gtrsim 1$ galaxies \citep{elmegreen05,ceverino10}. However, the structural details of these ``clumps", and hence their importance in galaxy evolution, have been intensely debated for well over a decade \citep[e.g.,][]{tamburello15,oklopcic17}. 
\subsection{Kilo parsec-scale star-forming clumps}
Initial \textit{HST} observations revealed that many $z \gtrsim 1$ star-forming galaxies exhibit irregular morphologies dominated by giant $\sim \rm kpc$-scale star-forming clumps \citep{cowie95,abraham96,vandenbergh96,elmegreen04,elmegreen05,elmegreen07,elmegreen09}.  Spectroscopic studies using ionized gas tracers showed that the characteristic sizes and masses of giant clumps (derived from UV and optical images) are consistent with the Toomre scales of their host disks, supporting an interpretation of giant clumps forming as a result of gravitational instability-driven fragmentation in gas-rich galaxies \citep{genzel08,genzel11, forsterschreiber09,forsterschreiber11,wisnioski12,mieda16}. Detailed statistical studies further consolidated this idea of kpc-scale structures, with SED-based stellar mass values approaching $\sim 10^{9.5}\,\rm M_{\odot}$ in the most massive galaxies \citep{guo12,guo15,shibuya16,kalita25a,kalita25b, sok25, zhu26}.  
 
In parallel, theoretical works began exploring possible effects these massive structures, often assumed to be gravitationally bound, could have on the morphological transformation of their host galaxies. Through the exchange of angular momentum with the surrounding disk via dynamical friction, the clumps were predicted to migrate towards the galaxy barycenter, leading to bulge growth and disk thickening \citep{noguchi99,immeli04,bournaud07,dekel09,ceverino10,mandelker14,mandelker17,fensch21, dekel22}.  However, other works suggested that giant clumps may be disrupted by stellar feedback on timescales shorter than those required for migration toward the galaxy center \citep{genel12,oklopcic17}. 
 
\subsection{Clumps at hundred parsec-scales}
Over time, a more nuanced picture began emerging. Through the use of gravitational lensing, multiple works detected star-forming clumps down to sizes of $\sim 100\,\mathrm{pc}$ \citep{jones10,livermore12,olmstead14,livermore15}. These structures were often termed $\mathrm{HII}$ regions owing to their observed scaling between luminosity and size. Through hydrodynamical simulations, \cite{tamburello15,tamburello17} matched the results of lensed studies and concluded that the majority of star-forming clumps have characteristic sizes of $\sim 100-400\,\mathrm{pc}$ and stellar masses of $\sim 10^{7} - 10^{8}\,\mathrm{M_{\odot}}$.  Hence, giant clumps were suggested to instead be blends of these $\sim 100\,\mathrm{pc}$ clumps, with over-inflated size and mass measurements reflecting the limitations of telescope resolution. Along similar lines, other simulation-based efforts also termed giant clumps as ``clump clusters" \citep{behrendt16,behrendt19}, with these structures having measured properties similar to those of kpc-scale clumps. 
 
This picture has been further consolidated by a series of observational studies over the recent years \citep{dessaugeszavadsky17,cava18, vanzella21, messa22, messa24, claeyssens23, claeyssens25}, with minimum clump sizes reaching star-cluster-like scales of $\sim 10\,\mathrm{pc}$.  In large sample studies like that of \cite{claeyssens25}, some clumps are still found to be approaching the kpc-scale regime. However, the implicit conclusion has been that this regime is a minority, reflected by the clump stellar mass function with a characteristic slope of $\sim 2$ \citep[][]{dessaugeszavadsky18,claeyssens25}.  Such a distribution makes a clump with a stellar mass of $10^9\,\mathrm{M_{\odot}}$ is $\sim 10^{4}$ times less likely to form than a $10^7\,\mathrm{M_{\odot}}$ clump.  It is qualitatively assumed that the massive clumps are either one of these extreme objects, or more likely, a blend of multiple smaller clumps.  
 
 
\subsection{The much needed bridge}
\label{subsec:bridge}

\begin{figure*}
    \centering
    \includegraphics[width=\textwidth]{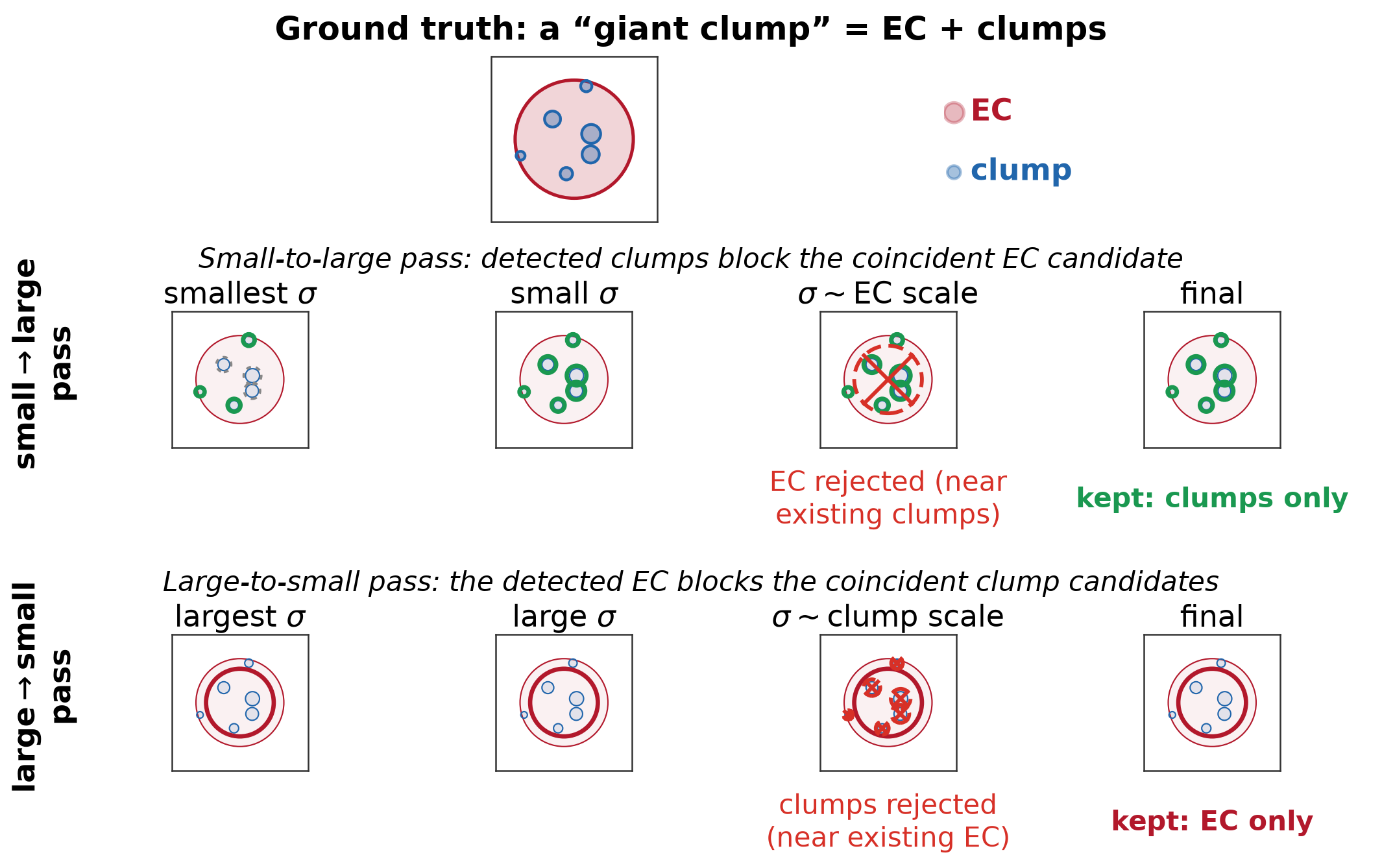}
    \caption{Schematic our detection algorithm. (Top) A giant clump, as it would appear at $\sim$kpc (unlensed) resolution, is assumed to be the blend of one EC with several clumps. (Middle) In the small-to-large pass, we first detect the smallest structures (clumps), while rejecting extended features (ECs) as they typically overlap spatially with the clumps. This pass therefore yields the clump catalogue. (Bottom) In the large-to-small pass, the EC is detected first, at the largest scales; the clump candidates found later, coincident with the EC's position, are then rejected as duplicates of the already-detected EC. This pass therefore yields the EC catalogue.}
    \label{fig:detection_pass_schematic}
\end{figure*}

Given the wide array of efforts towards understanding these star-forming structures within specific windows of spatial resolution and depth, the field is now mature enough to establish a single framework connecting them across scales. Such a hierarchy would link the smallest stellar structures like star clusters (tens of parsecs), the lensed clumps at $\approx 50-400\,\mathrm{pc}$ , and the giant clumps at $\sim \mathrm{kpc}-$scales. We therefore present SCALES (Star-Forming Clumps As Layered Emergent Structures), which is a series of works that will attempt at quantifying this hierarchy. This work, the first of the series, builds a methodological framework to simultaneously model two components at these scales: a \emph{clump}, the $\approx 50-400\,\mathrm{pc}$ compact structure resolved individually in our lensed sample (elsewhere termed a lensed clump, but simply ``clump'' throughout this work); and an \emph{EC} (extended component), the $\sim$kpc-scale emission surrounding it. We stress at the outset that, within this paper, the EC is not interpreted as a singular physical structure. It is a scale-specific morphological component that collects \emph{all} flux at scales larger than the compact clumps, but below scales of a few kpc \citep[informed by giant clump sizes measured in][]{kalita25b}, and may be attributed to truly diffuse emission, clumps below our detection threshold, and/or the overlapping wings of the detected clumps. We reserve the term \emph{giant clump} for the combination of a clump and its EC together, which is how such a structure would appear at the $\sim$kpc resolution of unlensed surveys; the physical interpretation of giant clumps as clump+EC associations is deferred to the next paper in this series (SCALES-II; \citealt{kalita26}). This work rather focuses on presenting the details of the new methodology as well as the measurement of key properties of clumps. Since this clump/EC distinction underlies everything that follows, we outline it up front: Fig.~\ref{fig:detection_pass_schematic} illustrates the two-pass detection scheme used throughout this work to separate the two, with the full details provided in Sec.~\ref{subsec:det_flux_fit}.

Sec.~\ref{sec:methods} deals with the lensed galaxy dataset we use, as well as the novel methodology we implement. Sec.~\ref{sec:results} compiles the key results, and Sec.~\ref{sec:discussions} interprets these in order to provide conclusions regarding the star-forming properties and internal sub-structure of clumps. Finally in Sec.~\ref{sec:summary} summarizes the key findings of this work.  Throughout this paper, we adopt a concordance $\Lambda$CDM cosmology, characterized by  $\Omega_{\rm m}=0.3$, $\Omega_{\Lambda}=0.7$, and $\rm H_{0}=70~\rm km\, s^{-1}\,\rm Mpc^{-1}$. All images are oriented such that north is up and east is left.  Use of the $\log$-scale always refers to $\log_{10}$.  All SED-based properties implement the \cite{kroupa01} initial mass function.
 
\section{Methodology}
\label{sec:methods}
In this section we detail our full methodological framework.

\subsection{Data and Sample} \label{subsec:data_sample}
We draw our sample from the largest study of lensed galaxies to date, \citep[][referred to as C25 from here on]{claeyssens25}.  The study targets the core of the Abell 2744 cluster, which is covered by two separate JWST programs.  The UNCOVER program \citep[GO 2561; PIs: Rachel Bezanson and Ivo Labbé;][]{bezanson24} provides short-wavelength (SW) imaging in F115W, F150W, and F200W  ($0.02^{\prime\prime}/\rm pixel$), together with long-wavelength (LW) imaging in the broad-band filters F277W, F356W, and F444W ($0.04^{\prime\prime}/\rm pixel$), as well as the medium-band filter F410M . In addition, the core of Abell 2744 was observed during Cycle 2 as part of the MEGASCIENCE programme \citep[GO 4111; PI: Katherine A. Suess;][]{suess24}. 
 
These observations covered the cluster region in 10 medium-band filters (F140M, F162M, F182M, F210M, F250M, F300M, F335M, F360M, F430M, F460M, and F480M), along with the broad-band filters F070W and F090W. All filters shorter than F210M have a pixel scale of $0.02^{\prime\prime}/\rm pixel$, while the rest have a scale of $0.04^{\prime\prime}/\rm pixel$.  The empirical point spread functions (PSFs) for each of the filters are also provided as part of the MEGASCIENCE and UNCOVER datasets. Part of the UNCOVER dataset is the galaxy catalog providing key parameters like redshift (photometric/spectroscopic),  stellar mass and star-formation rate (SFR) for all sources covered by the image mosaics \citep{weaver24,wang24}.
 
Using a sample of 476 galaxies, spanning a wide range of redshifts ($z = 0.7-9$) and magnifications ($\mu = 1.5 - 300$), C25 carried out a detailed analysis of 1956 star-forming clumps.  However, the majority of the galaxies contain $\lesssim 5$ clumps each (Fig.~4 in C25).  Hence given the goal of our work to combine the hundred pc to kpc scales, we limit our sample to galaxies with $>10$ clumps. This threshold returns a total of 28 galaxies.  We reject 5 galaxies  with nearby companions,  as they might interfere with the accurate analysis of the clumps. Hence we are left with 23 galaxies in our final sample (Table~\ref{tab:host_galaxies}; Fig.~\ref{fig:galaxy_collage}). These span a redshift range $z = 1-4$ and a magnification range of $\mu = 2.0-9.1$ (median per-clump magnification of each galaxy). Increasing the threshold of the required number of ``C25 clumps'' (a nomenclature we will follow to refer to clumps detected in the C25 study) results in a rapid shrinking of our sample size along with the lowering of the upper boundary of the redshift range. Lowering the threshold instead, on the other hand, does not change the resulting sample drastically, so long as it remains above $\sim5$ clumps per galaxy. Hence we settle on the somewhat arbitrary threshold of 10 C25 clumps per galaxy. It is worth noting that there are a total of 430 C25 clumps within these galaxies which makes up $22\%$ of their total clump sample, even though the 23 galaxies make up only $5\%$ of their host galaxy sample.
 
The redshift values for the majority (18/23) of our sample are spectroscopic, drawn from \cite{richard21}. For the remaining five galaxies (IDs 15832, 19355, 22864, 23491, and 25706), no spectroscopic redshift is available, and we instead adopt the photometric SED-fitting redshift from \cite{weaver24}, which has an uncertainty $<0.15$ in all such cases; these five are marked with an asterisk in Table~\ref{tab:host_galaxies}.  For simplicity,  during our analysis, we treat the galaxy redshift as a fixed parameter unless stated otherwise.  C25 provides the galaxy photometry and SED-based stellar mass and star-formation rates, the latter of which we use throughout this work (Table~\ref{tab:host_galaxies}). The only exception is the comparison of the clumps and their hosts with the star-forming main sequence (Sec.~\ref{subsec:resolved_sfms}), for which we re-do the SED fitting of the hosts with the same photometry, but with the SED fitting prescription of the clumps in our study (described in Sec.~\ref{subsec:sed_fitting}), and correct the results for magnification. We do acknowledge that the clump prescription might be too simplistic for galaxies. The purpose of this choice is to not over-interpret differences in star-formation rates between clumps and their hosts that can be attributed to SED fitting prescription.

The lensing corrections in this work were performed using the publicly available lensing maps released by the UNCOVER collaboration for Abell 2744 \citep{furtak23,price25}, specifically the convergence ($\kappa$) and shear ($\gamma_1$, $\gamma_2$) maps. These are scaled to the source redshift of each galaxy and combined to give the per-clump magnification $\mu$, which we use to correct all measured fluxes and integrated quantities. The magnification of each clump is evaluated at its position in the best-fit lens model. Its uncertainty is derived from the set of 50 lens-model realisations (also released by the UNCOVER collaboration) drawn from the final MCMC chain \citep{furtak23}. For each clump we first take the fractional 16th--84th percentile spread of $\mu$ across the realisations, which is a multiplicative term. This however does not account for the difference between the median of the realisations and the best-fit value. This is calculated separately. Both these terms are then added in quadrature to get the final magnification uncertainties, which have a median of $\approx4\%$ across our sample, with $90\%$ of the clumps below $\approx15\%$. We note that the largest errors occur in the most strongly magnified galaxies. These uncertainties are propagated into the stellar masses and SFRs (Sec.~\ref{subsec:sed_fitting}) and into the completeness estimate (Appendix~\ref{app:completeness}). The same maps additionally yield the maximum lensing-induced ellipticity at each clump position, which is used to define the 2D models. The derivation of the magnification and ellipticity is detailed in Appendix~\ref{app:lensing}. 
 
 
\begin{table*}
\caption{Host galaxies in the final sample. Stellar masses and SFRs are in log$_{10}$ units ($M_\odot$ and $M_\odot\,\mathrm{yr}^{-1}$); $\mu$ is the median of the per-clump lensing magnifications of each galaxy (Sec.~\ref{subsec:data_sample}). $N_\mathrm{clump}$ is the number of small clumps with a measured stellar mass and $N_\mathrm{EC}$ the number of ECs. These however only include those with robust SED measurements, and hence would be a subset of the clumps and ECs shown in Fig.~\ref{fig:galaxy_collage}. Quoted uncertainties of parameters are the catalog 16th/84th-percentile bounds. Redshifts marked with an asterisk ($^{*}$) are photometric rather than spectroscopic.}
\label{tab:host_galaxies}
\begin{tabular}{l c c c c c c c c}
\hline\hline
ID & R.A. & Dec. & $z_\mathrm{spec}$ & $\mu$ & $\log M_{*}$ & $\log \mathrm{SFR}_{100}$ & $N_\mathrm{clump}$ & $N_\mathrm{EC}$ \\
   & (deg) & (deg) & & & $[M_\odot]$ & $[M_\odot\,\mathrm{yr}^{-1}]$ & & \\
\hline
4304 & 3.59977 & -30.41392 & $1.77$ & $3.44$ & $9.47^{+0.05}_{-0.00}$ & $1.58^{+0.48}_{-0.47}$ & 23 & 5 \\
7775 & 3.59755 & -30.40388 & $1.69$ & $6.30$ & $9.62^{+0.17}_{-0.04}$ & $0.47^{+0.35}_{-0.08}$ & 22 & 10 \\
8136 & 3.58285 & -30.40283 & $1.89$ & $8.89$ & $9.09^{+0.11}_{-0.06}$ & $0.21^{+0.25}_{-0.03}$ & 21 & 7 \\
9198 & 3.57731 & -30.40159 & $2.58$ & $3.81$ & $9.43^{+0.06}_{-0.04}$ & $-0.88^{+0.18}_{-0.14}$ & 7 & 6 \\
10570 & 3.56888 & -30.39553 & $1.01$ & $1.96$ & $8.41^{+0.03}_{-0.01}$ & $0.44 \pm 0.00$ & 10 & 8 \\
11188 & 3.57784 & -30.39386 & $1.16$ & $3.71$ & $8.97 \pm 0.08$ & $0.02^{+0.23}_{-0.04}$ & 19 & 7 \\
11670 & 3.56848 & -30.39230 & $1.54$ & $2.46$ & $9.94^{+0.02}_{-0.01}$ & $1.38^{+0.85}_{-0.88}$ & 28 & 9 \\
12575 & 3.59146 & -30.38950 & $1.37$ & $3.40$ & $8.95^{+0.08}_{-0.12}$ & $0.86^{+0.47}_{-0.13}$ & 21 & 11 \\
12800 & 3.56988 & -30.38891 & $3.79$ & $3.79$ & $9.30^{+0.04}_{-0.03}$ & $1.39^{+0.04}_{-0.01}$ & 12 & 3 \\
13428 & 3.58697 & -30.38700 & $1.86$ & $4.00$ & $8.58^{+0.07}_{-0.13}$ & $0.66^{+0.23}_{-0.04}$ & 8 & 3 \\
14002 & 3.56501 & -30.38542 & $2.68$ & $3.56$ & $8.46 \pm 0.04$ & $0.51^{+0.01}_{-0.03}$ & 14 & 3 \\
14601 & 3.55667 & -30.38410 & $2.58$ & $3.58$ & $10.03^{+0.04}_{-0.03}$ & $2.10^{+0.05}_{-0.01}$ & 6 & 2 \\
15439 & 3.54260 & -30.38268 & $1.41$ & $2.25$ & $10.65^{+0.04}_{-0.03}$ & $0.68 \pm 0.10$ & 76 & 18 \\
15832 & 3.56231 & -30.38118 & $3.49^{*}$ & $4.58$ & $8.91^{+0.05}_{-0.11}$ & $0.97^{+0.01}_{-0.07}$ & 28 & 4 \\
16673 & 3.59169 & -30.37959 & $1.80$ & $2.18$ & $8.36^{+0.03}_{-0.04}$ & $0.38^{+0.01}_{-0.00}$ & 13 & 4 \\
17172 & 3.58441 & -30.37872 & $2.36$ & $2.60$ & $9.62^{+0.04}_{-0.06}$ & $1.69^{+0.08}_{-0.05}$ & 21 & 6 \\
18098 & 3.54063 & -30.37719 & $1.88$ & $4.03$ & $10.10^{+0.05}_{-0.04}$ & $1.77^{+1.87}_{-1.78}$ & 13 & 5 \\
19355 & 3.58157 & -30.37488 & $1.85^{*}$ & $2.42$ & $9.28 \pm 0.06$ & $1.38^{+1.11}_{-1.03}$ & 10 & 2 \\
19708 & 3.55820 & -30.37451 & $2.41$ & $7.40$ & $9.90^{+0.04}_{-0.12}$ & $1.99^{+0.01}_{-0.02}$ & 10 & 9 \\
20054 & 3.54041 & -30.37505 & $2.58$ & $9.05$ & $10.06 \pm 0.05$ & $0.04^{+0.18}_{-0.19}$ & 23 & 10 \\
22864 & 3.54791 & -30.36882 & $1.52^{*}$ & $4.38$ & $10.18^{+0.06}_{-0.03}$ & $1.05^{+0.10}_{-0.13}$ & 40 & 13 \\
23491 & 3.54102 & -30.36718 & $3.49^{*}$ & $6.15$ & $9.79^{+0.07}_{-0.00}$ & $1.88 \pm 0.00$ & 6 & 3 \\
25706 & 3.53720 & -30.36574 & $1.36^{*}$ & $3.17$ & $10.16^{+0.04}_{-0.01}$ & $2.24^{+0.01}_{-0.00}$ & 20 & 8 \\
\hline
\end{tabular}
\end{table*}
 
\subsection{Detection and multi-scale fitting} \label{subsec:det_flux_fit}
 
\begin{figure*} 
    \centering
    \includegraphics[width=\textwidth]{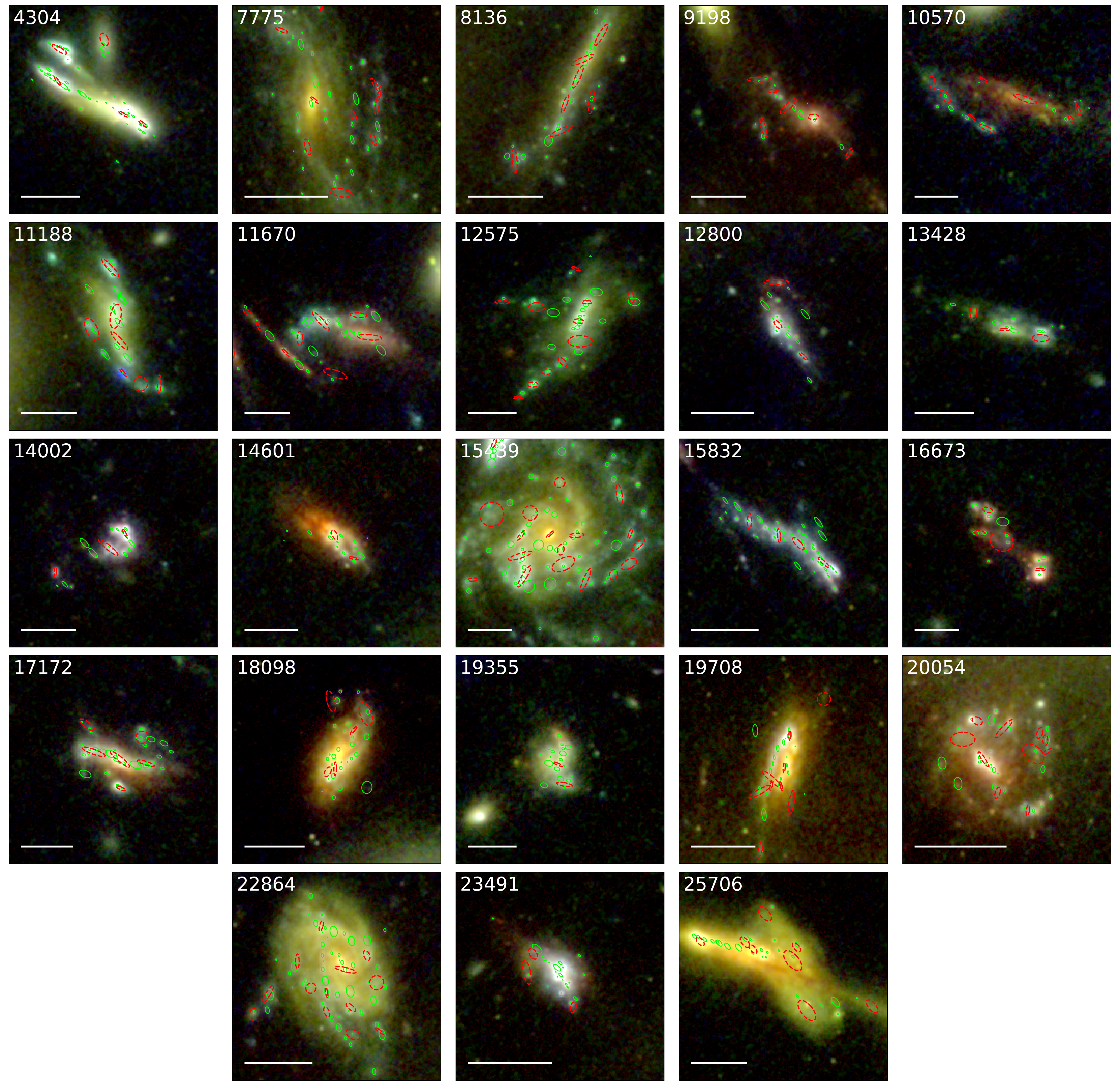}
    \caption{RGB (F070W/F115W/F200W) images of our sample galaxies, with the detected clumps (in cyan; all clumps detected inside the F444W host segment) and ECs (in red; those with a robust SED fit). All F070W and F115W images are downconvolved to the F200W resolution using a Gaussian kernel. The PSF matching follows the method used in \cite{kalita25b}. Since these images are generated only for visualization purposes and are not pertinent for the analysis in this work, we do not provide further details of the PSF matching procedure. In each panel, we show a scale bar of length $5\times \sqrt{\mu} \,\mathrm{kpc}$ at the bottom left.  For galaxy ids: 7775, 15439, and 22864, we show a 251x251 pixel cutout given the extent of the galaxy (as discussed in Sec.~\ref{subsec:det_flux_fit}), which exceeds the cutout size of 201x201 pixels adopted for all other galaxies in the figure.}
    \label{fig:galaxy_collage}
\end{figure*}
 
\begin{figure*} 
    \centering
    \includegraphics[width=0.8\textwidth]{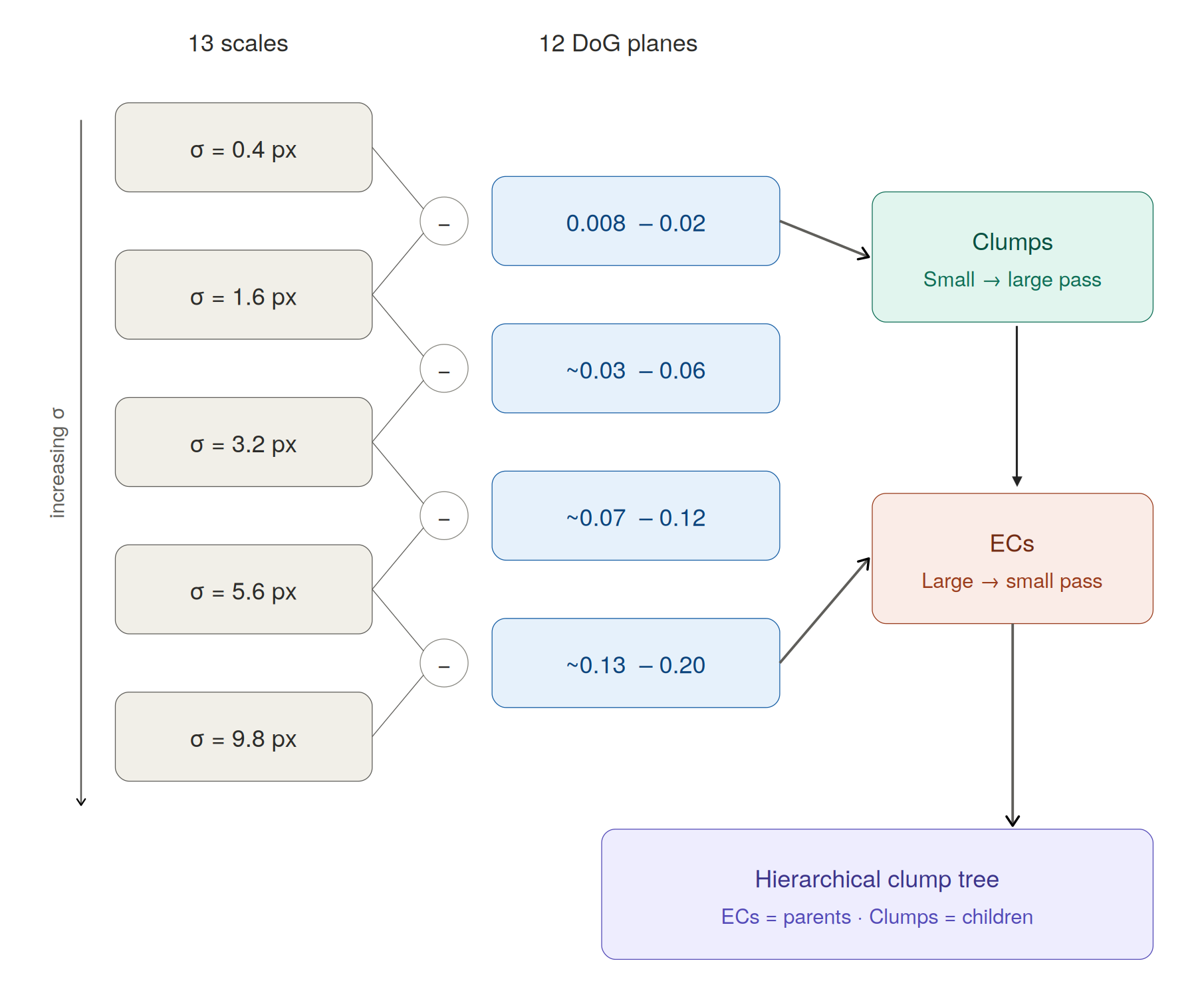}
    \includegraphics[width=\textwidth]{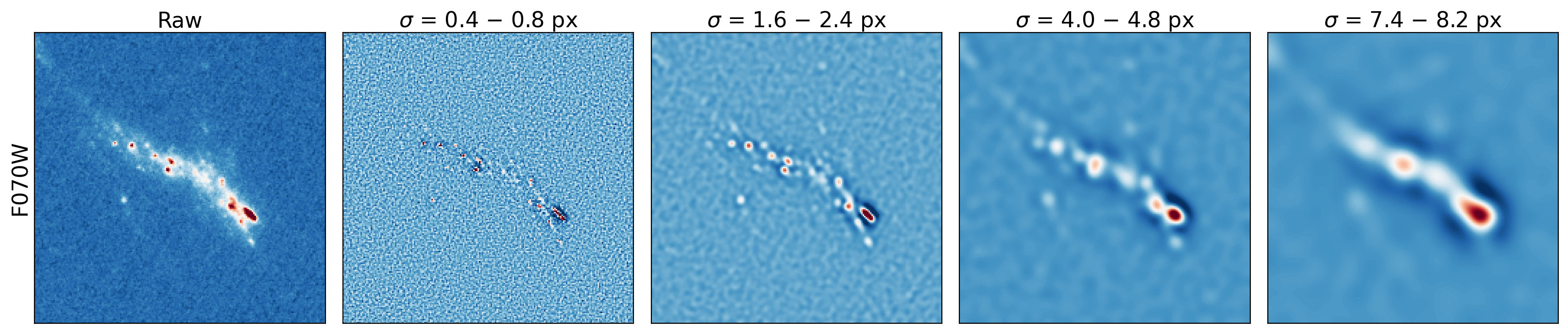}
    \caption{(Top) The flowchart describing our Difference-of-Gaussian (DoG) approach which divides each galaxy image into scale specific sub-images (shown in the series of images in the lower panel), which are then used to detect (compact) clumps and ECs.}
    \label{fig:pipeline}
\end{figure*}

\begin{figure*}
    \centering
    \includegraphics[width=\textwidth]{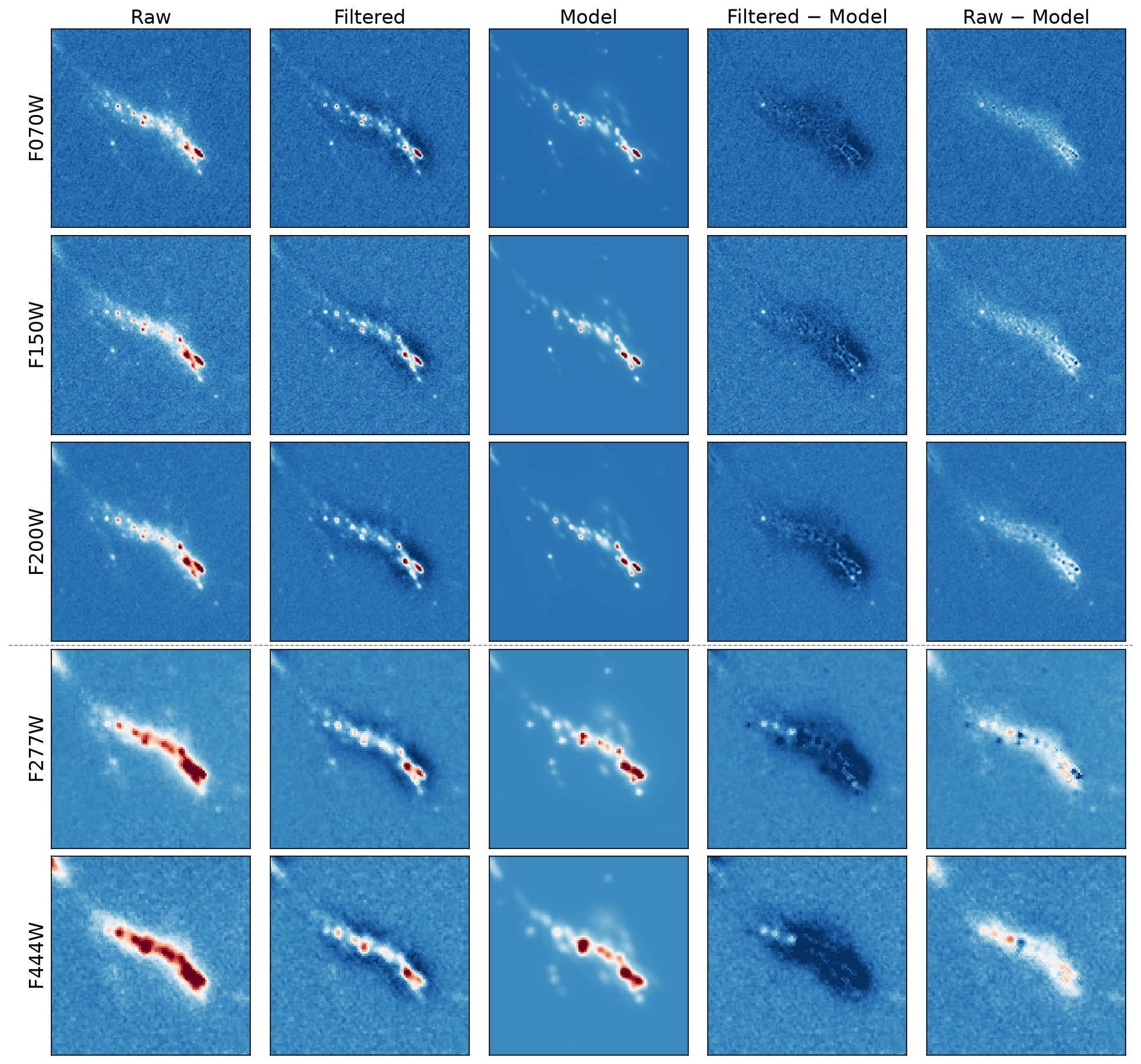}
    \caption{We show the fitting results of galaxy id: 15832 using our multi-scale fitting pipeline for 5 out of the 13 filters used in our study. The cutout sizes are the same as in Fig.~\ref{fig:galaxy_collage} for the SW filters F070W, F150W and F200W ($201\times201$ pixel cutout at $0.02\arcsec$/pixel). The LW filters F277W and F444W cover the same sky footprint, but due to their coarser pixel scale ($0.04\arcsec$/pixel), correspond to $101\times101$ pixel cutouts. The first column shows the image, with the dynamic range set by the 0.1th and 99.95th percentiles of the high-pass filtered image (second column), and this same range is applied across all columns for a given filter. The second column shows the high-pass filtered
image, obtained by subtracting a Gaussian-smoothed version of the image at the largest detection scale of our DoG grid. The third column shows the best-fit PSF-convolved composite Sérsic ($n=0.5$) model, and the fourth and fifth columns show the filtered$-$model and raw$-$model residuals, respectively.}
    \label{fig:image_fitting}
\end{figure*}
 
In order to detect the clumps in each of the galaxies, we use the F070W filter which has the highest resolution (point spread function full width half maximum; PSF FWHM $= 0.029^{\prime\prime}$) as well as depth (based on the $5\,\sigma$ point source sensitivity). As this filter covers the rest-frame UV wavelengths $\sim 1250 - 3900$~\AA\ across our sample, we choose to not shift to other filters to match the rest-frame wavelength window in favor of retaining similar resolution in our detection band. This choice is at the expense of introducing a wavelength-dependent bias in the clump detection rather than having to deal with a resolution-dependent bias. This prevents us from detecting clumps that might be bright in the rest-frame optical or near-IR but faint in the rest-frame UV, due to older ages and/or dust attenuation. But we show that this incompleteness predominantly affects clumps below our mass completeness limit (Appendix~\ref{app:completeness}), and hence does not affect our main results. Furthermore, any such undetected clumps would be absorbed into the EC component, and hence would not affect the measured properties of the detected clumps.

The detection of clumps within each galaxy is carried out using a multi-scale difference-of-Gaussian (DoG) decomposition (Fig.~\ref{fig:pipeline}).  This method involves iterative subtraction of an image convolved with a Gaussian from another that has been convolved with a narrower Gaussian. The DoG method is widely used as a band-pass filter in image analysis which can be used in `blob-detections' in specific spatial scales \citep{lowe04}.  The use of difference images, with at least one being convolved with a Gaussian has already been extensively implemented to detect clumps in galaxies \citep[e.g.,][]{elmegreen04, guo15, kalita24,sok25}. The DoG approach can hence be considered an extension.
 
We first create the F070W cutout for each galaxy, except three (discussed later), with size of $201\times201~$pixels. This choice of a fixed cutout size is motivated by the difficulty of defining a size-based cutout for lensed galaxies, whose extent is hard to measure reliably. A fixed, sufficiently large\footnote{visually ensured to incorporate any nearby source that is modeled during the fitting} cutout avoids this challenge and ensures every source samples the same background statistics and injection area when estimating uncertainties and completeness (discussed later). In three cases, however, this cutout size is not sufficient, as evidenced by the spatial distribution of the C25 clumps in these sources (IDs: 7775, 15439, 22864). For these sources, we generate cutouts of $251\times251~$pixels. Each cutout is convolved with Gaussian kernels over a grid of 14 smoothing scales, $\sigma = {0.4, 0.8, 1.6, 2.4, 3.2, 4.0, 4.8, 5.6, 6.4, 7.4, 8.2, 9.0, 9.8, 10.6}$ pixels (relative to the original image), and successive smoothed images are subtracted from the preceding one to produce 13 band-pass planes, each sensitive to structures on a characteristic scale ($\sim \sqrt{\sigma^2_{i+1} - \sigma^2_{i}}$). At the $0.02''$ pixel scale of the JWST/NIRCam short-wavelength channels, this grid spans $\sim0.008''{-}0.196''$ in the observed frame. For our sample ($z=1{-}4$), this corresponds to an image-plane scale of $\sim0.13{-}1.7\,\mathrm{kpc}$. After accounting for the lensing magnification of the sample ($\mu\sim2{-}9$), the DoG grid probes source-plane physical scales from the hundred-parsec regime up to $\sim2\,\mathrm{kpc}$. 
 
A visual representation of clump-EC detection methodology is provided in Fig.~\ref{fig:detection_pass_schematic}, with the full logic described below. Sources are detected independently on every DoG plane above a $3\sigma$ threshold\footnote{the image sigma in each DoG plane is estimated using $(3-)$ sigma clipped statistics feature of \texttt{photutils}} over a minimum of 3 connected pixels and deblended using \texttt{photutils}.  The involved parameters are chosen based on visual inspection of the detected regions. The detection is performed in two passes: smallest-to-largest and largest-to-smallest scales. Candidates that are within a threshold distance of a detected source in an earlier step of each pass are rejected as duplicates.   The threshold distance is set to the average of the current and previous smoothing scale added to the F070W PSF FWHM ($0.029^{\prime\prime}$) in quadrature.  The clumps that are detected in small-to-large pass will be referred to as clumps, while those from the large-to-small pass will be the ECs, as defined in Sec.~\ref{subsec:bridge}. Fig.~\ref{fig:detection_pass_schematic} illustrates why the two passes recover these two different catalogues: since a clump and its parent EC are spatially coincident, whichever of the two is detected first at a given step of a pass causes the other to be rejected as a duplicate when the smoothing scale later reaches its size.  We show the clumps and ECs in each galaxy in Fig.~\ref{fig:galaxy_collage}. The resulting catalogue is organised into a hierarchical \emph{clump tree}, where the ECs act as parents and compact small clumps falling within the scale radius of a parent are assigned as its children, while small clumps without an enclosing parent are attached to the nearest large clump. We emphasise that this parent--child organisation is a methodological setup, not a physical hierarchy: the EC ``parent'' is simply the large-scale component that absorbs the residual, non-clump light in the neighbourhood of its assigned clumps, and its flux should be read as a local background rather than as an enclosing physical object.

Each galaxy is then modelled as a superposition of two-dimensional Sérsic profiles with fixed index $n=0.5$ (Gaussian profiles), one for each detected clump in the clump tree, convolved with the empirical F070W PSF. The composite model is fit to the galaxy image without the largest ($>2\,\mathrm{kpc}$) scales, obtained by subtracting a version of the cutout smoothed on the largest scale, through a three-stage progressively relaxed optimisation. In the first stage ($\leq1000$ iterations), only the amplitudes and centroids are varied. The second stage ($\leq2000$ iterations) fits the clump sizes in addition to the amplitudes and centroids, while during the final stage the ellipticity and position angle of the ECs are also allowed to be fit.  
 
The model for clumps in the source plane are kept circular for simplicity.  Hence, in the image plane where the fitting is done, we fix their ellipticity and phase angle based on the lensing shear (Appendix~\ref{app:lensing}).  Meanwhile the EC models are initialised using the same lensing shear-based model, but are allowed to vary during the fitting as mentioned above.
During the fitting, the sizes are bounded to $\sigma = 0.2{-}7.0$ pixels, before PSF convolution, for clumps ($\sim0.004''{-}0.14''$ or $50 - 1100\,\mathrm{pc}$,  making them effectively unbound) and $\sigma = 3.0{-}7.0$ pixels for ECs ($\sim0.06''{-}0.14''$ or $400 - 1100\,\mathrm{pc}$).  The bounding window for the ECs is derived from the results for giant clumps in \cite{kalita25b}. To prevent the models from drifting away from the structures they were detected on, the centroids are confined to a box around their detection positions that is kept fixed through all three stages. Each clump may therefore move by at most the sum of the F070W PSF and the characteristic scale length at which it is detected.  Each stage is solved using a bounded L-BFGS-B minimiser, with the solution from one stage initialising the next.  The $\chi^2$ is measured separately in each of the 13 DoG planes, all of which are then added to get a combined $\chi^2$ value\footnote{$\chi^2_{\mathrm{final}} = \sum\limits_{\mathrm{planes}} \mathrm{weight}_{i} \times \chi^2_{i}$} that is used for the minimisation. The residual on each plane is normalised using a sigma-clipped ($3\sigma$) estimate of its noise from the corresponding sub-image. 
 
The contribution from each scale is weighted differently for the clumps and ECs. Clumps are constrained using a near-uniform weighting across all planes, which mimics a single $\chi^2$-optimisation based fitting on the original image. Meanwhile the ECs are weighted using a Gaussian-shaped profile centred at $\sigma=5$ px ($\sim0.10''$), with the two smallest-scale planes suppressed.  The width of this Gaussian is such that the $\pm 2$ profile-$\sigma$ are at $3$ and $7$ pixels (which are the size constraints for the EC profile set earlier). This weighting scheme allows clumps and ECs to be optimised simultaneously.  Results for a single galaxy are shown in Fig.~\ref{fig:image_fitting}. We find the DoG approach to outperform a classical multi-component fitting on the original image, through tests that are detailed in Appendix~\ref{app:method_comparison}.
 
The pipeline is run for each of the 23 galaxies in our sample.  Once complete, we then use the results to measure the spectral energy distributions of all detected clumps across the remaining 13 JWST/NIRCam filters spanning both the short- and long-wavelength channels: the broad bands F070W, F090W, F115W, F150W, F200W (SW) and F277W, F356W, F444W (LW), together with the medium bands F140M, F162M, F182M, F210M (SW) and F250M (LW).  We skip the use of the rest of the medium bands due to lower depths and resolution which results in inaccurate flux measurements in multiple galaxies. 

The clump positions, sizes, ellipticities (for the ECs), position angles, and the hierarchical clump tree are all fixed to the values obtained from the F070W structural fit, and only the amplitudes of the clumps are refit independently in each filter. This ensures that the photometry in all bands corresponds to the same physical structures, while also preventing flux variations across bands being affected by band-to-band changes in clump structure.  All SW filters (including F070W) share the same pixel scale of $0.02^{\prime\prime}/\mathrm{pixel}$.  However, the LW filters have a coarser imaging grid ($0.04^{\prime\prime}/\mathrm{pixel}$).  Hence, the structural parameters are scaled accordingly during fitting of these bands.
 
In each filter, the amplitude refit minimises the same multi-scale, dual-weighted DoG $\chi^2$ used in the structural fitting (for F070W). Images of each filter are convolved with their own corresponding empirical PSF. The integrated flux of each clump in every filter is then obtained by summing  over the corresponding Sérsic $n=0.5$ profile. This procedure results in a 13-band photometric SED for every clump in every galaxy,  which will be used for the stellar population analysis presented in Sec.~\ref{subsec:sed_fitting}.
 
As can be seen in Fig.~\ref{fig:image_fitting}, we model all clumps and ECs within the image cutout for a robust combined modelling of the flux. However, as can be observed in Fig.~\ref{fig:galaxy_collage}, not all sources are marked as clumps.  In our analysis, we only include sources (that were detected in F070W) within the visual extent of the galaxy emission in the F444W filter.  We select this filter due to its lower resolution and long wavelength coverage (hence best approximation of the stellar distribution) relative to all other filters.  In order to remove contamination from globular clusters in Abell 2744, we further remove any of the sources with flux distributions that agree better with an SED of a single stellar population at the lensing cluster redshift during the SED-fitting procedure that will be later discussed in Sec.~\ref{subsec:sed_fitting}.  We do not remove clumps that may correspond to a central bulge, since such an identification depends on the galaxy in the image plane resembling a disk--bulge configuration, which is required for such an assessment. We instead discuss the effect of removing visually identified bulge candidates in the relevant sections.
 
\subsection{Uncertainty estimates} \label{subsec:flux_uncertainties}
 
We estimate the uncertainties on the clump (and EC) measurements through an injection--recovery method.  For the F070W filter, where the clump amplitudes and sizes are fit simultaneously, we inject the best-fit model back in to the host galaxy image at a random position within $30$~SW~pixels ($0.6\arcsec$, $\approx20\times$ the F070W PSF FWHM) of its true location. We ensure that the position does not fall within $1.5$~pixels ($\approx1\times$ the PSF FWHM) of an existing real clump or within $7.5$~pixels ($\approx5\times$ the PSF FWHM) of the image edge.  The clump is then re-fit with the amplitudes and sizes. We do the same procedure for the ECs, but keeping the sizes fixed to the best-fit value. The reason being ECs are constrained structures without a clear morphological identity. Hence in this work, we do not claim to measure their sizes.

The current clump and EC models already in the galaxy before the injection stay fixed, since the corresponding structures are still present in the image.  We repeat this process 10 times for each clump and EC. Comparing the injected and recovered parameters over the ten realisations provides the distribution of fractional residuals in amplitude and size (and hence flux). The F070W flux and size uncertainties are taken as the standard deviations of these fractional residuals, converted to absolute units.  
 
For the remaining 12 filters, where only the amplitudes are fit while the sizes are held fixed, we follow a similar approach while determining the uncertainties. In each filter, the clumps and ECs are injected at their measured amplitudes in that band. For the clumps, the injected size is perturbed by a value drawn from that clump's own F070W fractional size-uncertainty distribution. However, during the amplitude-only fitting that follows, the model size is held at the original best-fit value.  This discrepancy in size ensures the size uncertainties in the F070W fitting is propagated into the flux error estimations in all other filters.  The ECs are injected and fit with their sizes held fixed throughout, without any perturbation. This procedure yields a per-clump, per-filter flux uncertainty for every clump and EC which will be used in the SED-fitting procedure.
 
Finally, we also estimate the 5$\sigma$, 90\%-completeness limits for each of the filters through a separate point-source injection–recovery analysis.  For a given filter, we injected artificial PSF convolved point sources of known flux (within their effective radius) at random positions into the sky patch.  We then performed a \texttt{photutils}-based detection with a threshold of $5\sigma$. An injected source was considered recovered if a detection was found within a fixed matching radius of the input position. Repeating this procedure over 100 realizations at each of 15 logarithmically spaced flux levels produces a completeness curve for every filter. We then interpolate in log-flux space to determine the flux corresponding to 90\% completeness. These 90\%-completeness limits are adopted as the effective $5\sigma$ upper limits for non-detections in the corresponding filters.
 
\subsection{Spectral energy distribution fitting}
\label{subsec:sed_fitting}
 
\begin{figure*} 
    \centering
    \includegraphics[width=\textwidth]{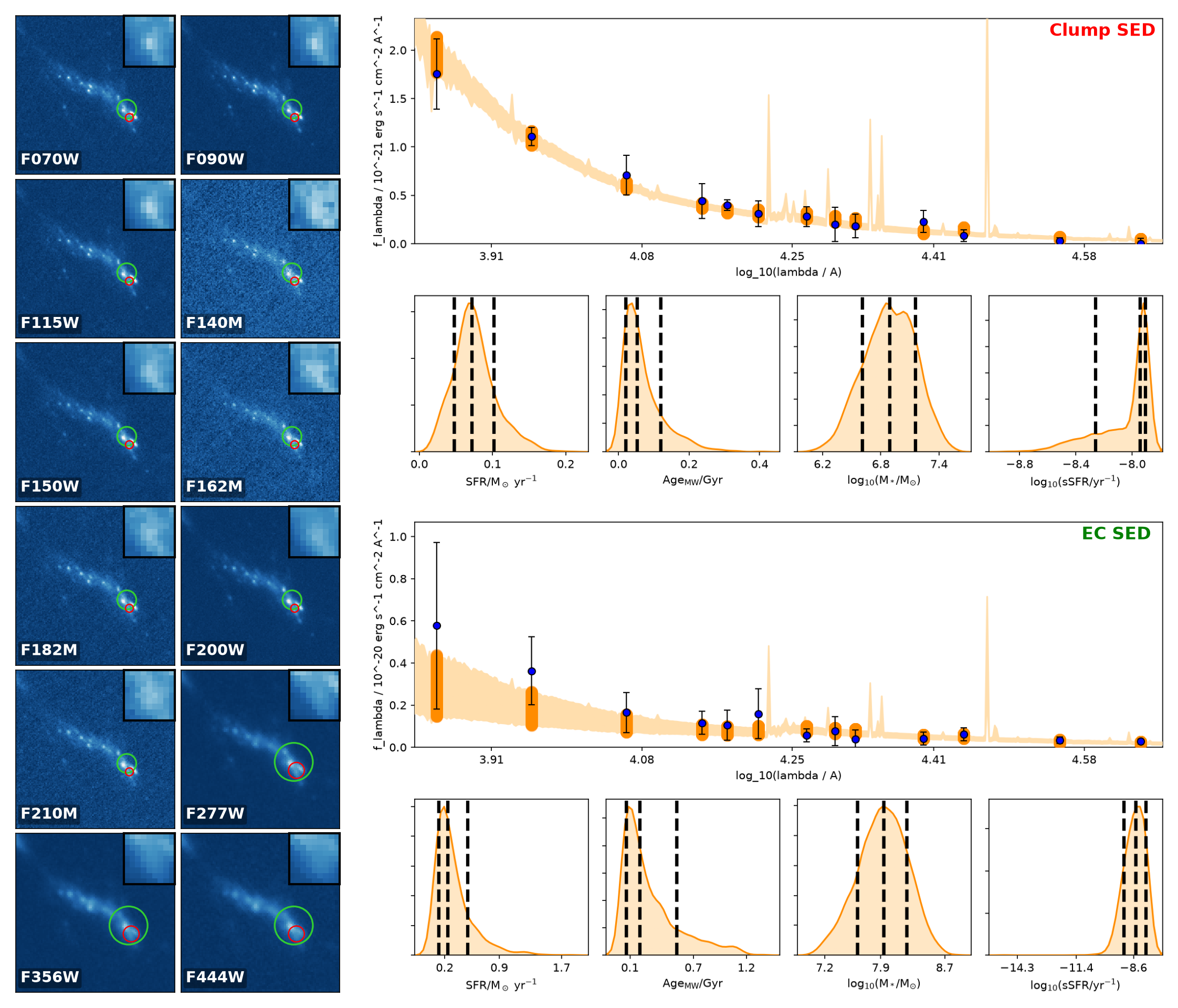}
    \caption{SED fitting results for one of the clumps in galaxy id: 15832, marked in red in each of the per-filter panels showing the corresponding galaxy image. The lower extended panel shows the best-fit SED together with the photometric data points in each filter (blue markers with errors) and the model fluxes at the same effective wavelengths. The four panels in the bottom row show the posterior distributions for the estimated parameters: SFR, mass-weighted age (denoted $t_{\mathrm{age}}$ throughout this work), stellar mass, and specific SFR.}
    \label{fig:sed_collage}
\end{figure*}
 
Using the flux measurements described in the previous sections (Sec.~\ref{subsec:det_flux_fit},\ref{subsec:flux_uncertainties}), we derive the physical properties of the individual lensed clumps by fitting their spectral energy distributions (SEDs) using \texttt{BAGPIPES} \citep{carnall18}, following C25. Example of a clump and EC SED-fitting is shown in Fig.~\ref{fig:sed_collage}.  We will be discussing the analysis for the ECs in the next subsection. For each clump, we construct photometry in 13 JWST/NIRCam bands. Since the clumps lie in the strongly lensed field of Abell 2744, all fluxes and associated uncertainties are corrected for magnification by dividing by the magnification at the location of the clump (Sec.~\ref{subsec:data_sample} and Appendix~\ref{app:lensing}).  For filters where the measured flux is $<$ $0.5\times$ the corresponding error, we treat them as non-detections and set the flux to zero and the error to the maximum of either the sum of the measured flux and the uncertainty, or the $5\sigma$ upper limit derived in the previous section (Sec.~\ref{subsec:flux_uncertainties}). This essentially treats the measurements as upper-limits.

The SED fitting prescription broadly follows C25. Each clump is modelled with an exponentially declining star-formation history (SFH; Eq.~\ref{eq:exp_sfh} below), with stellar ages spanning 0.01–2 Gyr and e-folding timescales $\tau$ between 0.001 and 1 Gyr.  
\begin{equation} \label{eq:exp_sfh}
\mathrm{SFR}(t) \propto e^{-t/\tau}
\end{equation}
Stellar metallicity is left free with a log-uniform prior between $10^{-4}$ and $3\,Z_{\odot}$ (as in C25), and dust attenuation is defined using the \cite{calzetti00} law with $A_V$ in the range $0-2$ mag. Nebular emission is included via Cloudy templates, with the ionization parameter $\log \mathrm{U}$ allowed to vary between $-4$ and $-2$. The redshift is fixed to that of the host galaxy for all galaxies, including the five with photometric redshifts (as in C25). In total, six free parameters (age, $\tau$, formed mass, metallicity, $A_V$, and $\log U$) are fitted independently for each clump, with posterior sampling performed using the nested-sampling implementation in \texttt{BAGPIPES}. From the resulting posteriors, we derive the measured parameters of stellar mass, mass-weighted age and SFR (integrated over $100~\mathrm{Myr}$). The values we shall be quoting are the medians with 16th–84th percentile uncertainties. We propagate the magnification uncertainty into these posteriors. Since lensing is achromatic, $\mu$ only rescales the SED: the stellar mass and SFR scale as $\mu^{-1}$, while the age, dust attenuation, metallicity and sSFR are unaffected. For every sample of a clump's posterior, we therefore draw a magnification from that clump's uncertainty distribution (Appendix~\ref{app:lensing}) and rescale the sample's stellar mass and SFR by the ratio of the best-fit magnification to the drawn value. The quoted uncertainties hence include both the photometric and the lens-model terms. We implement the same procedure for the ECs.  

In some cases, a clump is not detected in a fraction of the filters.In such cases, the photometry becomes limited primarily to the rest-frame UV wavelengths, thereby losing constraints on the stellar mass (due to the lack of rest-frame optical and near-IR photometry), as well as on the age-sensitive $4000$\,\AA\ break for galaxies at $z \gtrsim 2.5$. In Appendix~\ref{app:min_filters}, we empirically verify that six filters is the minimum number needed for robust stellar-mass estimation, which is the key parameter we discuss in this work. Hence, for any clump without a detection in at least six filters, its SED results are discarded ($\sim 43\%$).  Consequently, we do not include that clump in the rest of the analysis.
 
In Appendix~\ref{app:completeness}, we estimate the stellar mass completeness of our clump sample empirically, since the rest-frame UV mass-to-light ratio does not allow an analytic estimate.  We rescale the best-fit SED of each clump to a range of forced stellar masses, re-inject it into the host galaxy image, and measure the detection fraction, giving a $90\%$ completeness limit of $\log(M_\star/\mathrm{M_\odot}) = 7.10^{+0.06}_{-0.06}$. We also check that this limit is not undermined by the minimum 6 filters cut for the SED-fitting: we approximate the stellar mass of the clumps discarded by this requirement by comparing them to the clumps with robust SEDs in a color-color parameter space, and find that the fraction of clumps with robust SEDs first drops below $90\%$ at $\log(M_\star/\mathrm{M_\odot}) = 7.38$, only $0.28\,\mathrm{dex}$ above the completeness limit. The six-band requirement therefore introduces only a modest additional incompleteness, concentrated just above the completeness limit (Appendix~\ref{app:completeness}). Finally, for these clumps we check how their detectability would vary had they been at stellar ages of $0.3-1.1\,$Gyr: we use the best-fit SED of each clump to predict how it would fade and redden as its stellar population ages, inject these aged clumps into the F070W imaging, and measure the recovery fraction as a function of stellar mass. We find that the $90\%$ completeness limit of $\log(M_\star/\mathrm{M_\odot}) = 7.10$ applies to the clumps at their observed ages (median $0.28\,\mathrm{Gyr}$), and rises to $7.46$, $7.77$, $7.99$, $8.18$ and $8.33$ if they were aged to $0.3$, $0.5$, $0.7$, $0.9$ and $1.1\,$Gyr, respectively. The definition of the stellar age is shown in Eq.~\ref{eq:mass_weighted_age}, later in this paper.

\subsection{Properties of the extended components}
\label{subsec:extended_clump_masses}

We apply the same SED fitting procedure described in Sec.~\ref{subsec:sed_fitting} to the ECs in our sample, with their flux uncertainties estimated through the injection-recovery procedure outlined in Sec.~\ref{subsec:flux_uncertainties}. Given the scale-specific nature of the EC (Sec.~\ref{subsec:det_flux_fit}),  we reiterate that SED-based stellar mass (the only measurement we will be quoting here) is a conditional background measurement without a clear physical definition. We find that for ECs, the non-detections are not necessarily in the longest wavelength bands (as is the case for the clumps). We therefore determine the minimum number of filters required for a robust stellar-mass estimate by dropping filters randomly (in Appendix~\ref{app:min_filters}), and find it to be 4. This removes $11\%$ of our sample. The use of the EC stellar masses is deferred to SCALES-II \citep{kalita26}.
 
\subsection{Clump surface densities and redshift bias}
\label{subsec:surface_densities}
 
The stellar-mass surface densities ($\Sigma_{M_\star}=M_\star/\pi r_{\rm eff}^2$) and SFR surface densities ($\Sigma_{\rm SFR}=\mathrm{SFR}/\pi r_{\rm eff}^2$) are measured for each clump in our sample using the SED-based measurements as well as the F070W sizes ($r_{\rm eff}$). Additionally, the integrated stellar mass and SFR and the projected area are magnified by the same factor. Hence surface densities are, to first order, invariant under lensing magnification, and so the magnification uncertainty cancels in them to first order. The sizes are corrected with the magnification of each clump as $r_{\rm eff}\propto\mu^{-1/2}$.  The uncertainties on $M_\star$  and SFR provided by \texttt{BAGPIPES}, along with the size uncertainties (Sec.~\ref{subsec:flux_uncertainties}), are propagated to the $\Sigma_{\rm M_\star}$ and $\Sigma_{\rm SFR}$ measurements.
 
In addition to the surface density measurements, we also estimate the detectability of each clump as a function of redshift, which serves as a proxy for the redshift-dependent selection bias in our sample. This is particularly important given that we assess the evolution of the surface densities as a function of redshift in Sec.~\ref{subsec:sigma_over_redshift}.  For each such clump, we identify the best-fit model SED, and use it to predict the observed-frame F070W flux if the same stellar population were placed at $z = 3.5$, with all other parameters held fixed. We then inject this predicted flux into the F070W image cutout of the host galaxy at 100 random positions and re-run the detection pipeline (Sec.~\ref{subsec:det_flux_fit}) on each realisation.  During the injection, we ensure that the new clump does not overlap with any already present source within $2\times$ the F070W PSF FWHM. The fraction of these 100 realisations in which the clump is successfully recovered is recorded as the redshifted detectability fraction, $f_{\rm det}$.
 
\begin{figure} 
    \centering
    \includegraphics[width=0.49\textwidth]{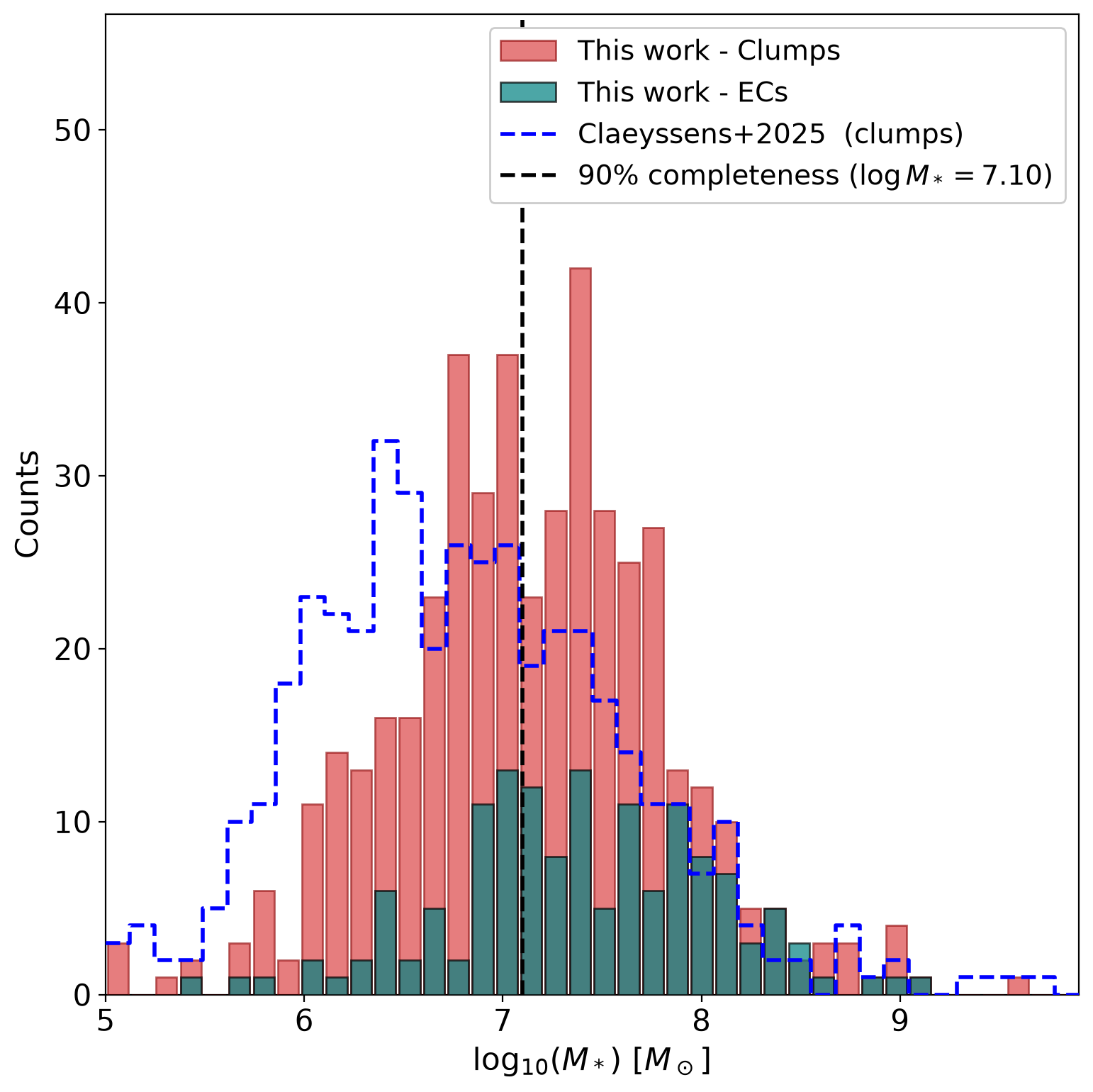}
    \caption{Stellar mass distribution of the clumps (red) and ECs (green) in our sample.  C25 clumps are shown in dark blue. The vertical dashed line marks the 90\% completeness limit (as defined in Sec.~\ref{subsec:sed_fitting}).}
    \label{fig:mass_histogram}
\end{figure}

\begin{figure} 
    \centering
    \includegraphics[width=0.48\textwidth]{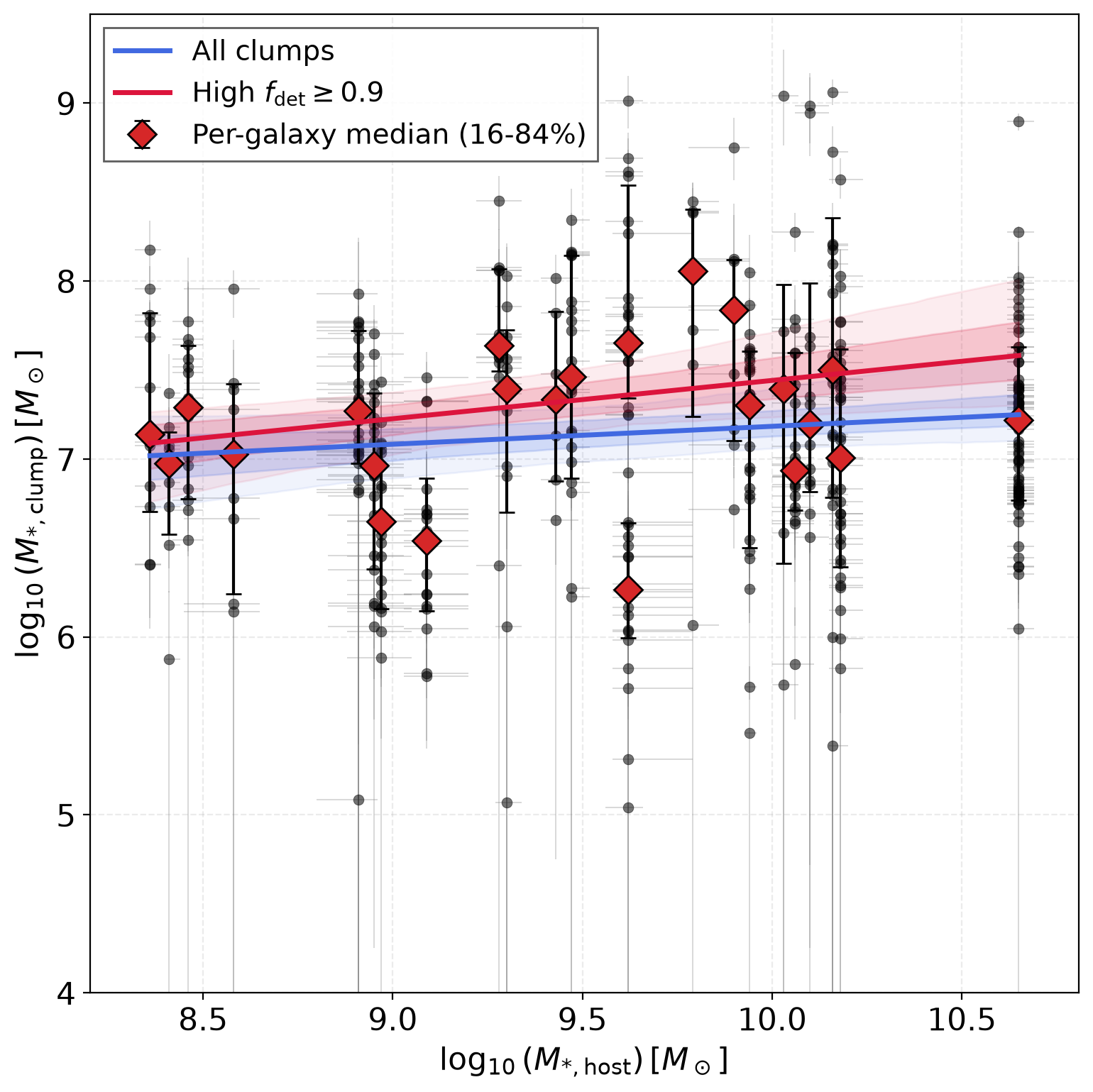}
    \caption{The stellar mass of clumps plotted against their respective host stellar mass. The latter values are based on measurements from C25.}
    \label{fig:clump_mass_host_mass}
\end{figure}

\begin{figure} 
    \centering
    \includegraphics[width=0.49\textwidth]{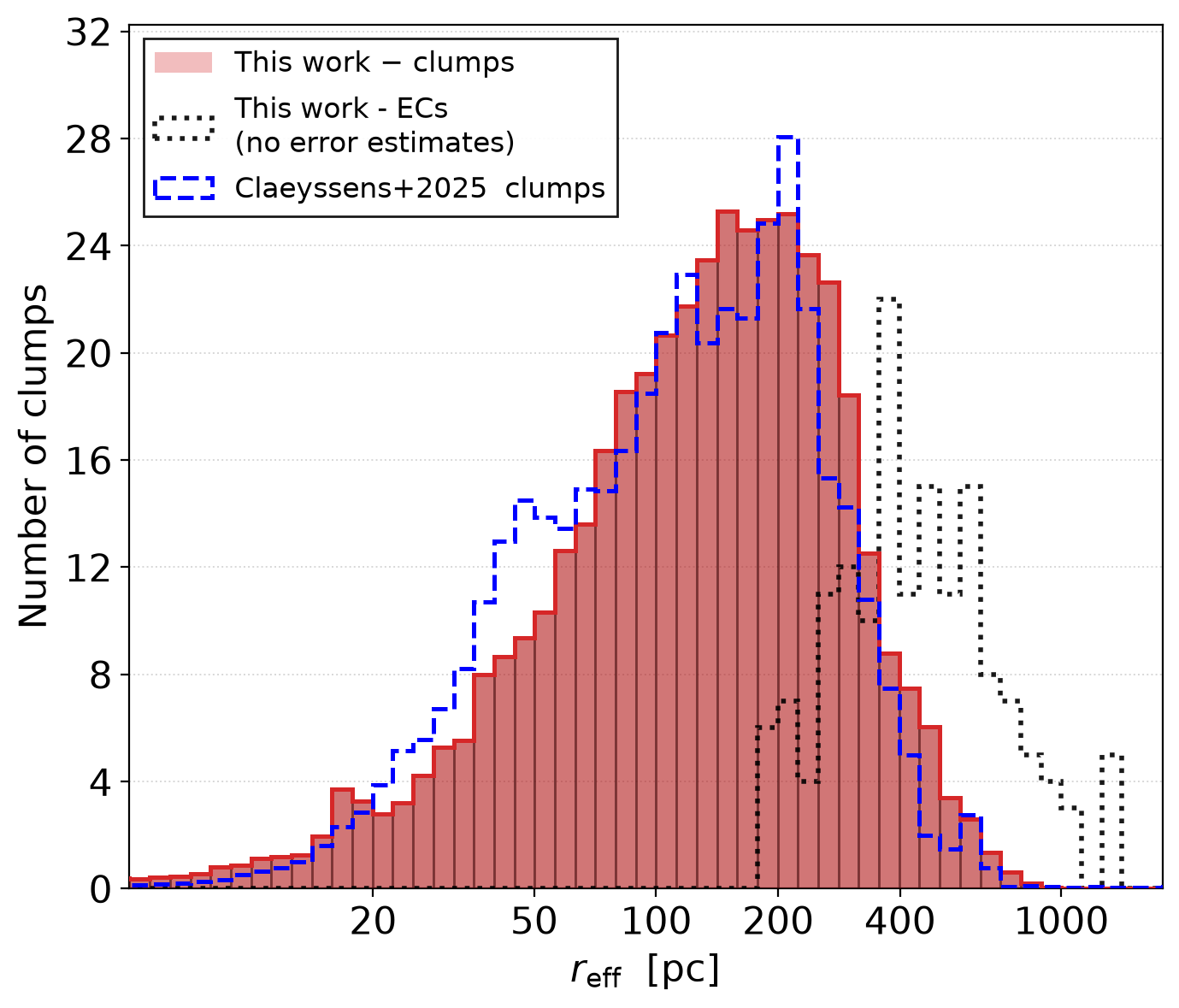}
    \caption{The $r_{\mathrm{eff}}$ distribution for the clumps in our study, as well as those of the C25 clumps.}
    \label{fig:radius_hist}
\end{figure}
 
\begin{figure} 
    \centering
    \includegraphics[width=0.49\textwidth]{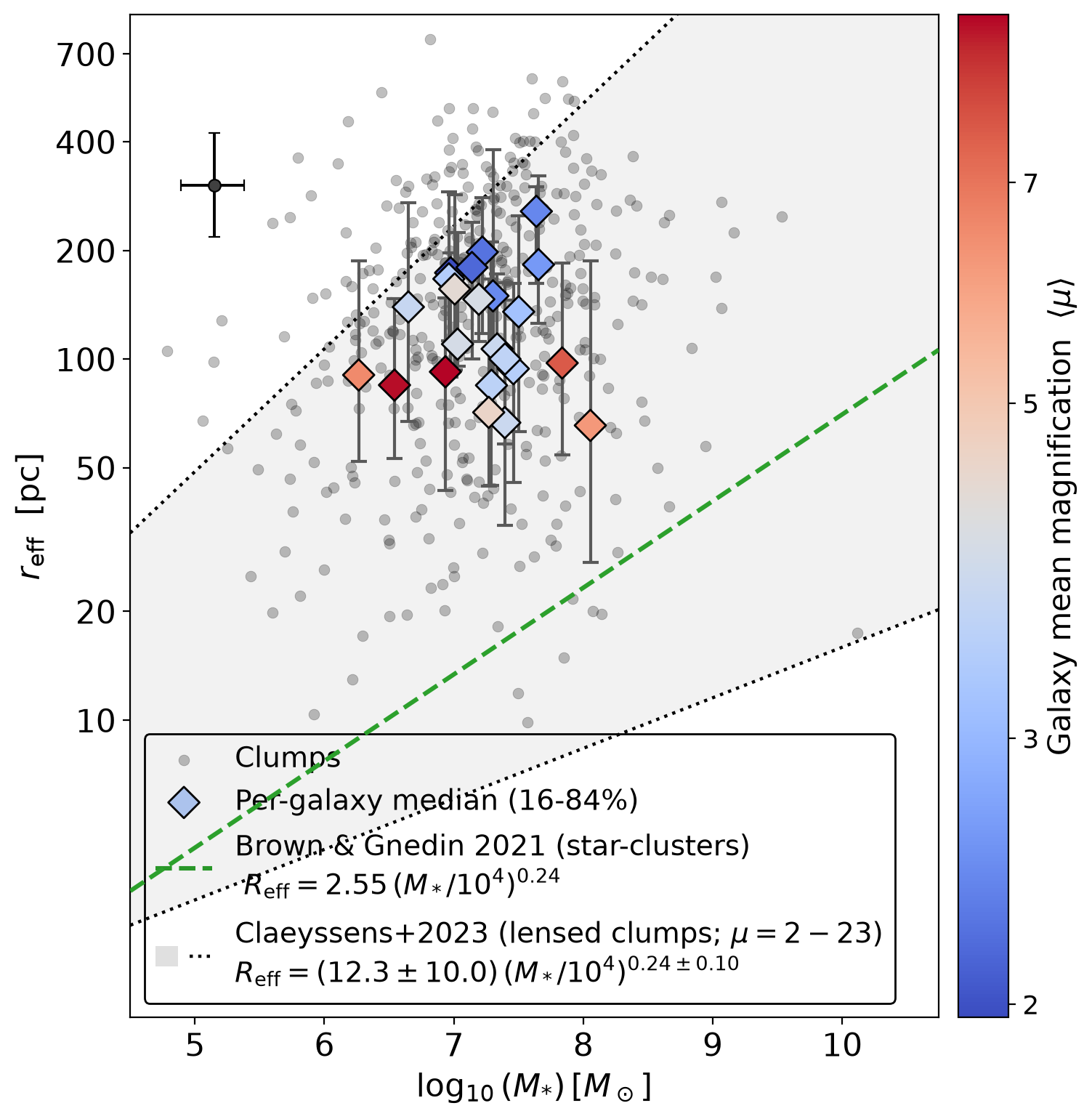}
    \caption{The distribution of clump stellar mass vs. $r_{\mathrm{eff}}$. The red diamonds give the per-galaxy median.}
    \label{fig:radius_mass}
\end{figure}
 
\begin{figure} 
    \centering
    \includegraphics[width=0.49\textwidth]{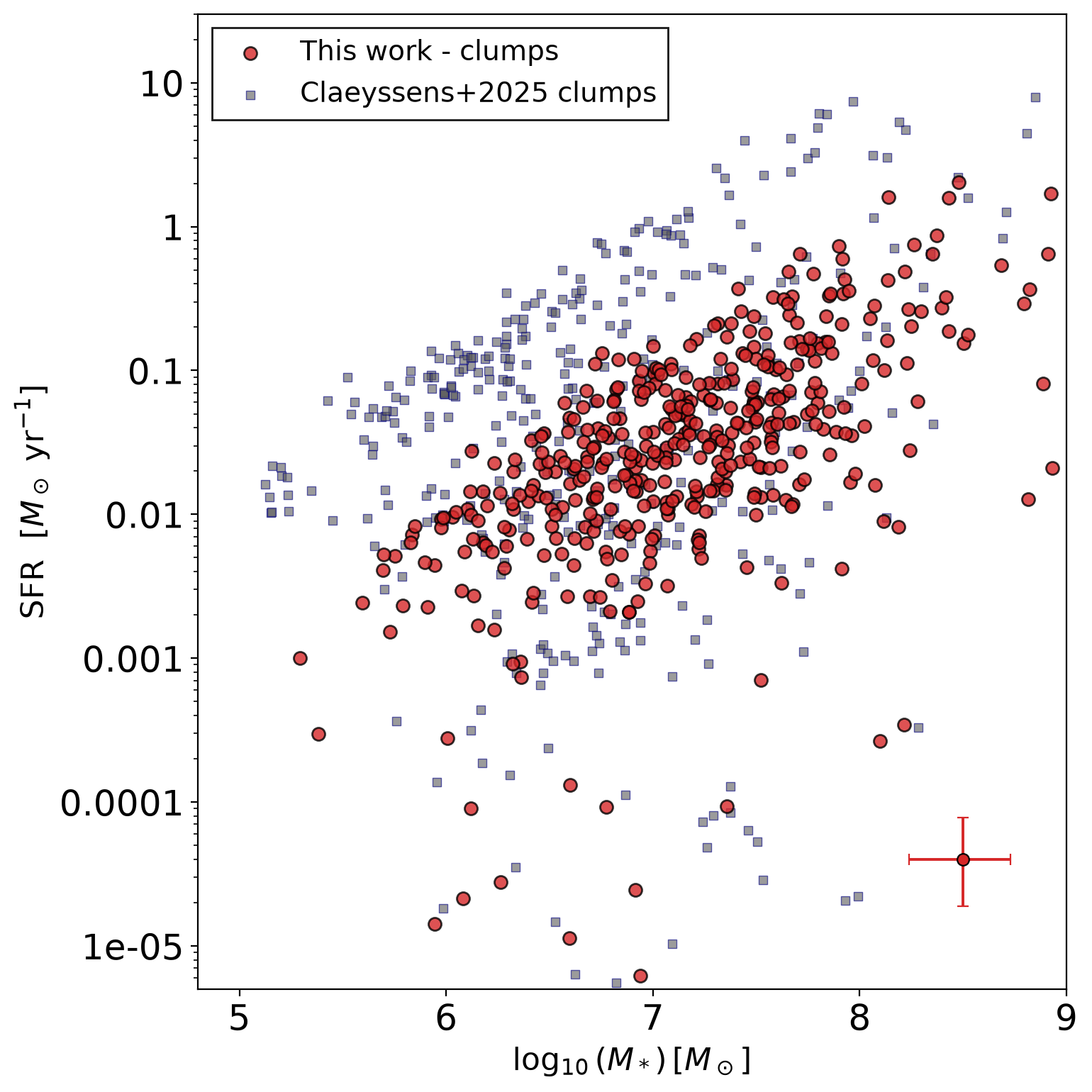}
    \caption{Stellar mass versus SFR for the clumps, with each clump resampled once from its asymmetric posteriors. Direct-SED clumps (red), SED-scaled clumps (open orange), and C25 clumps (navy squares) are shown.}
    \label{fig:sfr_mass}
\end{figure}

\subsection{Uncertainties due to lens model}

In this work we have used the lens model provided by the UNCOVER survey team. The statistical uncertainties of this model are propagated into our measurements through its MCMC realisations (Sec.~\ref{subsec:data_sample}). It needs to be acknowledged however that systematic errors in the model (e.g.\ from the choice of modelling approach) can introduce additional offsets in measured clump properties. Surface densities remain unaffected to first order since the magnification on both parameters of the ratio cancel out. The majority of our results and discussions deal with relative effects within our sample. Hence, one can consider our results to be true for the magnification maps being used. The only case where uncertainty in the lens map is worth discussing is while comparing our measured clump stellar mass, SFR and size to those in C25.  

\section{Results}
\label{sec:results}
 
Our pipeline yields, for each galaxy in the sample, a set of clumps with measurements of stellar mass, SFR, and effective radii ($r_{\rm eff}$), and SED-based ages. We detail the resulting distributions and properties in the following subsections. As noted in Sec.~\ref{subsec:extended_clump_masses}, the ECs are mainly dealt with in SCALES-II \citep{kalita26} and future works, since they require a more dedicated analysis to establish a physical interpretation. In this work, we
only quote their stellar mass and size for reference in a few instances.
 
\subsection{Clump properties}
\label{subsec:mass_prop}

In our sample of 23 galaxies, we have 451 clumps and 156 ECs. Both clumps and ECs span similar mass ranges of $\approx 10^{6}$ and $10^{9}\,\mathrm{M_{\odot}}$. Hence, the ECs individually are a non-negligible fraction of the clump-EC mass budget. Focusing on the clumps, the massive end of our mass distribution agrees closely with that of C25 for the same galaxies (Fig.~\ref{fig:mass_histogram}). The most massive clumps reach $10^{9.6}$ and $10^{9.7}\,\mathrm{M_{\odot}}$ respectively, with four clumps above $10^{9}\,\mathrm{M_{\odot}}$ in each sample. Above our $90\%$ stellar mass completeness limit ($\log(M_\star/\mathrm{M_\odot}) = 7.10^{+0.06}_{-0.06}$), the two distributions are statistically indistinguishable (two-sample Kolmogorov--Smirnov $p\approx0.24$, with a median of $10^{7.6}\,\mathrm{M_{\odot}}$ in both). They differ only below this limit, where a larger fraction of the C25 clumps reside ($66\%$, against $49\%$ of ours). We discuss this comparison further in Sec.~\ref{subsubsec:detection_comparison}.

In Fig.~\ref{fig:clump_mass_host_mass}, we show the clump stellar mass as a function of host stellar mass. We find no significant evolution of the clump-to-host mass ratio across our sample, both in the full clump sample and the high-detectability ($f_{\rm det} \geq 0.9$) subset. This indicates that individual clump masses are largely decoupled from their host galaxy's stellar mass. The sizes ($\mathrm{r_{\rm eff}}$) measured in our procedure for the clumps cover a range of $\sim 11\,\mathrm{pc} - 650\,\mathrm{pc}$. In Fig.~\ref{fig:radius_hist}, we show the distribution of sizes in our sample, while incorporating the measurement uncertainties by treating each clump as a Gaussian distribution of 100 samples whose width is set by the uncertainty range in $r_{\rm eff}$ (estimated in Sec.~\ref{subsec:flux_uncertainties}).  We find the sizes to be in agreement with the C25 sample, but do not observe any correlation with stellar mass (Fig.~\ref{fig:radius_mass}). However, given the uncertainties in the sizes (from tens of parsec to $\sim 100\,\mathrm{pc}$), we cannot rule out the existence of a weak correlation. However, the points are in agreement with the tentative relation found in \cite{claeyssens23},  albeit, the scatter in their relation reflects the weak correlation, if any, expected between the clump mass and size. It is worth noting however that this assessment should not be confused with the mass-size relation found in giant clumps \citep{kalita25b}. The connection between \emph{giant} clumps and the clumps we refer to in this work is studied in SCALES-II \citep{kalita26}. Finally, the clumps in our sample are unlikely to be star-clusters, evidenced by their disagreement with the \cite{brown21} relation, also shown in the plot. 
 
Using the SED-based SFRs, we find the clumps distributed within a large range: $0.7\,\mathrm{M_{\odot}/yr}$ to $10^{-4}\,\mathrm{M_{\odot}/yr}$.  We then plot the SFRs as a function of stellar mass in Fig.~\ref{fig:sfr_mass}. To avoid overcrowding the figure with error bars, each data point is drawn from an asymmetric Gaussian distribution centred on the measured value, with the lower and upper widths set by the respective $1\sigma$ uncertainties of each parameter. This way, the scatter in the plot reflects the measurement uncertainties. We observe a clear correlation with the stellar mass, in principle similar to that shown by galaxies \citep[the star-forming main sequence or SFMS;][]{speagle14, santini17}.  However, a direct comparison should not be made, since the SFMS was established and calibrated on the integrated properties of whole galaxies \citep{brinchmann04, noeske07, speagle14, whitaker14}, and cannot be directly applied to sub-galactic scales.  We do note an increased density along a thin range within the scatter. These are the clumps with ages short enough (discussed below) to be treated as single stellar populations, for which the SFR is a function of the stellar mass by design. However, these are not solely responsible for the observed correlation.

In Fig.~\ref{fig:sfr_mass}, we also show the distribution of the C25 clumps within our sample of galaxies for reference.  We only highlight the overall consistency of the distributions, since the slightly larger scatter in the C25 values is likely driven by their SFR being integrated over $10\,\mathrm{Myr}$ rather than $100\,\mathrm{Myr}$. We adopt the longer timescale because, in the absence of spectroscopic emission-line measurements (which trace star formation on $\sim5-10\,\mathrm{Myr}$ timescales), the rest-frame UV continuum available to us traces star formation averaged over longer timescales of up to $\sim100\,\mathrm{Myr}$ \citep{floresvelazquez21}, so we cannot reliably resolve shorter-timescale variability.

\begin{figure} 
    \centering
    \includegraphics[width=0.49\textwidth]{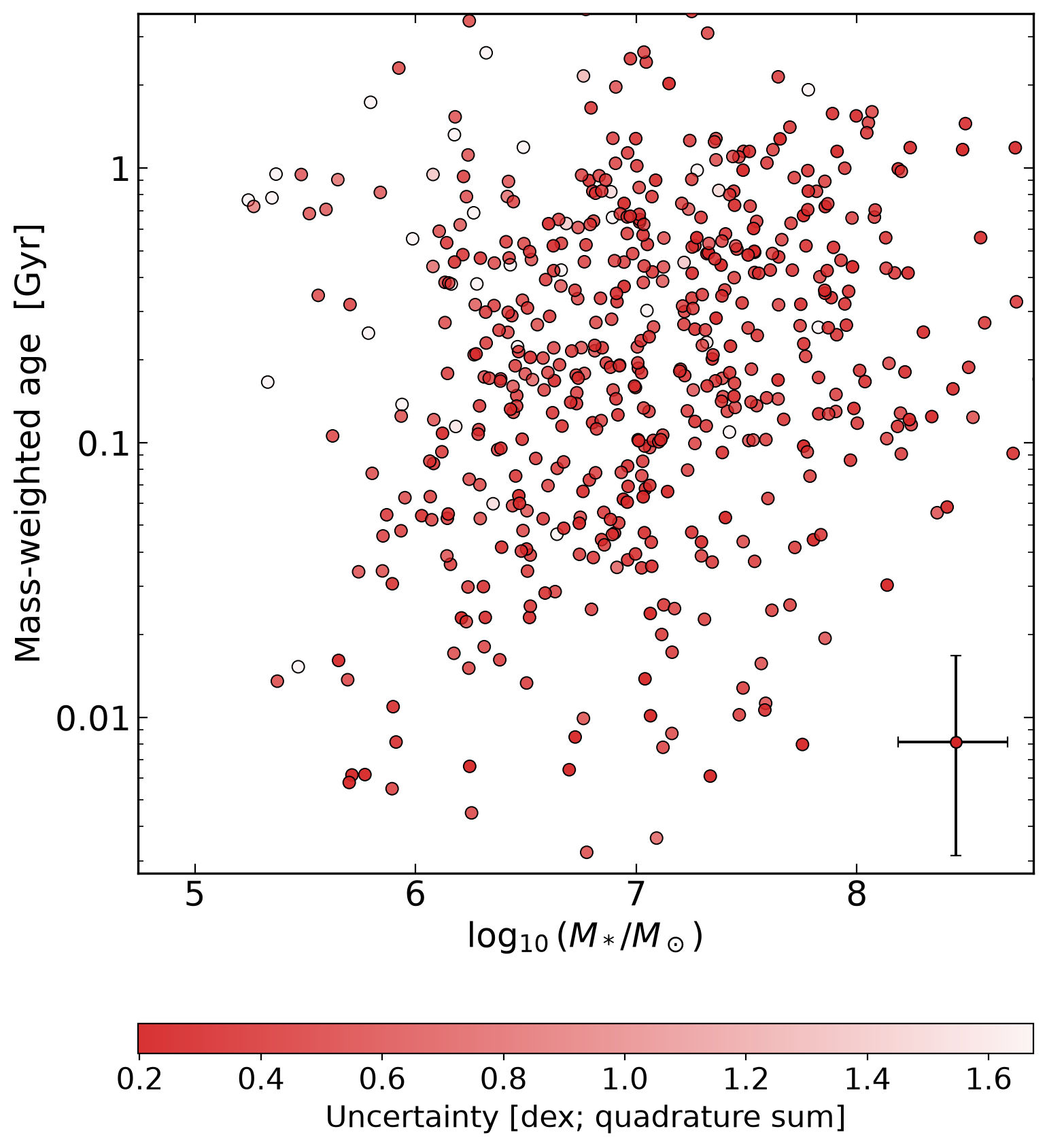}
    \caption{Clump stellar mass against mass-weighted age. Each clump is treated as 100 points drawn  from its asymmetric mass and age posteriors.}
    \label{fig:age_mass_distribution}
\end{figure}
 
The final key parameter we derive from the SED fitting of the clumps is the mass-weighted age ($t_{\mathrm{age}}$). This is determined by \texttt{BAGPIPES}, using the formula below \citep[Eq.~11 in][]{carnall18}:
\begin{equation} \label{eq:mass_weighted_age}
t_{\mathrm{age}} = \frac{\int^{t_{\mathrm{obs}}}_{0}t\,\mathrm{SFR}(t)}{\int^{t_{\mathrm{obs}}}_{0}\mathrm{SFR}(t)}
\end{equation}
The $t_{\mathrm{obs}}$ is the time elapsed from the beginning of the Universe to the source redshift, while $\mathrm{SFR}(t)$ is dictated by the exponentially declining star-formation history we use (Eq.~\ref{eq:exp_sfh}). The $\tau$ in this equation is the actual measured parameter, besides the source redshift, which is fixed to the host galaxy value during our procedure.  We find the $t_{\mathrm{age}}$ for most clumps are within the wide range of $80-640\,\mathrm{Myr}$, although those with the highest ages are also the most uncertain (errors $\sim 480\,\mathrm{Myr}$ for median ages $\sim 710\,\mathrm{Myr}$).  We do not detect any clear correlation between the stellar mass and age of the clumps (Fig.~\ref{fig:age_mass_distribution}). Had this been detected, one could assume that the clumps are in a star-forming equilibrium state. However, we can rule out any strong evidence of such behaviour. The lack of such an age-mass correlation could point to the clumps not being single monolithic units, as will be discussed later in Sec.~\ref{subsec:survival_substructure}.

 
\subsection{Stellar mass function}
 
\begin{figure*} 
    \centering
    \includegraphics[width=\textwidth]{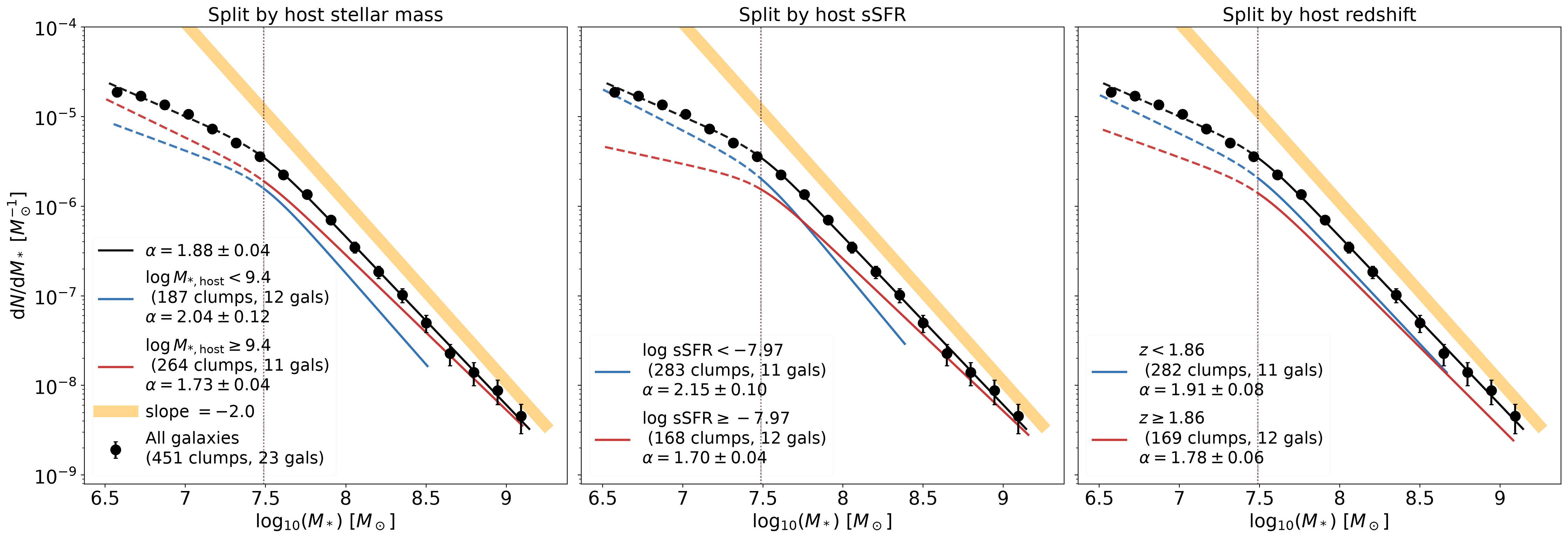}
    \caption{Clump stellar mass function, split three different ways by host-galaxy properties. In each panel, black points and the fitted relation are for the full sample, with the turnover mass (log $M_{\mathrm{b}}$) left free, and the orange band shows the fiducial slope $=2$ expected for scale-free star-forming regions. The blue and red lines show fits to two subsamples, with $M_{\mathrm{b}}$ fixed to the full-sample value: (Left) split by host stellar mass at $\log M_{*,\mathrm{host}} = 9.4$; (Middle) split by host sSFR at $\log(\mathrm{sSFR}) = -7.97\,\mathrm{yr^{-1}}$; (Right) split by host redshift at $z = 1.86$. The number of clumps and galaxies in each subsample, and the corresponding measured high-mass slope $\alpha$, are given in each panel's legend.}
    \label{fig:clump_smf}
\end{figure*}
 
Using the clump mass distribution, we construct a clump stellar mass function (CSMF) by plotting the gradient of clump count ($\mathrm{dN/dM}$) as a function of clump stellar mass ($M$).  Given that the clump mass measurement includes a $68\%$ percentile upper and lower limit, we treat each clump as a Gaussian distribution of 100 points with the same limits. This treatment allows us to account for the uncertainty in the mass estimates.  As shown in Fig.~\ref{fig:clump_smf} we then fit a broken power law of the functional form below:
\begin{equation} \label{eq:csmf_full}
\frac{dN}{dM} = A \left(\frac{M}{M_b}\right)^{-\beta} \left[ 1 + \left(\frac{M}{M_b}\right)^{1/\delta} \right]^{(\beta - \alpha)\,\delta}
\end{equation}
Here,  $M_b$ is the stellar mass at which the break appears, and $\delta$ sets its sharpness: the logarithmic slope changes from $\beta$ to $\alpha$ over $\log(M/M_b)\approx\pm\delta$. Since the CSMF is constructed by replacing each clump with 100 points drawn from within its stellar mass uncertainty, even an intrinsically sharp break would appear smoothed by the mass uncertainty. We therefore fix $\delta$ to $0.24$, the median stellar mass uncertainty (in dex) of the clumps in the range $10^{7.0-7.6}\,\mathrm{M_{\odot}}$, which we visually find to be where the break lies. On either side of $M_{b}$, eq.~\ref{eq:csmf_full} reduces to:
\begin{equation}
\frac{dN}{dM} \propto M^{-\beta} \quad (M \ll M_b)
\end{equation}
\vspace{-0.8em}
\begin{equation} \label{eq:csmf_high_mass}
\frac{dN}{dM} \propto M^{-\alpha} \quad (M \gg M_b)
\end{equation}
 
We find $M_b = 10^{7.49 \pm 0.04} \,\mathrm{M_{\odot}}$ (Fig.~\ref{fig:clump_smf}) with the conclusion that the mass range below this range is likely affected by a combination of detection incompleteness (below the limit of $\sim 10^{7.10}\,M_{\odot}$) or a physical turnover.  Meanwhile, above this stellar mass range, we expect the clumps to be complete at least up to a stellar age of $300\,\mathrm{Myr}$ (Fig.~\ref{fig:aged_detectability}).  Above $M_b$ (eq.~\ref{eq:csmf_high_mass}), we measure a slope $\alpha = 1.88 \pm 0.04$, while below it the slope is $\beta = 0.78 \pm 0.05$. We also visually identify clumps coincident with a possible central bulge (7 clumps in galaxies 7775, 15439, 20054 and 22864, four of which are the most massive clumps in their hosts). Excluding them steepens the slope to $\alpha = 1.94 \pm 0.05$ (with $M_b = 10^{7.53 \pm 0.05}\,M_{\odot}$), a change within $\approx1\sigma$ that does not affect the conclusions of this work.
 
We further split the sample into two bins by host stellar mass, sSFR and redshift (Fig.~\ref{fig:clump_smf}). Clumps in hosts with lower stellar mass ($<10^{9.4}\,M_{\odot}$) and lower sSFR ($\log(\mathrm{sSFR}/\mathrm{yr^{-1}})<-8$) show steeper slopes ($2.04$ vs $1.73$, and $2.15$ vs $1.70$), while the two redshift bins are consistent ($1.91$ vs $1.78$). The mass and sSFR splits are largely independent, with only 7 of the 23 galaxies falling on the same side of both. However, only the sSFR offset is significant ($4.2\sigma$, compared to $2.5\sigma$ for host mass and $1.3\sigma$ for redshift). Given that these offsets are based on only $11-12$ galaxies per bin, we consider it tentative.
 
 
\begin{figure*} 
    \centering
    \includegraphics[width=0.49\textwidth]{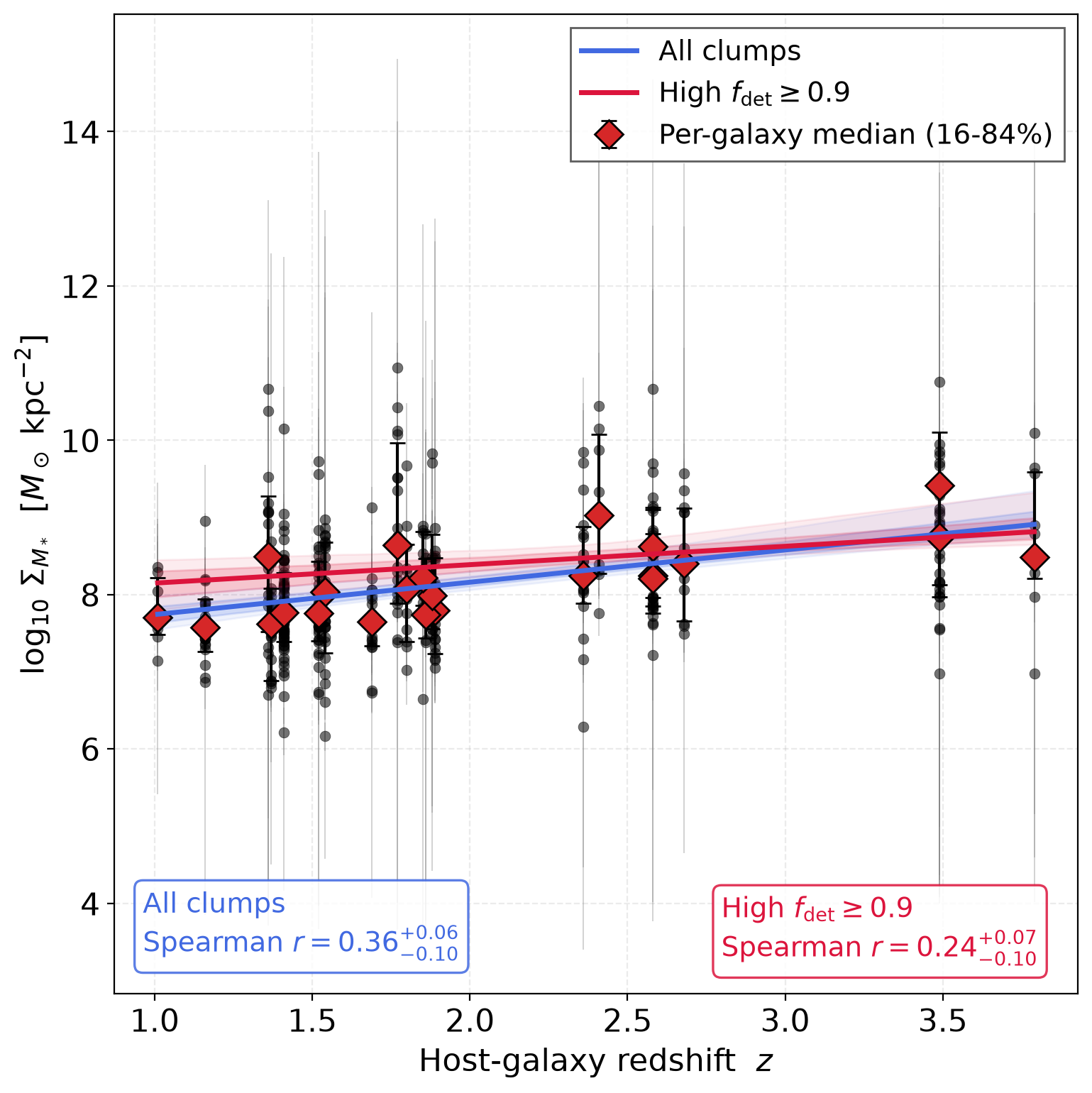}
    \includegraphics[width=0.49\textwidth]{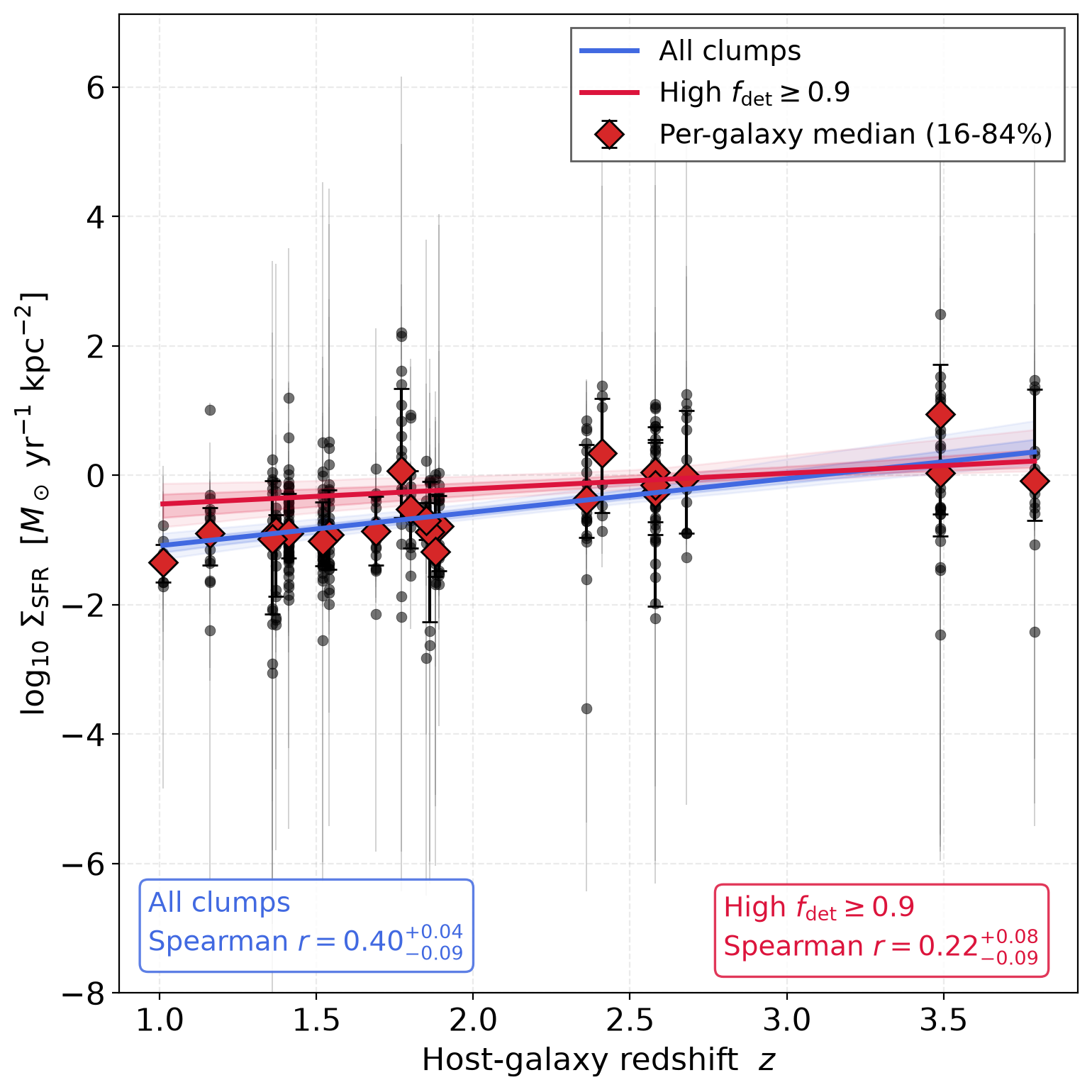}
    \includegraphics[width=0.49\textwidth]{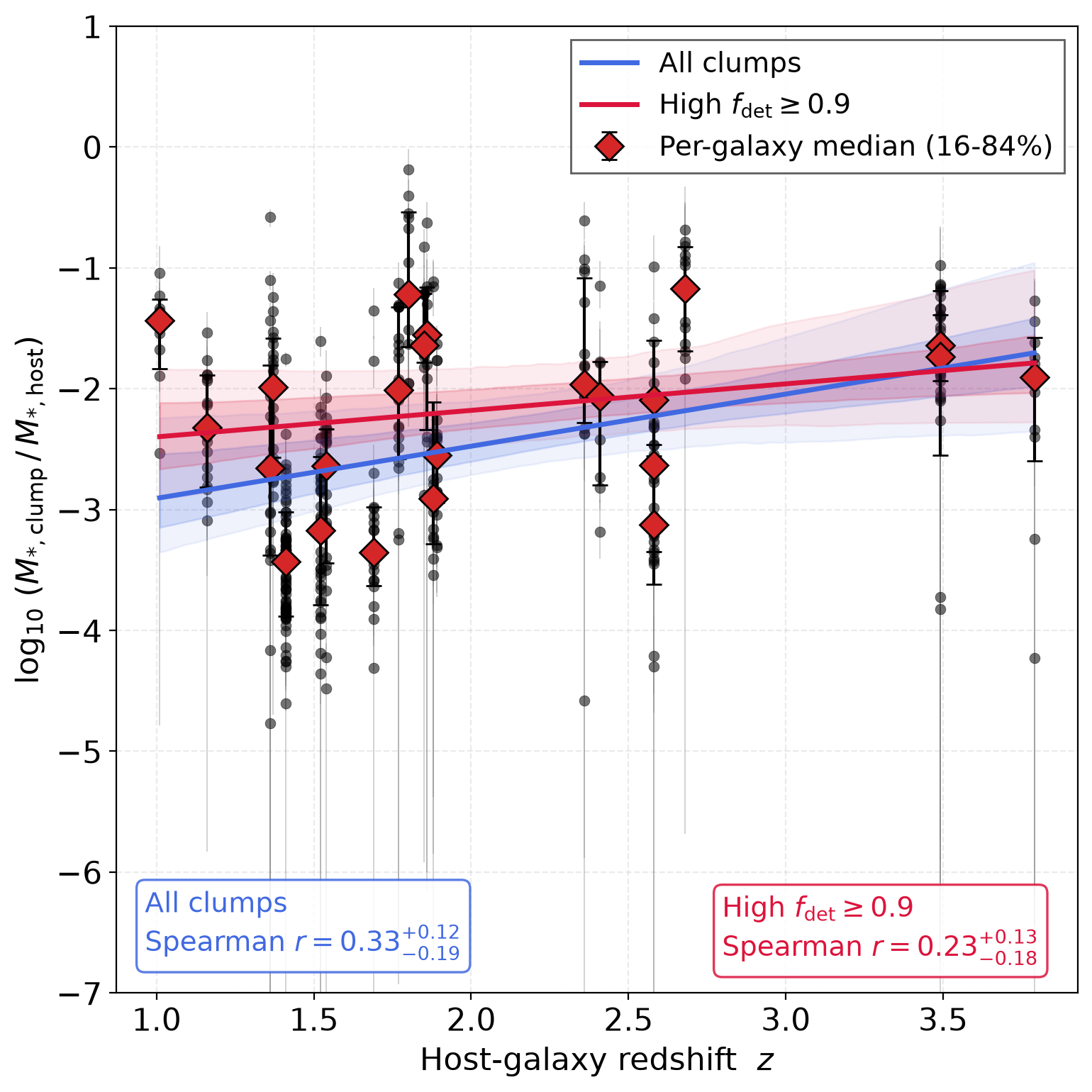}
    \caption{Clump surface densities vs. host redshift. (Left) The stellar-mass surface density $(\Sigma_{M_*})$ and (right) the SFR surface density $(\Sigma_{\mathrm{SFR}})$. The marker opacity scales with the redshifted detectability fraction $f_{\mathrm{det}}$ (Sec.~\ref{subsec:sed_fitting}).Red diamonds are per-galaxy medians ($16$--$84\%$) for the clumps with $\mathrm{f_{det}} > 0.9$. Each panel shows two
galaxy-level linear fits with the functional forms: $\log_{10}\Sigma = \log A + Bz$. The solid line is the best fit for all clumps, while that for the high-detectability subset ($f_{\mathrm{det}} \geq 0.9$) is shown with a dashed line. Spearman rank correlation coefficients ($\rho$) for both sets are given in the inset box.}
    \label{fig:sigma_z}
\end{figure*}
 
\subsection{Surface density across redshift}
\label{subsec:sigma_over_redshift}
The stellar mass surface densities span $\Sigma_{M_*} \approx$ $10^{7.1}$--$10^{10.9}$$\,\mathrm{M_\odot\,\mathrm{kpc}^{-2}}$, while the SFR surface densities cover $\Sigma_{\rm SFR} \approx$ $10^{-1.6}$--$10^{2.5}$$\,\mathrm{M_\odot\,yr^{-1}\,\mathrm{kpc}^{-2}}$. We note that these ranges correspond to the clumps with $f_{\rm det} \geq 0.9$ (Sec.~\ref{subsec:surface_densities}). Including the remaining clumps extends the lower limits to $10^{6.2}$ and $10^{-2.9}$, respectively, i.e.\ by $\approx1\,\mathrm{dex}$. These are however only detectable at the lower redshifts in our sample.
 
In Fig.~\ref{fig:sigma_z}, we plot clump $\Sigma_{M_*}$ (top left) and  $\Sigma_{\rm SFR}$ (top right) as a function of the host redshift. If we take the full sample, we observe an intermediate-to-strong correlation for both $\Sigma_{M_*}$ (Spearman $r = 0.36^{+0.06}_{-0.10}$) and $\Sigma_{\rm SFR}$ ($r = 0.40^{+0.05}_{-0.09}$). However, in both cases, the correlations significantly weaken when considering clumps with $f_{\rm det} > 0.9$, to $r = 0.24^{+0.07}_{-0.10}$ and $0.22^{+0.08}_{-0.09}$ respectively.  A similar correction is also observed for the ratio of clump (stellar) mass and host mass, where the correlation weakens from $r = 0.33^{+0.12}_{-0.19}$ to $r = 0.23^{+0.13}_{-0.18}$.
 
The weakening of the correlations when we enforce this completeness suggests that any empirical trends of increasing density with redshift are inflated by a redshift-dependent selection bias. At higher redshifts, only the most intrinsically luminous and compact clumps are detectable, which naturally produces an apparent correlation between surface density and redshift without necessarily requiring a physical evolution. Nevertheless,  this does not rule out a true correlation, which one could speculate to be indicative of increasing star-forming gas densities \citep[as is observed when comparing resolved gas clouds at $z \sim 0$ to those at $z \sim 1$;][]{dessaugeszavadsky23}. However, with the current depths of JWST/NIRCam datasets, one cannot make any conclusive claims.

\section{Discussions}
\label{sec:discussions}
 
The key focus of our work has been to accurately model the star-forming clumps, while accounting for extended emission around them (ECs). To do so, we have presented the newly developed multi-scale DoG methodology which is efficient at separating the flux at different scales,  necessary to properly model the hierarchical nature of star forming structures.  We have primarily measured the stellar mass,  SFR, sizes and mass-weighted ages of 451 clumps. This section will discuss the physical implications of these measurements. 
 
 
\subsection{Clumps in our analysis}
\label{subsec:clump_prop_discussion}
Our methodology has provided a few revisions regarding the nature of lensed clumps at $z>1$, in comparison to the parent C25 study, and to the wider literature.

\subsubsection{A comparison of clump selection}
\label{subsubsec:detection_comparison}

\begin{figure*}[t]
    \centering
    \includegraphics[width=\linewidth]{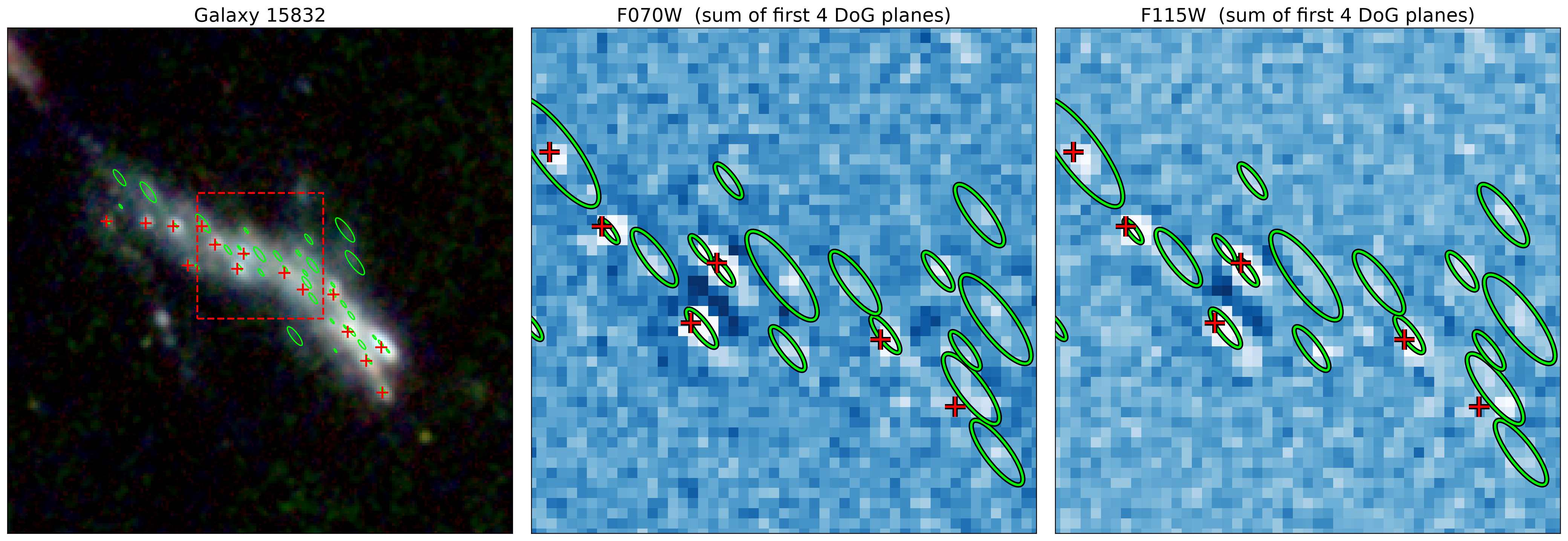}
    \caption{Comparison of clump selection between C25 and this work, for galaxy id: 15832. (Top) RGB image (as in Fig.~\ref{fig:galaxy_collage}) with C25 clump positions marked as red crosses and the clumps detected in this work marked as green ellipses (all clumps detected inside the F444W host segment), with size and orientation set by our best-fit structural parameters. The red dashed box marks the region shown in the zoomed panels. (Middle, Bottom) Zoomed F070W and F115W cutouts showing the sum of the first four DoG planes (i.e.\ emission on the finest detection scales), with the same clumps overlaid. Several individual C25 clumps resolve into two or more distinct, more compact clumps in our higher-resolution detection.}
    \label{fig:ours_vs_claeyssens}
\end{figure*}

As noted in Sec.~\ref{subsec:mass_prop}, the stellar masses of our clumps agree with those of C25 above our completeness limit, even though the two analyses differ in both the lens model and the methodology. The lens model affects the masses directly through $\mu$. We compared the median per-clump $\mu$ for each galaxy in our sample, derived from the UNCOVER maps (Sec.~\ref{subsec:data_sample}), to the mean magnification of the C25 clumps in the same galaxy. Our magnifications are higher in 21 of the 23 galaxies, by $\approx17\%$ for the median galaxy and by at most a factor of $\approx2.0$ (galaxy 19708). For individual galaxies, this can change the mass of the most massive clump substantially: in 19708, our most massive clump has $10^{8.75}\,\mathrm{M_{\odot}}$ against $10^{9.65}\,\mathrm{M_{\odot}}$ in C25, but would have $10^{9.05}\,\mathrm{M_{\odot}}$ with the C25 magnification. Across the sample, however, the different lens model lowers our masses by only by a median value of $\approx0.08\,\mathrm{dex}$.

The methodologies, however, differ substantially. While C25 detects clumps primarily in F150W for the majority of our redshift range, we opt for F070W to increase resolvability, with the PSF FWHM of the latter being smaller by a factor of $\sim 2$. Furthermore, C25 detects the clumps using Source Extractor based detection on the raw image, while we use the DoG-plane based approach. Fig.~\ref{fig:ours_vs_claeyssens} illustrates this qualitatively for one representative galaxy (id: 15832). Several C25 clumps (red crosses), particularly along the densely star-forming central regions of the disk, correspond to regions that resolve into two or more distinct, more compact clumps (green ellipses). Moreover, C25 models a background around each clump within a cutout size of $9\times9$ pixels (or $13\times13$ pixels in the case of clump pairs), while we instead absorb the underlying unresolved, undetected, and diffuse flux into the EC component. Finally, we fit all detected clumps simultaneously for each galaxy, while C25 uses a procedure where individual clumps or clump pairs are independently fit. 

Each of the differences listed above, combined with the influence of the different magnification values, result in a slightly different clump population, as is evident in Fig.~\ref{fig:mass_histogram}. Although the maximum mass of clumps recovered remains approximately similar, we note that the CSMF slope recovered using the C25 clumps in the same galaxies would be shallower as a result of less clumps in the $10^{6.5-7.5}\,\mathrm{M_{\odot}}$ range compared to our sample. However, given the extremely different prescriptions of error estimations in their work, we do not quote the shallower value.

\subsubsection{A scale-free CSMF at $\sim100\,\mathrm{pc}$ and $\sim\mathrm{kpc}$ scales}
\label{subsubsec:dyn_range_csmf}

Above $M_b$, we measure a CSMF slope of $1.88 \pm 0.04$ (Fig.~\ref{fig:clump_smf}). The commonly cited slope of $\approx 2$ reflects the absence of a characteristic mass in hierarchical star-forming structures: fragmentation proceeds self-similarly from large to small scales, with equal number of structures being present within each logarithmic mass bin \citep{elmegreen96, elmegreen02}. This slope is recovered in observations of young star cluster and giant molecular cloud (GMC) mass functions in the local Universe \citep[e.g.][]{zhang99, gieles06a, gieles06b, dowell08, larsen09, adamo17}. However, the reported values vary by $\sim \pm 0.2$ for star clusters \citep{krumholz19}, and over a wider range for GMCs \citep{rosolowsky05}, reflecting both observational systematics and genuine physical departures from a slope of 2. Nevertheless, assuming a fiducial slope as $2 \pm 0.2$, our measured slope at the high mass end (above the break at $10^{7.49}\,\mathrm{M_{\odot}}$) is consistent with this scale-free picture, as are those of previous lensed studies \citep[][C25]{dessaugeszavadsky18, claeyssens23, claeyssens26}. Furthermore, the shallower slope below the break can likely be attributed to incompleteness, as will be further investigated in SCALES-II \citep{kalita26}.

It is worth noting that our CSMF slope is in perfect agreement with the slope of $1.85 \pm 0.19$ measured by \cite{kalita25b} for $\sim0.1-1\,\mathrm{kpc}$ clumps in unlensed massive ($>10^{10.5}\,M_{\odot}$) main-sequence galaxies at $z\approx1.5$. Their fit covers clumps above $\approx10^{8.7}\,M_{\odot}$, more than an order of magnitude above our $M_b$, and structures at the scale of our giant clumps (clump+EC) rather than of our $\approx 50-400\,\mathrm{pc}$ clumps. The $\approx 50-400\,\mathrm{pc}$ clumps above $M_b$ and the kpc-scale structures detected in unlensed galaxies \citep[see also][]{kalita25a, zhu26} therefore follow the same scale-free distribution. The slope alone, however, cannot tell whether these kpc-scale structures are random blends of the smaller clumps, as suggested by simulations and lensed observations \citep{tamburello15, tamburello17, dessaugeszavadsky17}, or physically associated clump clusters \citep{behrendt16, behrendt19}. In SCALES-II \citep{kalita26}, we also find that the composition of the giant clumps (clumps+ECs) is consistent with both scenarios, and we will address their nature in an upcoming paper of the SCALES series.


It is worth noting that the blending of unresolved $\approx 50-400\,\mathrm{pc}$ clumps into single objects can move mass towards the high-mass end of the CSMF \citep{tamburello17, dessaugeszavadsky17}. Although our detection in F070W, the use of the EC as a background term, and the simultaneous fitting of all clumps (Sec.~\ref{subsec:det_flux_fit}) are designed to limit this effect, we cannot fully exclude it. However, using mock clump populations with random blending, \cite{dessaugeszavadsky18} showed that blending primarily removes clumps at the low-mass end, resulting in a break in the CSMF, while the slope above the break remains largely unaffected. Instead, they find the detection sensitivity to have a stronger flattening effect on the slope. This would also explain the agreement with \cite{kalita25b}, since despite their lower resolution of $\sim0.3\,\mathrm{kpc}$, their slope is measured only above their completeness limit of $10^{8.7}\,\mathrm{M_{\odot}}$, where the effect of blending on the slope is expected to be minimal.

\subsubsection{Manifestation of the galaxy $M_{*}$--SFR relation at clump scales}
\label{subsec:resolved_sfms}
\begin{figure*} 
    \centering
    \includegraphics[width=\textwidth]{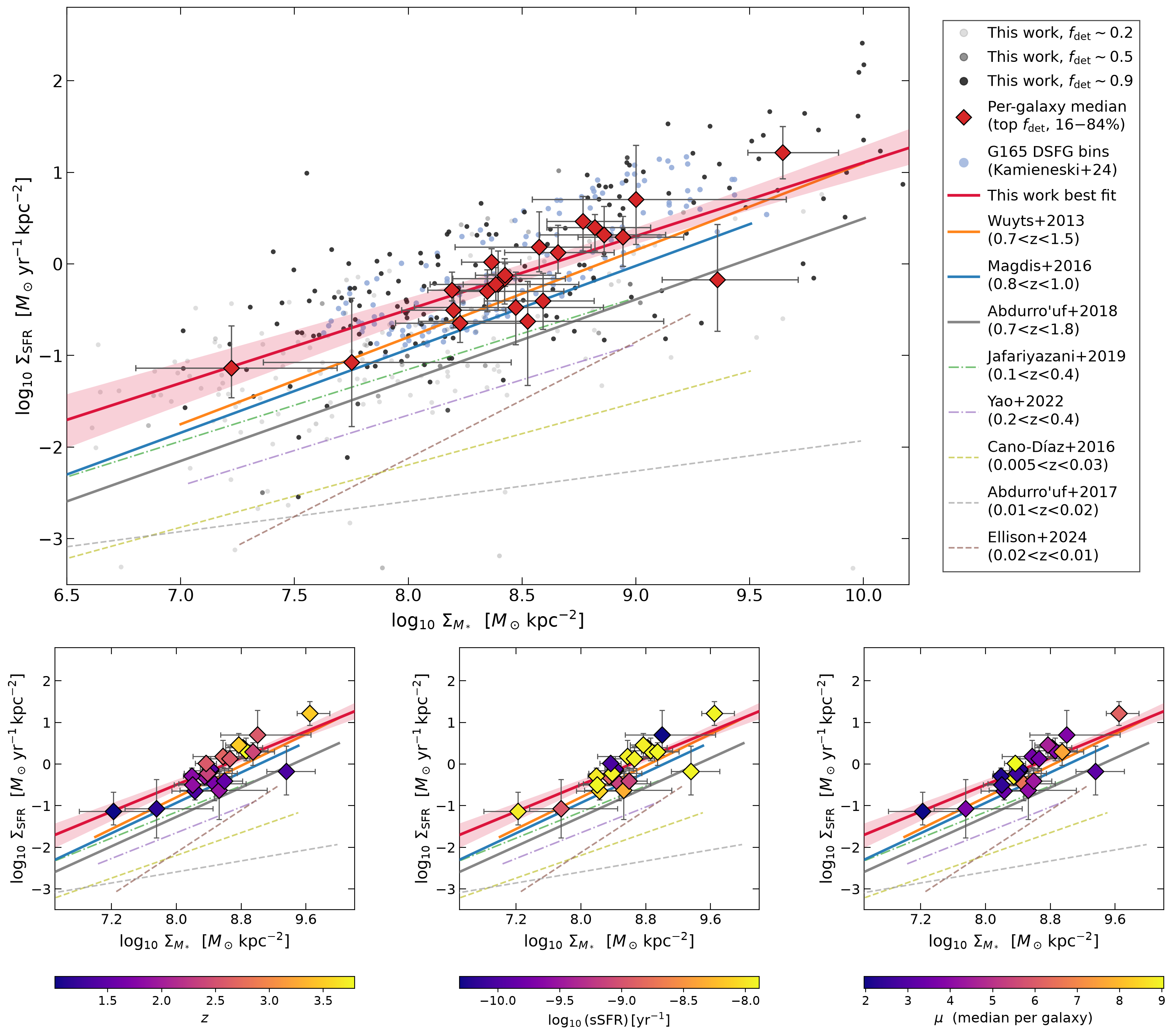}
    \caption{Resolved star-forming main sequence of the clumps,
$\Sigma_{\mathrm{SFR}}$ against $\Sigma_{M_*}$ (both per $\mathrm{kpc}^2$). Each clump is a single Monte-Carlo draw from its asymmetric $\Sigma_{M_*}$ and $\Sigma_{\mathrm{SFR}}$ posteriors (only clumps with $\log M_* \geq 5$ are shown). The opacity of each point is set by their respective $f_{\mathrm{det}}$. Light blue points are the $z\approx2.2$ resolved DSFG bins
of G165 from \citet{kamieneski24}. Coloured lines are literature rSFMS
relations (cited in the main text) over their calibrated $\Sigma_{M_*}$ ranges (solid: high-$z$; dash-dot: intermediate-$z$; dashed: local); the red line shows our best-fit relation to the high-detectability clumps ($f_{\mathrm{det}} \geq 0.9$), $\log\Sigma_{\mathrm{SFR}} = a\,\log\Sigma_{M_*} + b$. The large top panel shows all clumps with uncoloured per-galaxy medians (measured using the top-$f_{\mathrm{det}}$ tier); the three small panels repeat the per-galaxy medians coloured by host redshift $z$, specific SFR, and median magnification $\mu$.}
    \label{fig:resolved_sfrms}
\end{figure*}
 
\begin{figure} 
    \centering
    \includegraphics[width=0.49\textwidth]{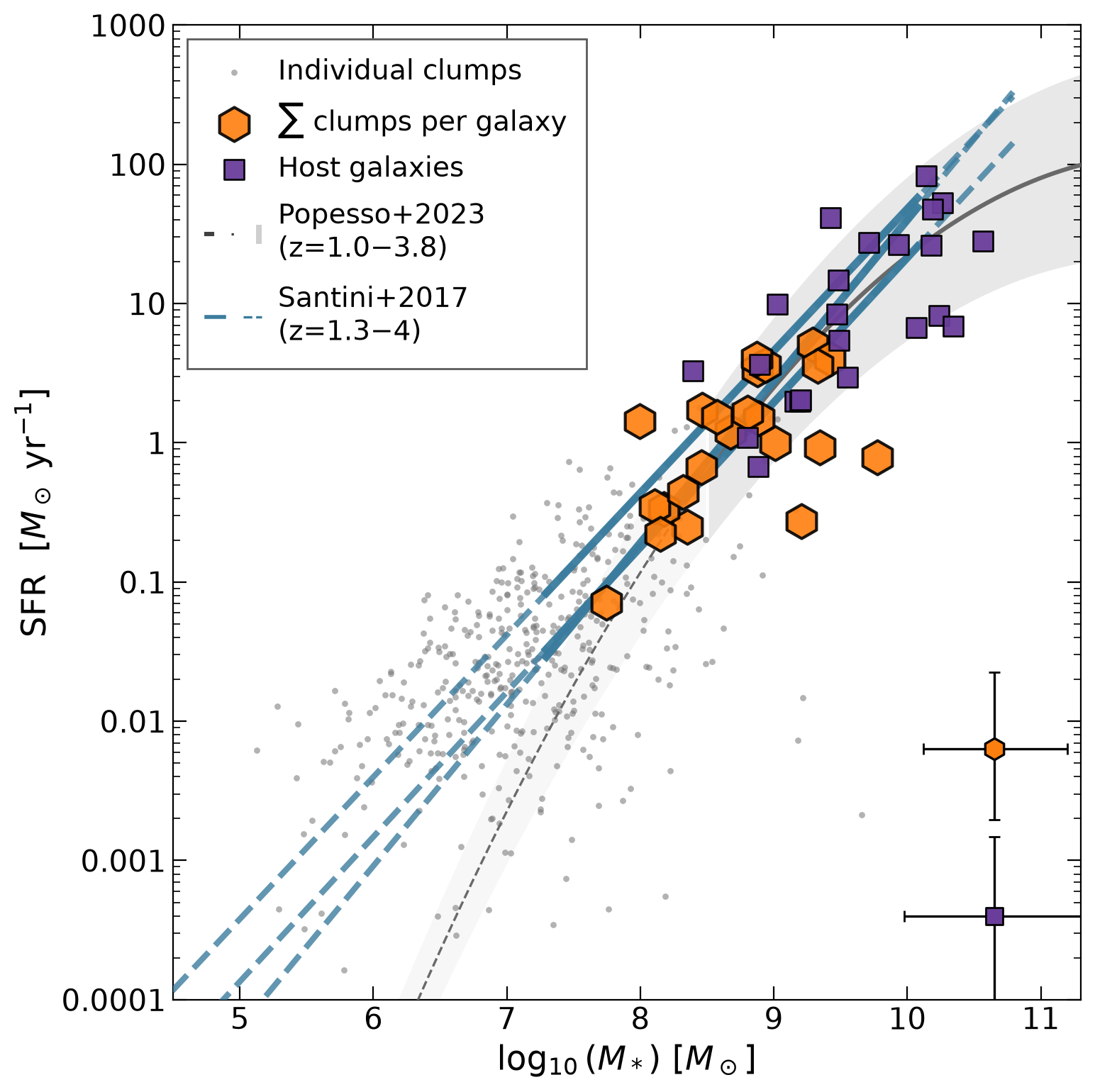}
    \caption{Stellar mass vs. SFR comparing host galaxies (purple squares, our remeasured SED fits) overlaid with  the summed clump stellar mass and SFR per galaxy (orange hexagons). Individual clumps make up the grey cloud. The \cite{popesso23} and \cite{santini17} star-forming main sequences are shown solid over their calibrated mass ranges and dashed where extrapolated.}
    \label{fig:sfms_sum_clumps}
\end{figure}
 
One key observational result in this work has been the clump SFR vs $M_{*}$ relation (Fig.~\ref{fig:sfr_mass}; Sec.~\ref{subsec:mass_prop}).  We cannot directly relate this to the star-forming galaxy main sequence (SFMS) since the latter is an empirical inference for galaxies, indicating an equilibrium of gas accretion, stellar mass build-up and gas expulsion through outflows.  However, at small (sub-kpc to kpc) scales a $\Sigma_{\mathrm{SFR}} - \Sigma_{M_{*}}$ correlation has also been found, both locally \citep{canodiaz16, abdurrof17,  jafariyazani19, ellison24} as well as in the $z \gtrsim 1$ Universe \citep{wuyts13, abdurrof18,yao22,kamieneski24}.  
 
Based on these works, a clear trend has emerged. As one progresses from $z\sim0$ to $z\sim1$,  $\Sigma_{\mathrm{SFR}}$ over sub-regions of galaxies increases by an order of magnitude or higher.  As we have already discussed in Sec.~\ref{subsubsec:dyn_range_csmf}, these regions are likely hierarchical, with the clumps in our study being at the hundreds of parsec regime.  Since hierarchical collapse would lead to denser structures at smaller scales, we find our star-forming clumps to consequently sit at an offset, but parallel to the $z\gtrsim 1$ relations (Fig.~\ref{fig:resolved_sfrms}, top panel), most of which are measured over kpc scales.  Moreover, since we are detecting the bright peaks in UV wavelengths rather than dividing the galaxy into sub-regions, we are further biased towards higher $\Sigma_{\mathrm{SFR}}$.
 
For the clumps with $f_\mathrm{det} > 0.9$, we get the following best-fit relation:
\begin{equation} 
    \log_{10}(\Sigma_{\mathrm{SFR}}) =
    (0.80^{+0.13}_{-0.12})\,\log_{10}(\Sigma_{M_*})
    - (6.92^{+1.15}_{-1.06}),
\end{equation}
We caution that the relation is driven by the SFR-stellar mass correlation shown in Fig.~\ref{fig:sfr_mass}. The sizes included in the surface density measurements rather decrease the correlation, as evidenced by the complete lack of a mass-size correlation in Fig.~\ref{fig:radius_mass}.
 
We additionally observe the median values for each galaxy in Fig.~\ref{fig:resolved_sfrms} to also be distributed along this relation, suggesting that clump location on this relation would be driven by the additional factors that affect the whole host galaxy.  We checked if this spread is driven by the host galaxy redshift, $\mathrm{sSFR}$ or $\mu$ and find gradients along it (Fig.~\ref{fig:resolved_sfrms}; lower panels), suggesting each of the parameters would have some contributing effect. However, disentangling the primary drivers would be challenging given our limited sample.  Moreover, we discuss in Sec.~\ref{subsec:surface_densities} (also shown in Fig.~\ref{fig:sigma_z}) that the correlation of $\Sigma_{\mathrm{SFR}}$ and $\Sigma_{M_{*}}$ with redshift within our sample is minimal.
 
We also compare our results to measurements of star-forming regions studied in \cite{kamieneski24}. Here, with the help of lensing they measure star-forming regions of scales from $150 - 1500\,\mathrm{pc}$ in two dusty star-forming galaxies at $z\approx 2.2$. Each of these has a $\log(\mathrm{sSFR}/\mathrm{yr}^{-1}) \gtrsim -7.7$,  putting them at a higher regime than our sample.  The Voronoi binning method likely isolates clump-like emission into single narrow spatial bins, while integrating over larger areas in other regions.  However, we see that most of our datapoints agree well with their results, indicating co-similarities of star-forming structures at these small scales.  However, as noted earlier, the size estimates of clumps are highly uncertain, and hence any underlying physical differences might be hence rendered undetectable.

 
 
Now, with this resolved understanding in mind regarding the star-forming galaxy main sequence, we return to the global relation \citep{popesso23, santini17}. We conclude that the key question to be asked here is: \emph{are the clumps in a galaxy collectively in agreement with the SFMS?} This is more relevant than their individual positions w.r.t. an extrapolation of the relation at stellar mass values that it has not been calibrated for.  We find that the collection of all clumps in a galaxy agrees with the SFMS \citep{popesso23}, with offsets ranging between $-0.50$ and $+0.28\,\mathrm{dex}$ (Fig.~\ref{fig:sfms_sum_clumps}). Here, we quote the 16--84th percentile range, estimated by drawing the summed stellar mass and SFR of each galaxy 100 times within their uncertainties. This is comparable to the scatter of the relation itself, and we find similar results using \cite{santini17}. To some extent this is expected, since the clumps trace the bulk of the UV-bright star forming regions and a non-negligible fraction of the mass of their hosts (on average, $11\%$ of the stellar mass and $14\%$ of the SFR). Hence, their integrated values track the position of their hosts, which also lie on the SFMS with offsets ranging between $-0.46$ and $+0.35\,\mathrm{dex}$ (Fig.~\ref{fig:sfms_sum_clumps}). This indicates that the clumps are integral to placing their host on the SFMS, rather than being sites of `starbursts',  a distinction that in any case cannot be made for the $<10^8\,M_{\odot}$ clumps using the SFMS.

\subsection{Connecting survival to clump substructure}
\label{subsec:survival_substructure}
 
The $t_{\mathrm{age}}$ derived for the clumps is a tracer of the average age of the stars within them, rather than a direct tracer of the dynamical age of the clump itself. Hence, the discussion of clump stability requires a more indirect approach, which we take over the following sections while deducing the internal structure of the clumps in the process. We use two diagnostics that find different fractions of the clumps to be stable, with the resulting discrepancy allowing us to infer the internal structure of the clumps in our sample.
 
\subsubsection{Virial crossing time}
\label{subsubsec:virial_crossing_time}
 
We begin with an estimate of stability based on the measured $t_{\mathrm{age}}$, following a technique established by \cite{portegies10} \citep[see also][]{gieles11} for local-Universe stellar associations and clusters. A structure with a stellar age greater than its crossing time, $t_{\mathrm{cross}} = R/\sigma$, can be assumed to be dynamically stable, i.e. not expanding at a constant velocity. Here $R$ is the size of the structure and $\sigma$ the stellar velocity dispersion. However, rather than relying on an empirical velocity dispersion, the technique derives a fiducial $t^{\mathrm{vir}}_{\mathrm{cross}}$ from the dispersion implied by the enclosed mass and size, $\sigma_{\mathrm{vir}} \sim (GM/r_{\mathrm{eff}})^{0.5}$, under the assumption that the structure is virialised. We note that this is the key distinction from the diagnostic in Sec.~\ref{subsubsec:virial_param}: $t^{\mathrm{vir}}_{\mathrm{cross}}$ is an ensemble-level quantity set entirely by $M$ and $r_{\mathrm{eff}}$, and makes no reference to the dispersion of any internal constituents. \cite{gieles11} provide the expression of $t^{\mathrm{vir}}_{\mathrm{cross}}$ for a Plummer profile:
\begin{equation} \label{eq:tcross_plummer}
t^{\mathrm{vir}}_{\mathrm{cross, plummer}} \approx 10\,\left(\frac{r_{\mathrm{eff}}^3}{G\,M}\right)^{0.5},
\end{equation}
with $r_{\mathrm{eff}}$ being the effective radius and $M$ the total mass. The choice of profile determines the ratio $f \equiv R/r_{\mathrm{eff}}$ of the structure radius to the projected $r_{\mathrm{eff}}$. With $R_{\mathrm{vir}} \equiv GM^{2}/|U|$, where $U$ is the total gravitational potential energy, a Plummer sphere has $U = -(3\pi/32)\,GM^{2}/a$ and $R_{\mathrm{eff}} = a$, giving $f_{\mathrm{Plummer}} = 32/(3\pi) \approx 3.4$. Since we model our clumps as Gaussians, $\rho(r) \propto \exp(-r^{2}/2a^{2})$ with a projected $r_{\mathrm{eff}} = a\sqrt{2\ln 2}$, we obtain $f_{\mathrm{Gaussian}} \approx 3.0$. Therefore, Eq.~\ref{eq:tcross_plummer} can be modified for a Gaussian profile as:
\begin{equation} \label{eq:tcross}
t^{\mathrm{vir}}_{\mathrm{cross}} \approx 10\left(\frac{f_{\mathrm{Gaussian}}}{f_{\mathrm{Plummer}}}\right)^{3/2}\left(\frac{r_{\mathrm{eff}}^3}{G\,M}\right)^{0.5} \approx 8.3\left(\frac{r_{\mathrm{eff}}^3}{G\,M}\right)^{0.5}.
\end{equation}
 
This fiducial $t^{\mathrm{vir}}_{\mathrm{cross}}$ is compared to the SED-derived stellar age through the parameter $\Pi = t_{\mathrm{age}}/t^{\mathrm{vir}}_{\mathrm{cross}}$. Given the assumption of virialisation, $t_{\mathrm{cross}}$ is overestimated for unbound structures, which further facilitates their distinction (based on $\Pi$) from their bound counterparts \citep{gieles11}, albeit with a gradual transition into boundedness over values of $\Pi \sim 1{-}5$. However, \cite{bastian12} adopted $\mathrm{log}(\Pi) \approx 0$ as the threshold, likely due to practicality, as have multiple recent works studying $z>1$ lensed clumps \citep[over scales $\lesssim$ ours;][]{vanzella21, vanzella22, messa22, claeyssens23, messa24}. For our sample, we find $\mathrm{log}(\Pi)$ to span a range of $-1.7$ to $2.8$ (Fig.~\ref{fig:mono_vs_subclump}), with $95\%$ of the clumps above our mass completeness limit of $10^{7.1}\,\mathrm{M_{\odot}}$ (Sec.~\ref{subsec:sed_fitting}) having $t_{\mathrm{age}} > t^{\mathrm{vir}}_{\mathrm{cross}}$. In the following, we quote all fractions for these clumps only. Taking $\mathrm{log}(\Pi) \approx 0$ at face value, the clumps above $10^{7.1}\,\mathrm{M_{\odot}}$ are therefore consistent with being dynamically stable over their stellar ages, i.e. for at least a few $100\,\mathrm{Myr}$. It should be noted, however, that the single profile we adopt, although valid for star clusters ($\sim$ tens of $\mathrm{pc}$), is likely an oversimplification at $\sim 100\,\mathrm{pc}$ scales. Hence, the $\mathrm{log}(\Pi)$ in our estimates should be regarded as an average over a range of internal dynamical states.
 
\subsubsection{The super-virial discrepancy}
\label{subsubsec:virial_param}
 
The crossing-time analysis concludes $95\%$ of the clumps above the mass completeness limit are dynamically stable. We now test this against a direct virial estimate, which employs a different assumption: rather than deriving the dispersion from the enclosed mass, it imports an externally measured one. We define the scaled virial parameter ($\alpha^{\mathrm{scaled}}_{\mathrm{vir}}$\footnote{not to be confused with the slope $\alpha$ of the CSMF}) as
\begin{equation} \label{eq:virial_param}
\alpha^{\mathrm{scaled}}_{\mathrm{vir}} = \eta\,\frac{\sigma^2\,r_{\mathrm{eff}}}{G\,M},
\end{equation}
where $\eta \simeq 9$ for a structure modelled as a 2D Gaussian (Appendix~\ref{app:virial_prefactor}). We use $M=$ stellar mass of the clumps, and defer the addition of gas to Sec.~\ref{subsubsec:bridging}. Eq.~\ref{eq:virial_param} also crucially contains the velocity dispersion ($\sigma$), for which we have no measurement. Stellar velocity dispersions are in fact highly challenging to measure beyond the local Universe, even for lensed systems \citep{hill16}. Hence, we make a debatable assumption:
\begin{equation} \label{eq:sigma_assumption}
\sigma_{\mathrm{stars}} \approx \sigma_{\mathrm{molgas}},
\end{equation}
which would be true if the stars forming out of the molecular gas inherit the velocity dispersion of the latter, \emph{with both components sharing similar physical scales}.
 
If the clumps we measure are formed as singular units of the measured sizes ($\approx 50-400\,\mathrm{pc}$; Fig.~\ref{fig:radius_hist}), we can use literature-based cold molecular gas velocity dispersion measurements at the same scales. \cite{dessaugeszavadsky19, dessaugeszavadsky23} study two lensed systems at $z\sim1$ and find their giant molecular clouds (GMCs), resolved down to $\sim30\,\mathrm{pc}$, to be mostly virialised and to consequently follow Larson's relation, $\sigma_{\mathrm{molgas}} \propto r_{\mathrm{eff}}^{0.5}$, with $\sigma$ varying within $\sim 10{-}40\,\mathrm{km\,s^{-1}}$ at $100\,\mathrm{pc}$. Using this scaling to assign each clump its expected $\sigma_{\mathrm{molgas}}$, we derive the $\alpha^{\mathrm{scaled}}_{\mathrm{vir}}$. 
 
We do note two empirical reasons why this assumption \emph{underpredicts} the $\alpha^{\mathrm{scaled}}_{\mathrm{vir}}$.  The average mass of the gas clouds at the clump scales matches the stellar mass of our clumps. However, we do not expect all the gas mass to be converted to stars \citep[efficiencies are reported to be $\sim 30\%$; ][]{dessaugeszavadsky23}, which would suggest that the size matching we use, would likely underpredict the velocity dispersion of the parent cloud.  Second, $\sigma_{\mathrm{stars}} \approx \sigma_{\mathrm{molgas}}$ is in detail a lower bound on the stellar dispersion, since where both are measured in the same local region, young stars are found to be marginally \emph{hotter} than their natal gas. In NGC~1333, \cite{foster15} measure an embedded stellar dispersion of $0.92\pm0.12\,\mathrm{km\,s^{-1}}$ against a dense-core dispersion of $0.51\pm0.05\,\mathrm{km\,s^{-1}}$, and in the Ophiuchus complex \cite{gupta22} find the young stellar group to be both hotter than its natal gas.
 
Classically, $\alpha_{\mathrm{vir}} \sim 1$ is a virialised system and $\alpha_{\mathrm{vir}} \sim 2$ marks the boundary at which the structure as a whole is bound. For stellar systems, however, this boundary is not sharp. N-body studies of young star clusters show that a bound core can survive the expulsion of residual gas even if the stars are left super-virial, up to a virial ratio corresponding to $\alpha_{\mathrm{vir}} \approx 3$ \citep{goodwin06, baumgardt07, shukirgaliyev17}. Moreover, the velocity dispersions measured in young clusters frequently indicate super-virial states, which can be biased high by binaries \citep{gieles10} or reflect clusters out of equilibrium after gas expulsion \citep{goodwin06}, while whether they are bound often remains ambiguous given the systematic uncertainties \citep{kuhn19}. We therefore treat $\alpha^{\mathrm{scaled}}_{\mathrm{vir}} < 2$ as bound and, allowing for these effects, $\alpha^{\mathrm{scaled}}_{\mathrm{vir}} < 10$ as bound or partially bound. The contours at $\alpha^{\mathrm{scaled}}_{\mathrm{vir}} = 2$, $10$ and $50$ in Fig.~\ref{fig:mono_vs_subclump} are shown for visual reference. Under the monolithic assumption of the clumps forming as single units (left panel), only $25\%$ of the clumps above $10^{7.1}\,\mathrm{M_{\odot}}$ have $\alpha^{\mathrm{scaled}}_{\mathrm{vir}} < 2$ and $55\%$ have $\alpha^{\mathrm{scaled}}_{\mathrm{vir}} < 10$. This would imply dispersal on crossing times of $\approx 2{-}15\,\mathrm{Myr}$ ($r_{\mathrm{eff}}/\sigma_{\mathrm{molgas}}$, with  $\sigma_{\mathrm{molgas}} = 10{-}40\,\mathrm{km\,s^{-1}}$), which is in clear tension both with the stellar ages reaching $\sim 700\,\mathrm{Myr}$ and with the $95\%$ of clumps above $10^{7.1}\,\mathrm{M_{\odot}}$ found to be stable by the virialised crossing-time analysis of the previous section. Given that $t^{\mathrm{vir}}_{\mathrm{cross}}$ depends only on $M$ and $r_{\mathrm{eff}}$, while $\alpha^{\mathrm{scaled}}_{\mathrm{vir}}$ additionally imports $\sigma_{\mathrm{molgas}}$ measured at $100\,\mathrm{pc}$, we conclude that the discrepancy points directly at the assumption in Eq.~\ref{eq:sigma_assumption}.

\subsubsection{Bridging the discrepancy: clumps as ensembles of sub-100\,pc structures}
\label{subsubsec:bridging}

\begin{figure*}
    \centering
    \includegraphics[width=\textwidth]{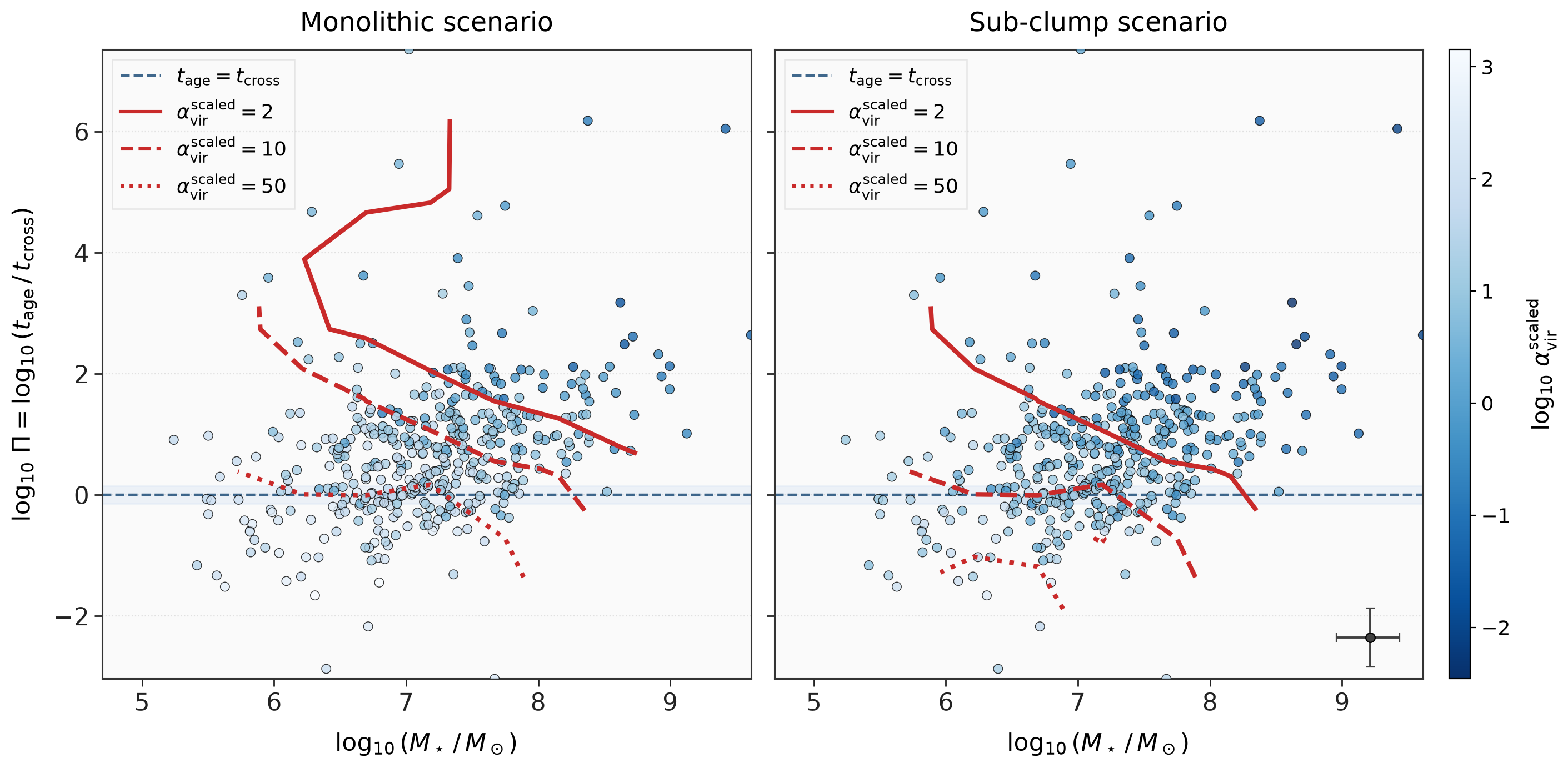}
    \caption{The ratio of mass-weighted age ($t_{\mathrm{age}}$) and crossing time ($t^{\mathrm{vir}}_{\mathrm{cross}}$) against clump stellar mass, with each clump coloured by its $\alpha^{\mathrm{scaled}}_{\mathrm{vir}}$, under the two interpretations of the clump's internal composition. (Left) The monolithic scenario, in which each clump inherits the $100\,\mathrm{pc}$-scale dispersion $\sigma_{\mathrm{gas,100}}$ of a single cloud (Sec.~\ref{subsubsec:virial_param}). (Right) The sub-clump scenario, in which the clump is an ensemble of star clusters formed in sub-clouds of size $R_{\mathrm{c}} = 20\,\mathrm{pc}$, lowering the ensemble dispersion to $\sigma_{*,\mathrm{tot}}$ via Eq.~\ref{eq:sigma_star_tot}. The blue dashed line marks $t_{\mathrm{age}} = t^{\mathrm{vir}}_{\mathrm{cross}}$, while the solid, dashed and dotted red contours mark $\alpha^{\mathrm{scaled}}_{\mathrm{vir}} = 2$, $10$ and $50$, shown for visual reference.}
    \label{fig:mono_vs_subclump}
\end{figure*}

We first ask whether the discrepancy can be resolved by accounting for additional mass, since both diagnostics depend on $M$. In the $\alpha^{\mathrm{scaled}}_{\mathrm{vir}}$ estimate we used only the stellar mass; including the gas via a factor $K_{\mathrm{gas}}$ multiplying $M$ reduces $\mathrm{log}(\alpha_{\mathrm{vir}}^{\mathrm{scaled}})$ by $\mathrm{log}(K_{\mathrm{gas}})$. With the $\sim30\%$ star-formation efficiency found by \cite{dessaugeszavadsky23}, $K_{\mathrm{gas}} = 1/0.3$ gives $\mathrm{log}(K_{\mathrm{gas}}) = 0.5$. Nevertheless, the remaining gas is known (at least in local galaxies) to be rapidly photo-dissociated by the stellar radiation and hence unlikely to remain in the systems \citep{whitworth79, hollenbach99, lada03, chevance20}.

Adding this mass, albeit unphysical, narrows the gap from both sides: the fraction of clumps with $\alpha^{\mathrm{scaled}}_{\mathrm{vir}} < 2$ ($< 10$) increases from $25\%$ ($55\%$) to $45\%$ ($84\%$), while the shorter $t^{\mathrm{vir}}_{\mathrm{cross}}$ (Eq.~\ref{eq:tcross}) raises the fraction with $\Pi > 1$ from $95\%$ to $97\%$. Crucially, however, the two cannot be reconciled in this way, since the extra mass shifts both diagnostics in the \emph{same} direction (toward boundedness); it can compress the offset but never eliminate it, and the modest $\sim0.5\,\mathrm{dex}$ of available mass is in any case split across the two parameters. The reduction is thus only marginal, and a residual discrepancy survives. Moreover, the same $z\sim1$ studies \citep{dessaugeszavadsky19, dessaugeszavadsky23} find the gas and $\sim100\,\mathrm{pc}$ stellar complexes to be spatially offset by $\sim 100\,\mathrm{pc}{-}1\,\mathrm{kpc}$, making it unlikely that the stellar clumps retain any appreciable gas at all.

The remaining parameter that can be modified is the velocity dispersion, and specifically the scale at which it is imposed, which acts on $\alpha^{\mathrm{scaled}}_{\mathrm{vir}}$ alone. Here it is worth being careful about what the gas observations do and do not establish. The resolved GMCs in these $z\sim1$ galaxies undoubtedly carry sufficient \emph{mass} to form the stellar clumps in situ, spanning $\sim10^{5.9}{-}10^{7.9}\,M_{\odot}$ \citep{dessaugeszavadsky19, dessaugeszavadsky23}, which is comparable to the stellar clump masses measured in the same systems \citep[$1.5\times10^{6}{-}6\times10^{9}\,M_{\odot}$ in the Cosmic Snake and A521;][]{cava18, messa22}, and these clouds are in fact explicitly interpreted as the parent structures of those clumps \citep{dessaugeszavadsky19, dessaugeszavadsky23}. Hence the mass budget is unlikely to be the obstacle here. 

Rather, the issue is that these clouds, when treated as \emph{single} units at the $\sim100\,\mathrm{pc}$ scale at which they are resolved, would be far too dynamically hot, since it is precisely their $\sim100\,\mathrm{pc}$-scale $\sigma_{\mathrm{molgas}}$ that drives our clumps super-virial in Sec.~\ref{subsubsec:virial_param}. A clump that inherits the dispersion of one such monolithic cloud would therefore be unlikely to survive. The resolution is hence not to find more mass, but rather to assume that both the gas and the stars are sub-divided: a sub-unit of a GMC, of size $\ll100\,\mathrm{pc}$, forms a sub-unit of a clump (a star cluster), and it is the \emph{ensemble} of these cooler sub-units that we observe as a single $\sim100\,\mathrm{pc}$ clump.

Here we invoke our understanding of local-Universe star-forming regions to make this concrete. Unlike the resolution-driven blending of $\sim$kpc clumps into $\sim100\,\mathrm{pc}$ constituents discussed in the literature \citep[e.g.][]{tamburello15, tamburello17, dessaugeszavadsky17, cava18, behrendt16}, we are dealing here with the internal composition of an individual, resolved $\sim100\,\mathrm{pc}$ clump. The dominant stellar structures (in terms of mass density) in local star-forming regions are star clusters, which are limited to scales of a few tens of $\mathrm{pc}$, and these are also the structures expected to remain gravitationally bound for at least a few $t^{\mathrm{vir}}_{\mathrm{cross}}$ (Sec.~\ref{subsubsec:virial_crossing_time}). However, these form out of density peaks within gas clouds, the latter surviving only a few Myr. This picture, in which a single resolved aperture encloses not one monolithic cloud but a full spectrum of self-gravitating gas structures, with smaller clouds nested within \citep[as observed locally in][]{keto05} is the generic expectation for a supersonically turbulent ISM \citep{hopkins12}.  We hence adopt this picture to propose that the clumps in our study consist of smaller star-cluster-like substructures.

We can assume that the random velocities of these clusters are $\sim \sigma_{\mathrm{gas}}$, the dispersion of their parent sub-cloud within a bounding box of size $100\,\mathrm{pc}$. This is because the cluster velocities would likely be drawn from the random motion of the gas particles in the parent cloud, the distribution of which has a standard deviation $= \sigma_{\mathrm{gas}}$. We re-emphasize that this inheritance is an assumption of our framework rather than a measured quantity. For simplicity, we consider a single cluster with velocity $v_{*,1} \sim \sigma_{\mathrm{gas,1}}$ forming out of one cloud, and then take multiple generations of clouds, each contributing a cluster to our $100\,\mathrm{pc}$ box. We therefore end up with $[v_{*,1}, v_{*,2}, \ldots, v_{*,n}] \sim [\sigma_{\mathrm{gas,1}}, \sigma_{\mathrm{gas,2}}, \ldots, \sigma_{\mathrm{gas,n}}]$. Since these velocities are oriented at random in the inertial frame of the box, their mean would be $\approx 0$, and hence the dispersion of the ensemble is the root-mean-square,
\begin{equation} \label{eq:sigma_star_tot}
\sigma_{*,\mathrm{tot}} = \sqrt{\frac{1}{n}\sum_{i=1}^{n} v^2_{*,i}} \sim \sqrt{\frac{1}{n}\sum_{i=1}^{n} \sigma^2_{\mathrm{gas},i}}.
\end{equation}
Now, each of these sub-clouds would always be $\lesssim 100\,\mathrm{pc}$ in size, and by the size--linewidth relation of a turbulent medium, $\sigma_{\mathrm{gas}} \propto R^{(p-1)/2}$ with $p>1$ \citep{larson81, hopkins12}, would have systematically lower dispersions than a cloud filling the $100\,\mathrm{pc}$ box (whose dispersion we call $\sigma_{\mathrm{gas,100}}$). Since $\sigma_{*,\mathrm{tot}}$ is a root-mean-square of quantities each $\lesssim \sigma_{\mathrm{gas,100}}$, it is itself bounded by $\sigma_{\mathrm{gas,100}}$, and is strictly lower whenever the constituent clouds are appreciably smaller.

It is worth emphasising that this dispersion \emph{only} enters $\alpha^{\mathrm{scaled}}_{\mathrm{vir}}$. We continue to evaluate Eq.~\ref{eq:virial_param} at the observed ensemble $r_{\mathrm{eff}}$ and $M$, replacing only the imported $\sigma_{\mathrm{gas,100}}$ with the cooler ensemble value $\sigma_{*,\mathrm{tot}}$. Taking the canonical Larson slope $p\approx2$ ($\sigma_{\mathrm{gas}} \propto R^{0.5}$) and a single characteristic constituent size $R_{\mathrm{c}}$, Eq.~\ref{eq:sigma_star_tot} gives $\sigma_{*,\mathrm{tot}}^2 \sim \sigma_{\mathrm{gas,100}}^2\,(R_{\mathrm{c}}/100\,\mathrm{pc})$, so that $\alpha^{\mathrm{scaled}}_{\mathrm{vir}} \propto \sigma^2$ is reduced by the factor $(R_{\mathrm{c}}/100\,\mathrm{pc})$. Normalising to $\sigma_{\mathrm{gas,100}} = 25\,\mathrm{km\,s^{-1}}$ (the midpoint of the $10{-}40\,\mathrm{km\,s^{-1}}$ range above), a constituent size of $R_{\mathrm{c}} = 25\,\mathrm{pc}$ gives $\sigma_{*,\mathrm{tot}} \approx 12.5\,\mathrm{km\,s^{-1}}$, and hence a $\sim4\times$ reduction in $\alpha^{\mathrm{scaled}}_{\mathrm{vir}}$, while $R_{\mathrm{c}} = 20\,\mathrm{pc}$ gives $\approx 11\,\mathrm{km\,s^{-1}}$ and a $\sim5\times$ reduction. These values of velocity dispersion are remarkably consistent with measured values of young star-clusters \citep{ho96a,ho96b}. 

We stress that this is a reduction in the \emph{dispersion} and not the mass, so it lowers $\alpha^{\mathrm{scaled}}_{\mathrm{vir}}$ uniformly by $-(p-1)\log_{10}(100\,\mathrm{pc}/R_{\mathrm{c}})$ in log scale, without altering the $M$ at which any individual clump sits. A $\sim5\times$ reduction increases the fraction of clumps with $\alpha^{\mathrm{scaled}}_{\mathrm{vir}} < 2$ from $25\%$ to $55\%$, and that with $\alpha^{\mathrm{scaled}}_{\mathrm{vir}} < 10$ from $55\%$ to $91\%$. This is shown in Fig.~\ref{fig:mono_vs_subclump}: relative to the monolithic case (left panel), the sub-clump dispersion (right panel, for $R_{\mathrm{c}} = 20\,\mathrm{pc}$) shifts the whole sample to lower $\alpha^{\mathrm{scaled}}_{\mathrm{vir}}$ values, while every point retains its position in the $\Pi$--mass plane. The substructure correction therefore brings the fraction of bound or partially bound clumps close to the $95\%$ found by the crossing-time diagnostic, which the mass argument alone could not. Only about half of the clumps would be fully bound, however, with the remainder consistent with ensembles that retain a bound core while losing part of their stellar mass \citep{goodwin06}. We note that for a realistic range of constituent sizes, the effective reduction would be set by the mass-weighted mean of $\sigma_{\mathrm{gas}}^2$, and is hence somewhat weaker than this single-size estimate.

We highlight that this resolution is asymmetric. The crossing-time diagnostic refers to the clump as a single virialised ensemble of mass $M$ and size $r_{\mathrm{eff}}$, and never invokes the dispersion of the internal constituents. Hence Eq.~\ref{eq:tcross}, and with it the $95\%$ fraction of stable clumps, remains \emph{unchanged} whether the enclosed mass is monolithic or an association of sub-structures. The substructure framework hence leaves this diagnostic exactly as measured, and revises only $\alpha^{\mathrm{scaled}}_{\mathrm{vir}}$, which is the diagnostic that imports a scale-dependent external dispersion. 

The two would therefore approach each other not because both are adjusted, but rather because the substructure picture corrects the single assumption responsible for the discrepancy, while leaving the ensemble-level argument intact. In this framework, the star-cluster-like substructures  are observed as single objects at our resolution while having a low enough ensemble dispersion to remain bound, and the mass of a clump would consequently be a tracer of the \emph{number} of constituent sub-$100\,\mathrm{pc}$ clusters that make it up. A higher number of such members would then lead to a higher probability of the whole structure being bound, which would naturally explain why the most massive clumps already approach virialisation even under the monolithic $\sigma_{\mathrm{gas,100}}$ assumption (Fig.~\ref{fig:mono_vs_subclump}, left panel).

\section{Summary}
\label{sec:summary}

In this work, we use a sample of 23 strongly lensed galaxies at $z=1-4$, to investigate the hierarchical nature of star-forming regions. The key focus of this work has been the development of a novel methodological framework which implements a multi-scale difference-of-Gaussian (DoG) decomposition. This approach simultaneously models compact $\approx 50-400\,\mathrm{pc}$ clumps and the extended $\sim$kpc-scale structures (ECs), fit jointly across 13 JWST/NIRCam bands.  It is worth noting that the ECs do not have a clear physical interpretation, and are rather used as a local background flux sink for diffuse emission, undetected clumps, the overlapping PSF wings of clustered clumps, or a combination of these. Our main results are as follows.

\begin{enumerate}[leftmargin=*]

 
    \item We detect 451 clumps across our 23 galaxies, with sizes ($r_{\mathrm{eff}}$) of $\approx 50-400\,\mathrm{pc}$ and stellar masses of $\approx 10^{6}-10^{9}\,\mathrm{M_{\odot}}$. The clump stellar mass function (CSMF) has a high-mass slope of $1.88 \pm 0.04$, consistent with the fiducial scale-free star-forming region slope, and near-identical to the slope measured for $\sim$kpc-scale clumps in unlensed galaxies \citep{kalita25b}.
    
    \item We find that the apparent increase of clump $\Sigma_{M_*}$, $\Sigma_{\mathrm{SFR}}$ and clump mass to host mass ratio with redshift (within our redshift range) weakens substantially once we account for the redshift-dependent selection bias. Hence we show that these trends are highly affected by redshift biases, and any evidence of real physical evolution is tentative.
    
     \item Our individual clumps sit offset above, but parallel to, the resolved $\Sigma_{\mathrm{SFR}}-\Sigma_{M_*}$ relations measured at $z\gtrsim1$ over kpc-scales, consistent with the clumps being the densest peaks in star-forming regions. Furthermore the summation of clumps in each galaxy agrees with the global SFMS, with offsets of $-0.50$ to $+0.28\,\mathrm{dex}$ (16--84th percentiles) that are comparable to the scatter of the relation itself, indicating that the clumps are integral to placing their host galaxies on the SFMS rather than acting as distinct starburst sites. 

    \item We assess the dynamical stability of the clumps through two independent diagnostics.  We first use the ratio of the stellar age and a crossing time with assumed virialization.  We then estimate a separate virial parameter that imports literature based $z\sim1$ molecular-gas velocity dispersion ($\sigma_{\mathrm{molgas}}$) at $\sim100\,\mathrm{pc}$ scales. We find these diagnostics to be in tension: while $95\%$ of the clumps above the mass completeness limit ($10^{7.1}\,\mathrm{M_{\odot}}$) are stable according to the former, only $25\%$ ($55\%$) are bound (bound or partially bound) according to the latter.  We largely resolve the discrepancy by proposing that $\sim100\,\mathrm{pc}$ clumps are themselves an ensemble of sub-$100\,\mathrm{pc}$ star-cluster-like substructures (mirroring the framework of clump ensembles making up giant clumps), formed from smaller gas clouds. 


\end{enumerate}

\begin{acknowledgments}
We would like to thank Oleg Gnedin, Reinhard Genzel and Giulia Tozzi for the highly valuable discussions. BSK was supported by the China Postdoctoral Science Foundation (2025M773186), the Beijing Natural Science Foundation (IS26015), and the National Science Foundation of China (W2633004). LCH was supported by the National Science Foundation of China (12233001) and the China Manned Space Program (CMS-CSST-2025-A09). Finally, BSK would like to thank Xueling Wang for her constant support, without whom this work would not have been possible. 

\textit{Disclosure of Delegation to Generative AI:} The authors declare the use of generative AI in the research and writing process. According to the GAIDeT taxonomy (2026), the following tasks were delegated to GAI tools under full human supervision:
\begin{itemize}[noitemsep]
    \item Code generation
    \item Code optimization
    \item Process automation
    \item Proofreading and editing
\end{itemize}
The GAI tool used was: Claude Sonnet. Responsibility for the final manuscript lies entirely with the authors. GAI tools are not listed as authors and do not bear responsibility for the final outcomes. Declaration submitted by: Boris S. Kalita.

\end{acknowledgments}

\begin{contribution}
BSK carried out the analysis and prepared the manuscript. AC provided supporting datasets from a separate analysis of the parent sample. AC, LCH, MDZ, NMFS and ED also provided key insights during the planning phase of the project. All co-authors provided valuable comments during the preparation of the manuscript.



\end{contribution}

%
\facilities{JWST}



\appendix

\section{Magnification and lensing-induced ellipticity from the $\kappa$/$\gamma$ maps}
\label{app:lensing}

Here we describe how we get the magnification ($\mu$) and the maximum
lensing-induced ellipticity at each clump position, referred to in
Sec.~\ref{subsec:data_sample} and Sec.~\ref{subsec:flux_uncertainties}.
The steps below are based on \cite{narayan96}. We start with the publicly available UNCOVER v2.0 lensing model of Abell 2744 \citep{furtak23,price25}, which gives us the convergence $\kappa$ and the two shear components $\gamma_1$, $\gamma_2$ in the form of maps sharing the same WCS as the image mosaics. The convergence is effectively the surface mass density in units of its critical value, $\kappa = \Sigma/\Sigma_{\mathrm{cr}}$, and both $\kappa$ and $\gamma_{1,2}$ come from second derivatives of the same lensing potential $\psi$ \citep[][their Eq.~80]{narayan96}. The maps are by default normalised to a source at infinity ($D_{\mathrm{ds}}/D_{\mathrm{s}} = 1$). Throughout, we adopt a flat
$\Lambda$CDM cosmology ($H_0 = 70\,\mathrm{km\,s^{-1}\,Mpc^{-1}}$, $\Omega_{\mathrm{m}} = 0.3$) with the cluster at $z_l = 0.3072$ \citep{furtak23}.

\paragraph{Scaling to the source redshift.}
Before using the lensing map values at the location of a clump, we have to rescale them to that clump's (and hence its host's) source redshift $z$.  In order to do so, we require the following ratio
\begin{equation} \label{eq:beta}
\beta(z) = \frac{D_{\mathrm{ds}}(z_l, z)}{D_{\mathrm{s}}(z)},
\end{equation}
where $D_{\mathrm{ds}}$ is the angular-diameter distance from lens to source and $D_{\mathrm{s}}$ that from observer to source.  We can then rescale the $\kappa$ and $\gamma_{1,2}$ maps using the same factor $\beta$ \footnote{not to be confused with the $\beta$ used in the main text to denote the slope below the turnover mass ($M_{\mathrm{b}}$) in the CSMF.},
\begin{equation} \label{eq:scaled_kg}
\kappa_z = \beta\,\kappa, \qquad
\gamma_z = \beta\,\sqrt{\gamma_1^2 + \gamma_2^2}.
\end{equation}
The final (scaled) shear magnitude is defined as $\gamma_z$, and the scaled convergence as $\kappa_z$.

\paragraph{The lensing Jacobian and its eigenvalues.}
The way a small patch of the source region maps onto the corresponding image region is captured by the
Jacobian matrix $\mathsf{A} = \partial\vec{\beta}/\partial\vec{\theta}$. In terms of the redshift-scaled convergence and shear it can be expressed as
\citep[][their Eq.~58]{narayan96}
\begin{equation} \label{eq:jacobian}
\mathsf{A} =
\left(
\begin{array}{cc}
1 - \kappa_z - \gamma_{1,z} & -\gamma_{2,z} \\
-\gamma_{2,z} & 1 - \kappa_z + \gamma_{1,z}
\end{array}
\right),
\end{equation}
with $\gamma_{1,z} = \beta\gamma_1$ and $\gamma_{2,z} = \beta\gamma_2$. This
matrix is symmetric, so we can diagonalise it, and the two eigenvalues are
\begin{equation} \label{eq:eigenvalues}
\lambda_{\mathrm{t}} = 1 - \kappa_z - \gamma_z, \qquad
\lambda_{\mathrm{r}} = 1 - \kappa_z + \gamma_z .
\end{equation}
Physically, the convergence part $(1-\kappa_z)$ is axisymmetric (radial), while the shear $\gamma_z$ provides the  non-axisymmetric (tangential) component. 

\paragraph{Magnification.}
The magnification is given by one over the determinant of the Jacobian
\citep[][their Eq.~60]{narayan96},
\begin{equation} \label{eq:mu}
\mu = \frac{1}{\det \mathsf{A}}
    = \frac{1}{\left(1 - \kappa_z\right)^2 - \gamma_z^2}
    = \frac{1}{\lambda_{\mathrm{t}}\,\lambda_{\mathrm{r}}},
\end{equation}
and since the determinant of a symmetric matrix is the product of its eigenvalues, $\mu$ is just $1/(\lambda_{\mathrm{t}}\lambda_{\mathrm{r}})$, with per-axis magnifications $\mu_{\mathrm{t}} = 1/\lambda_{\mathrm{t}}$ and $\mu_{\mathrm{r}} = 1/\lambda_{\mathrm{r}}$. We read the lensing maps  at the location of each clump using the WCS, and the resulting per-clump $\mu$ is used to correct the measured fluxes and integrated quantities (Sec.~\ref{subsec:sed_fitting}).

\paragraph{Magnification uncertainty.}
To estimate the uncertainty on $\mu$, we use the 50 lens-model realisations drawn from the final MCMC chain of the UNCOVER model \citep{furtak23}, released at a pixel scale of $0.1''$ for error estimation. For each realisation $i$ we read $\kappa$ and $|\gamma|$ at the position of the clump, scale them to the source redshift with the same $\beta(z)$, and compute $\mu_i$ with Eq.~\ref{eq:mu}. The realisations share a common overall normalisation that differs from that of the best-fit maps; we therefore rescale all of them by a single factor, set by the median ratio of the best-fit to the realisation convergence over the region covering our sample, and use them only to measure the fractional scatter of $\mu$. With $P_{n}$ the $n$th percentile of the $\mu_i$ and $\mu$ the best-fit value, the lower and upper uncertainties are $\sigma_{\mu,\mathrm{lo}}^2 = [\mu\,(P_{50}-P_{16})/P_{50}]^2 + (P_{50}-\mu)^2$ and $\sigma_{\mu,\mathrm{hi}}^2 = [\mu\,(P_{84}-P_{50})/P_{50}]^2 + (P_{50}-\mu)^2$, where the second term accounts for offset between the best-fit value and the median of the realisations. The fractional uncertainties have a median of $4\%$, and $90\%$ of the clumps have uncertainties below $15\%$. These uncertainties are propagated into the stellar masses and SFRs as described in Sec.~\ref{subsec:sed_fitting}.

\paragraph{Maximum lensing-induced ellipticity.}
A source that is intrinsically circular gets stretched by $1/|\lambda|$ along
each eigen-direction, so its image is an ellipse whose axes are proportional
to $|\lambda_{\mathrm{t}}|^{-1}$ and $|\lambda_{\mathrm{r}}|^{-1}$
\citep[][their Eq.~59]{narayan96}.  The axis ratio (minor over major), which depends only on the
ratio of the two eigenvalues, is given by
\begin{equation} \label{eq:axis_ratio}
q = \frac{\min(|\lambda_{\mathrm{t}}|,\,|\lambda_{\mathrm{r}}|)}
         {\max(|\lambda_{\mathrm{t}}|,\,|\lambda_{\mathrm{r}}|)}.
\end{equation}
This is the value we use to define the maximum ellipticity that is fed into the injection-recovery analysis of Sec.~\ref{subsec:flux_uncertainties}.

\section{Comparison of the multi-scale fitting to single-scale fitting}
\label{app:method_comparison}

\begin{figure} 
    \centering
    \includegraphics[width=0.49\textwidth]{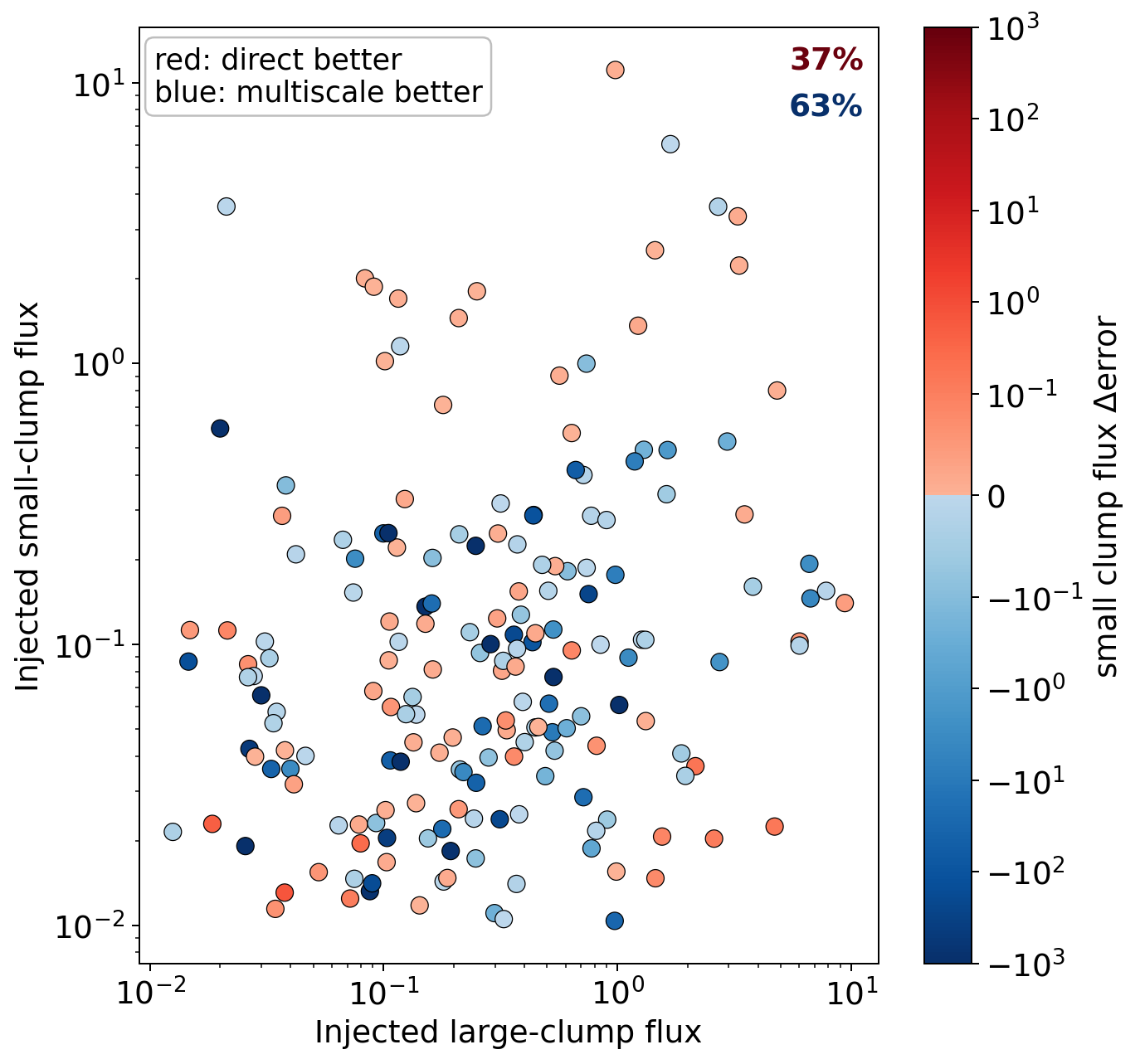}
    \includegraphics[width=0.49\textwidth]{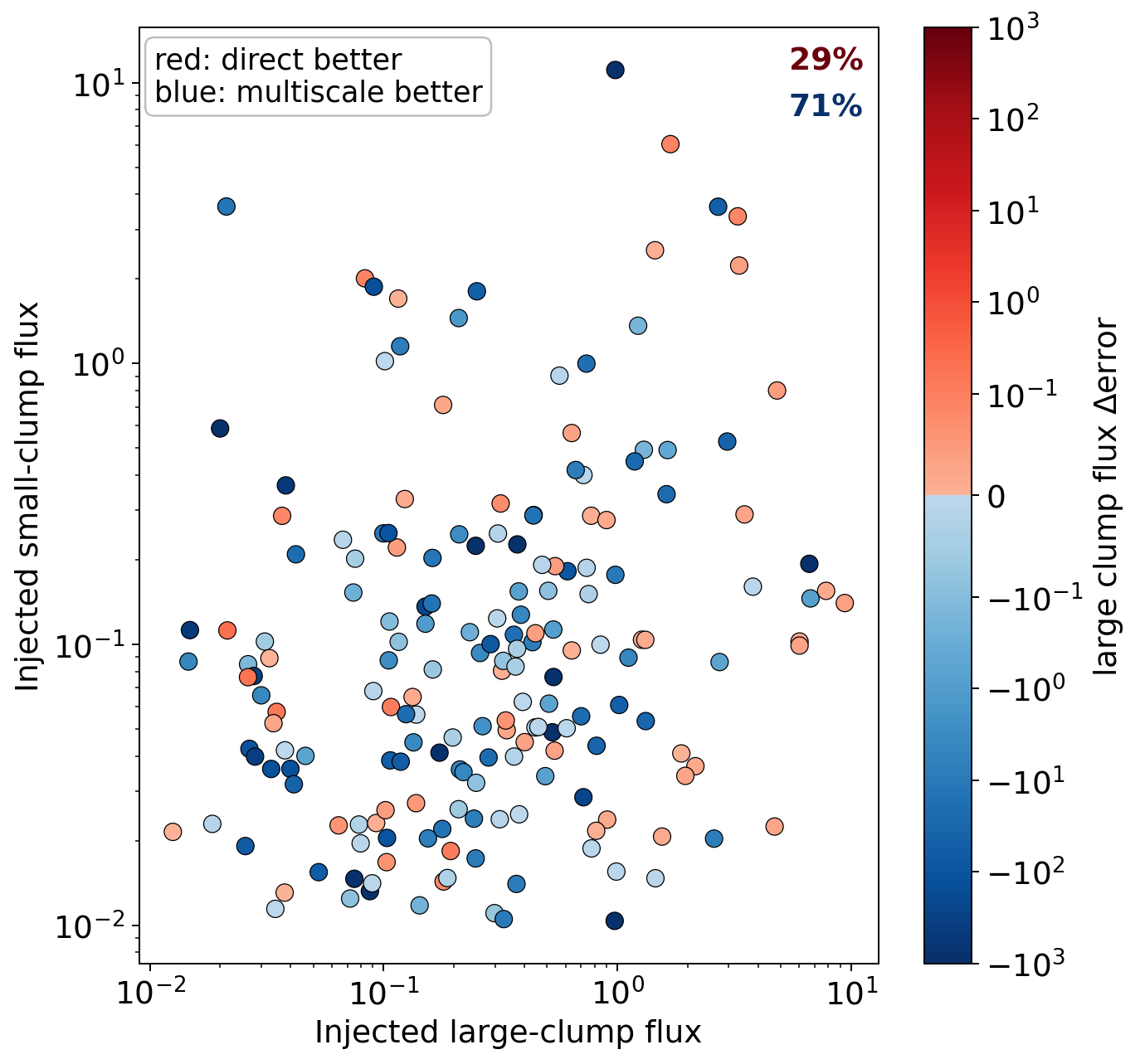}
     \includegraphics[width=0.49\textwidth]{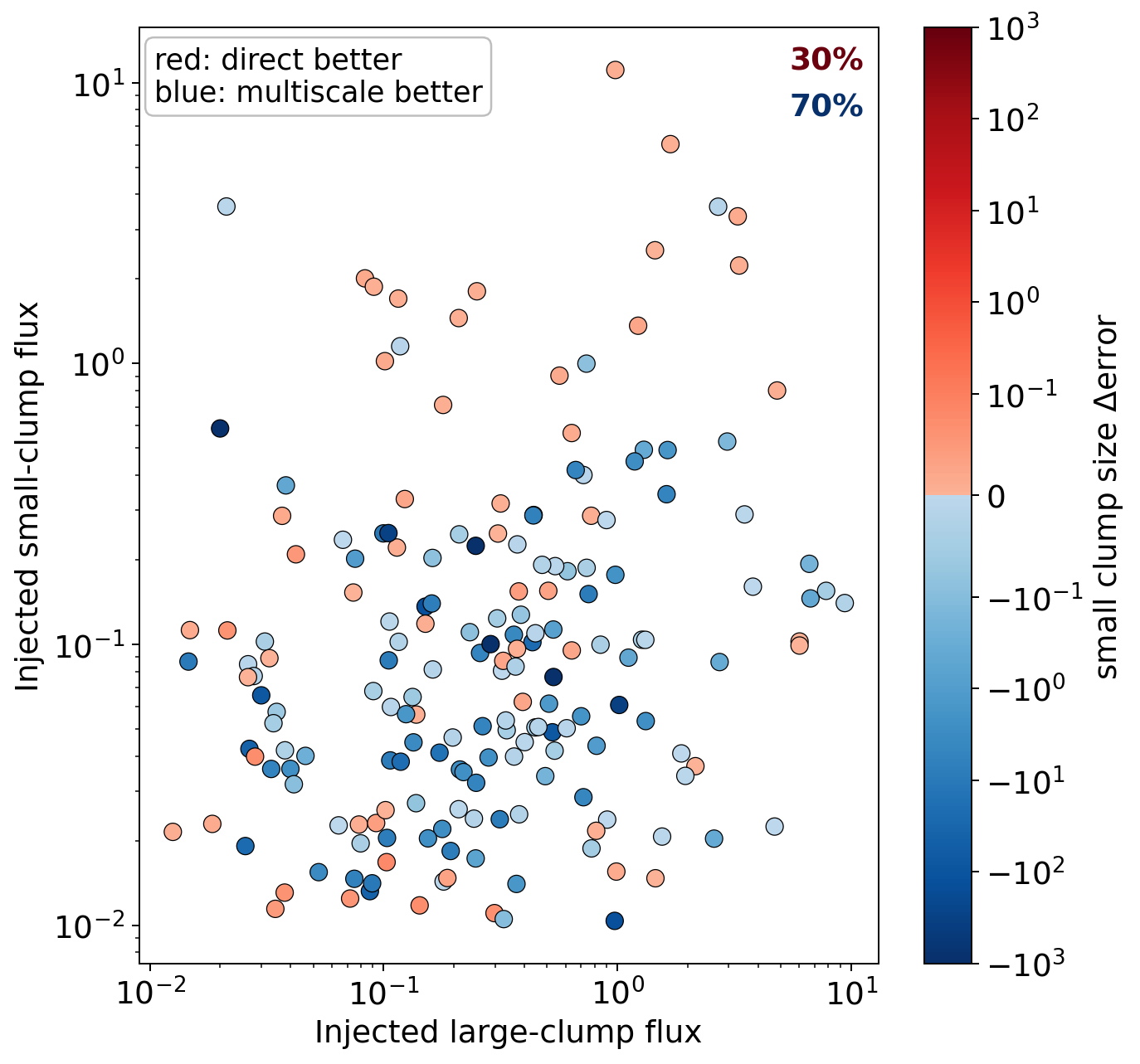}
    \caption{The three panels show the difference between the DoG and a classical model fitting method fractional errors in clump flux (top left), EC flux (top right) and clump size (bottom panel), plotted as a function of the injected EC and clump fluxes. Negative values (blue) mark instances in which the DoG method recovers the injected parameters more accurately, and positive values (red) those for which the classical fit does better.}
    \label{fig:method_compare}
\end{figure}

The DoG-based multi-scale fitting described in Sec.~\ref{subsec:det_flux_fit} is designed to fit clumps and ECs simultaneously by exploiting their differing characteristic scales. To verify that this offers a real advantage over a classical multi-component fit performed directly on the image, we compare the two approaches through an injection--recovery experiment carried out on the real galaxy cutouts.

For each galaxy, we collect the best-fit (F070W) models of clumps and ECs.  We then perform ten independent injection rounds per galaxy. In each round, one clump and one EC are drawn at random. We place the EC at a random position within the cutout (keeping it at least 8 pixels from the edges), and offset the clump from it by a random offset of length up to 4 pixels, thereby reproducing a realistic clump--EC pairing. We then fit the clump and EC using the same DoG pipeline (method A) used in this work, alongside a single plane (no DoG planes involved) model fitting (method B), both with identical initial parameters and constraints. 

We keep the current best-fit model for the galaxy image fixed throughout. The two methods differ only in the $\chi^2$ being minimised. Method A is the multi-scale, dual-weighted DoG $\chi^2$ of Sec.~\ref{subsec:det_flux_fit}. Method B is a classical inverse-variance-weighted pixel $\chi^2$
evaluated over the full unfiltered image, using the weight maps to construct the error
image.  For each injected source and each method we compute the total flux and effective radius analytically from the recovered amplitude, size, and ellipticity, and compare them against the injected truth via the fractional error $|X_\mathrm{rec} - X_\mathrm{inj}| / |X_\mathrm{inj}|$. This is shown as a function of the clump and EC flux in Fig.~\ref{fig:method_compare}.

We find that the small clump and EC fluxes are recovered better using method A for $63\%$ and $71\%$ of the trials respectively.  Meanwhile, method A also recovers the clump sizes in $70\%$ of cases. More critically however, in the minority of cases where method B outperforms method A, the fractional error difference never exceeds $\approx 1$. On the other hand, out of the cases where method A is better, the difference regularly reaches $\approx 10^3$ for each parameter being tested. We do not test the EC size recovery since they are constrained in both the methods based on literature-based expectations (Sec.~\ref{subsec:det_flux_fit}) and are hence not truly measured.

\section{Minimum filters for robust estimates}
\label{app:min_filters}

\begin{figure} 
    \centering
    \includegraphics[width=0.49\textwidth]{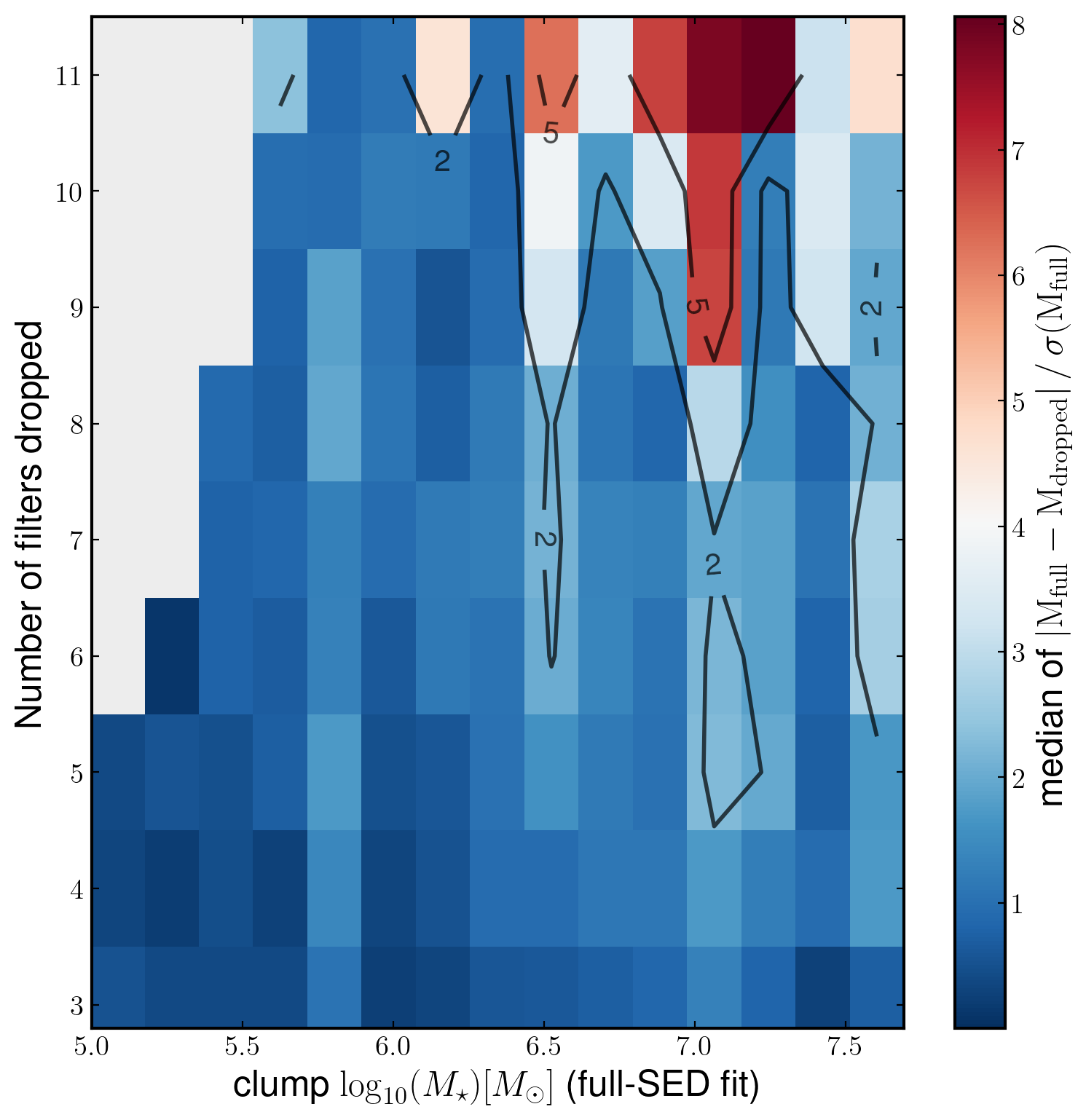}
    \includegraphics[width=0.49\textwidth]{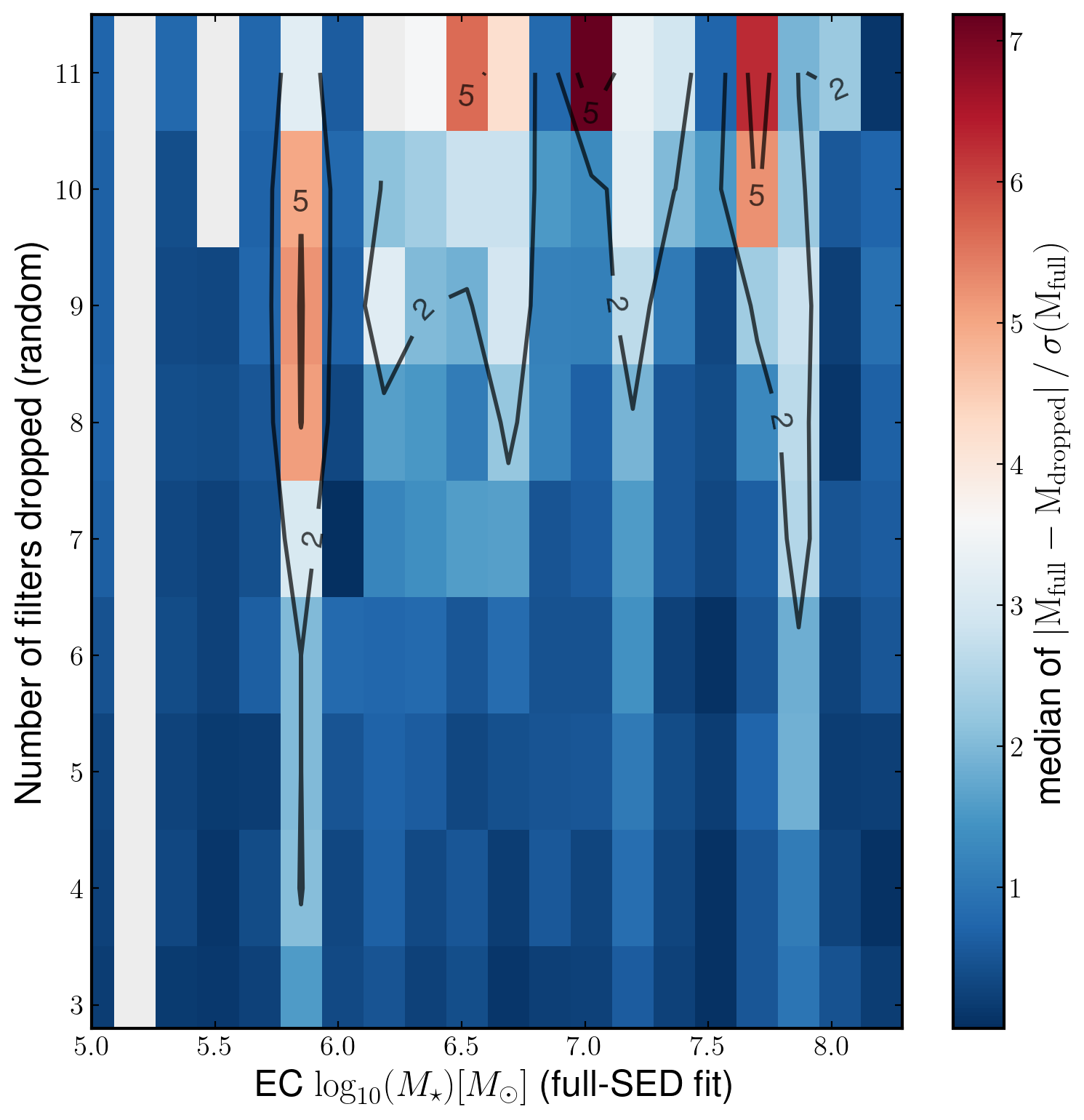}
    \caption{Sensitivity of the SED-derived stellar mass to the size of the photometric baseline, for clumps (left) and ECs (right). The colour shows the offset of the stellar mass measured from a reduced band set, relative to the value obtained from the full 13-band photometry and normalised by the uncertainty on the latter. The x-axis gives the full-SED stellar mass and the y-axis the number of filters dropped from the fit. }
    \label{fig:sed_drop}
\end{figure}

As part of the SED-fitting procedure, we have to introduce a limit on the minimum number of filters with robust photometry in order to accept the SED-based measurements as robust. Below, we empirically determine this threshold for both clumps and ECs. We do so by quantifying how sensitive the derived stellar masses (a key parameter estimated through the SED-fitting) are to the number of filters used. For this experiment we take the pool of clumps and ECs with a fully characterised SED, i.e. those detected in all filters.

We randomly pick 100 (clumps and ECs each) and re-fit the SED over multiple steps, removing one detected band at a time. All parameter choices --- priors, SFH parameterisation, and redshift treatment --- are held identical to the main run, so that the only difference is the iteratively decreasing photometry fed to the fit. At each step the stellar mass is remeasured, and its offset from the primary estimation is recorded. For the clumps, bands are dropped from the reddest to the bluest, mimicking the pattern of non-detection observed for the real clumps in our sample. Meanwhile, for the ECs, bands are instead dropped in random order, also based on the observed behaviour.

Using the full array of clumps and ECs, along with their step-wise SED re-measurements, we generate Fig.~\ref{fig:sed_drop} for the clumps (left panel) and ECs (right panel). The colour represents the magnitude of the offset from the stellar mass measured from the full SED, normalised by the uncertainty on that measurement. We observe that above the stellar mass completeness limit (Appendix~\ref{app:completeness}, below) of $\log(M_\star/\mathrm{M_\odot}) = 7.10$, the offset stays below $2\sigma$ down to a minimum of 6 (for clumps) and 4 (for ECs) remaining filters. We hence use these thresholds to determine which clumps and ECs are to be included in our study.

\section{Completeness}
\label{app:completeness}

\begin{figure} 
    \centering
    \includegraphics[width=\textwidth]{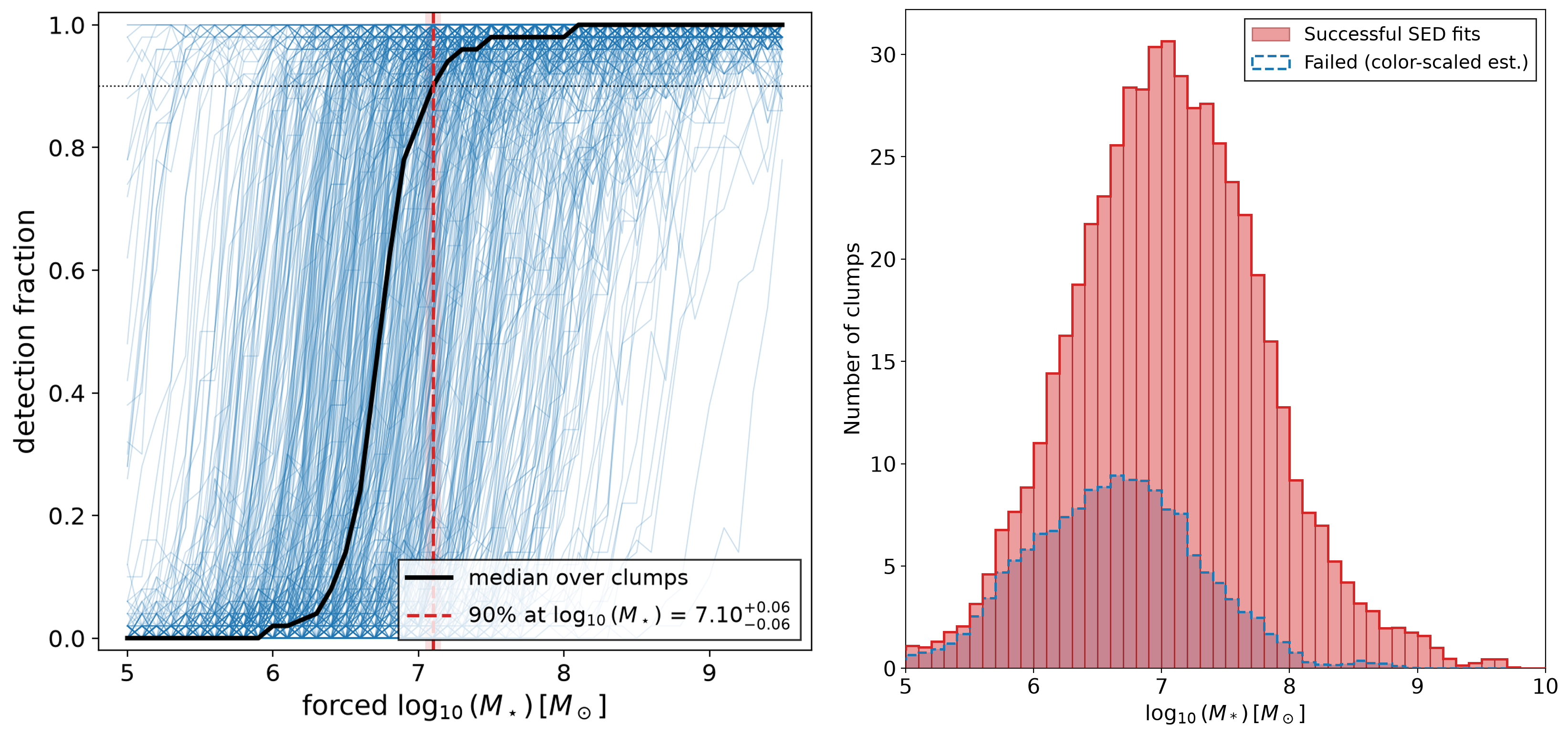}
    \caption{Stellar mass completeness of the clump sample. (Left) Detection fraction as a function of forced stellar mass. Each thin blue line is one of the 562 clumps, re-scaled to each mass in turn and re-injected into its host galaxy image 50 times; the black line is the median over all clumps. The dashed red line marks the $90\%$ completeness limit, $\log(M_\star/\mathrm{M_\odot}) = 7.10^{+0.06}_{-0.06}$, with the shaded band showing the $1\sigma$-equivalent interval from the galaxy-level block bootstrap. (Right) Stellar mass distribution of the clumps with a successful SED fit (red) and of those rejected by the six-band requirement, whose masses are estimated by colour-matching (blue).}
    \label{fig:mass_complete}
\end{figure}

\begin{figure} 
    \centering
    \includegraphics[width=\textwidth]{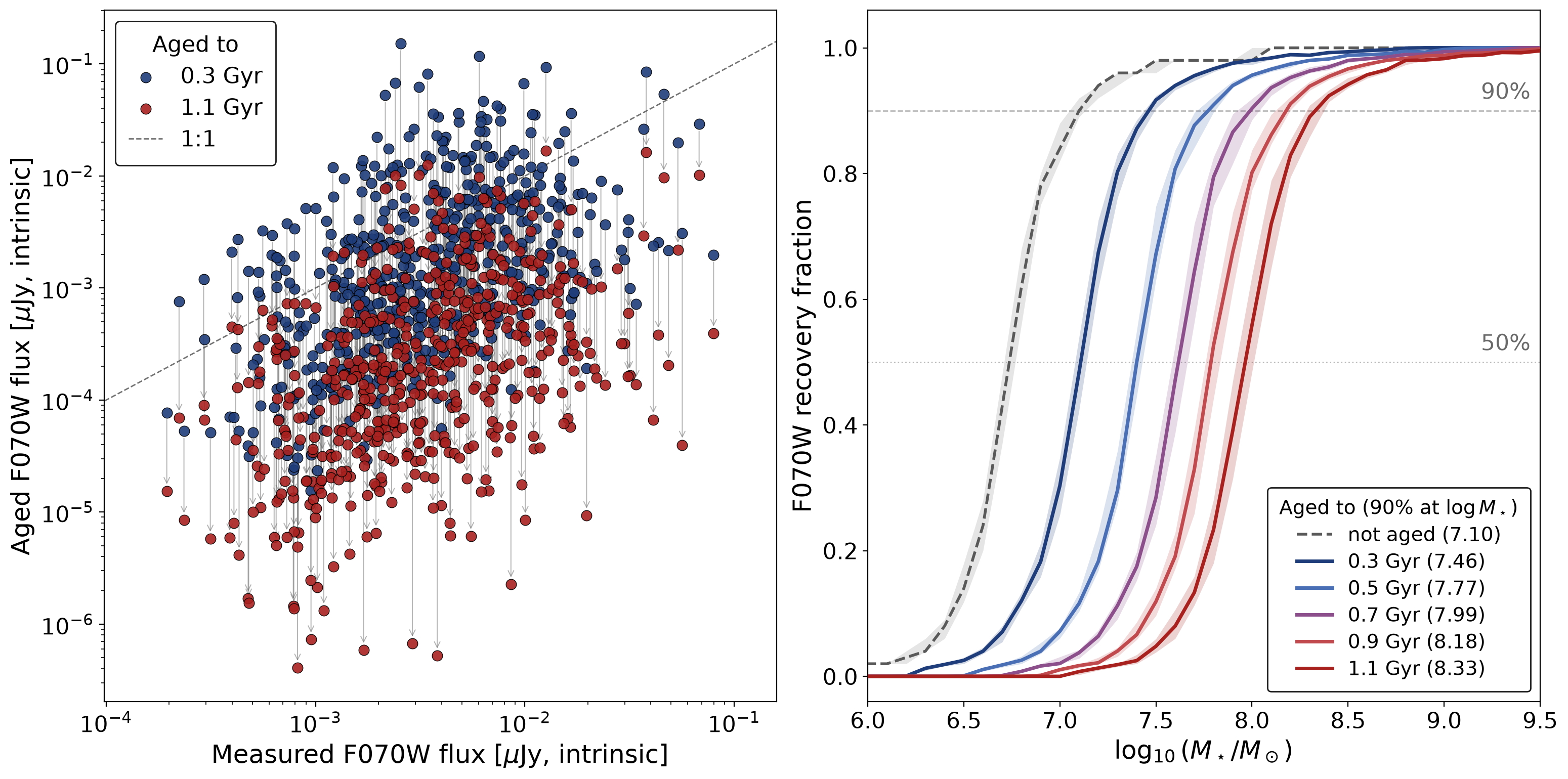}
    \caption{(Left) The measured vs artificially aged clump flux for the full robust SED clumps in our sample. We show only the aged to $0.3\,\mathrm{Gyr}$ and $1.1\,\mathrm{Gyr}$ runs.  (Right) Average detection fractions for our sample as a function of measured stellar mass (does not account for evolution of mass during the artificial aging). }
    \label{fig:aged_detectability}
\end{figure}

We detect the clumps at rest-frame UV wavelengths. As a result, the mass completeness cannot be assessed analytically from the image depths, since the mass-to-light ratio for UV emission is heavily influenced by dust attenuation as well as by the average age of the stellar population being analysed. We therefore employ an empirical determination of the completeness instead. We perform two complementary injection--recovery experiments, both built on the best-fit \textsc{bagpipes} model of each small clump with a successful SED fit (robust $\geq0.5\sigma$ detection in at least 6 filters).

In the first, we re-scale the SED of each clump to force its stellar mass to a value of our choice. This keeps all other parameters fixed (age, $\tau$, metallicity, $A_V$, ionisation parameter, and redshift) while only changing the normalisation. Stepping through an array of forced stellar mass values from $\log(M_\star/\mathrm{M_\odot}) = 9.5$ down to $5.0$ in $0.1$\,dex increments, we estimate the rest-frame UV (F070W) flux one would expect. At each step, this artificially rescaled clump is injected back into the respective galaxy image at 50 random locations within $20\times$ the F070W PSF FWHM (same as in Sec.~\ref{subsec:flux_uncertainties}). Each time, we use our DoG detection pipeline to check whether the new clump has been detected, giving us a detection fraction. Each of the 562 clumps in our sample provides a single detection fraction (out of 50 realisations) for every step in our forced stellar mass array (Fig.~\ref{fig:mass_complete}, left panel). We take the median over all clumps to get a representative detection fraction vs stellar mass curve, and define the $90\%$ completeness limit as the stellar mass at which this median curve first crosses a detection fraction of $0.9$, linearly interpolated in $\log M_\star$. We find this value to be $\log(M_\star/\mathrm{M_\odot}) = 7.10^{+0.06}_{-0.06}$. The uncertainty is estimated with a block bootstrap in which the galaxy, not the clump, is the independent unit --- clumps within a galaxy share the same image, PSF, and background. For each of $10^3$ realisations we resample the galaxies with replacement, redraw the magnification of each clump from its uncertainty (which shifts its detection curve along the intrinsic-mass axis), pool the clumps of the selected galaxies, rebuild the median curve, and recompute the crossing mass. The $16$th and $84$th percentiles of the resulting distribution give the asymmetric $1\sigma$-equivalent interval quoted on the limit.

We then compare the completeness limit against the minimum band threshold implemented during the SED fitting. A clump detected in F070W does not necessarily enter the SED analysis: we require a minimum of six detected bands (Sec.~\ref{subsec:sed_fitting}). To place the rejected clumps on the same mass axis, we estimate their masses by colour-matching. For each rejected clump we take its bluest, reddest, and an intermediate detected band, and compute the two resulting colours. We then identify the three successfully-fit clumps in the same galaxy lying closest in this colour--colour parameter space. The flux ratio between the rejected clump and the successfully-fit clump in the longest available filter is used to estimate the stellar mass of the former, and we take the average of the three values from the three matches. Clumps detected in fewer than three bands are excluded. Each clump in both populations is taken as $100$ Monte Carlo realisations constrained by the stellar mass errors of the successfully-fit clumps (which are also propagated to the rejected-clump mass estimates), and we show their distribution in Fig.~\ref{fig:mass_complete} (right panel). Scanning from the most massive end, we find that the fraction of clumps with robust SEDs first drops below $90\%$ at $\log(M_\star/\mathrm{M_\odot}) = 7.38$. This lies slightly above our $90\%$ mass completeness limit ($7.10$), albeit within the mass uncertainties (median $= 0.24\,\mathrm{dex}$). Furthermore, one of the key analysis these uncertainties might influence would be the CSMF. We note that the break above which the slope of the massive end is measured is actually higher than both the SED-fitting requirement based rejection threshold and the completeness limit. Nevertheless, we remeasure the CSMF after including the clumps without sufficient bands and using the aforementioned estimated mass using color-matching. We find the resulting CSMF high mass slope to be $1.92 \pm 0.04$, which is only $0.04$ higher than the value reported in the main text.


Finally, the forced stellar mass completeness estimation pools clumps of all stellar ages, and rescales each one at its own best-fit age. Hence, it does not allow for an assessment of the completeness as a function of increasing age, which would reduce the rest-frame UV flux at a constant stellar mass. One cannot simply split Fig.~\ref{fig:mass_complete} by clump age either: the observed age bins are themselves shaped by the incompleteness we are trying to map, since the oldest, faintest clumps are preferentially absent from the sample to begin with. We therefore conduct a second test that avoids this by evolving each detected clump forward in age, so that the same population is followed across ages rather than compared between differently-selected sub-samples. We use the best-fit model of each clump to predict its observed-frame F070W flux if the stellar population were evolved to a series of ages, $0.3$, $0.5$, $0.7$, $0.9$, and $1.1$\,Gyr (Fig.~\ref{fig:aged_detectability}, left panel). For each target age we regenerate the SED holding the formed mass, $\tau$, metallicity, $A_V$, and redshift fixed, while setting the stellar age to the target value (capped at $95\%$ of the cosmic age at the best-fit redshift where required). The ionisation parameter is not kept fixed but evolved with age following $\log U = \log U_\mathrm{best} - \zeta\log_{10}(\mathrm{age}/\mathrm{age_{best}})$, clipped to $[-5, -2]$. Here we assume $\zeta = 1$ as a fiducial round-number value, chosen to capture the expected decline of $\log U$ with population age, driven by the falling output of hydrogen-ionising photons as massive stars evolve off the main sequence. The clump with the re-estimated F070W flux is once again injected 50 times into the host galaxy image, following the same procedure as in the forced stellar mass method. We then show the resulting recovery fraction as a function of stellar mass for each stellar age in Fig.~\ref{fig:aged_detectability} (right panel). Based on this aged clump detectability test, we find that the mass completeness limit of $\log(M_\star/\mathrm{M_\odot}) = 7.10$ holds for the clumps at their observed ages, and rises to $7.46$ at $0.3\,$Gyr and $8.33$ at $1.1\,$Gyr. The CSMF above the break ($M_b = 10^{7.49}\,M_\odot$) is therefore complete for clumps younger than $\approx0.3\,$Gyr, while older clumps would be increasingly missed below $\sim10^{8.3}\,M_\odot$.

\section{Virial prefactor for a 2D Gaussian profile}
\label{app:virial_prefactor}

The scaled virial parameter used in the main text,
\begin{equation} \label{eq:virial_param_app}
  \alpha^{\mathrm{scaled}}_{\mathrm{vir}}
    = \eta\,\frac{\sigma^{2}\,r_{\mathrm{eff}}}{G\,M},
\end{equation}
has a dimensionless prefactor $\eta$ that depends on the assumed density profile and on the radius we adopt as $r_{\mathrm{eff}}$.  Here we estimate $\eta$ for our case, where each clump is modeled as a 2D Gaussian. Hence the $r_{\mathrm{eff}}$ is the projected half-light radius, and $\sigma$ is the one-dimensional (line-of-sight) velocity dispersion.

\paragraph{Relating $\eta$ to the gravitational energy.}
The virial parameter is twice the kinetic energy over the gravitational
energy. With $T = \frac{1}{2}M\sigma_{\mathrm{3D}}^{2}$, and using
$\sigma_{\mathrm{3D}}^{2} = 3\sigma^{2}$ to relate the (unobserved) 3D
dispersion to the line-of-sight value we actually measure,
\begin{equation} \label{eq:alpha_def_app}
  \alpha_{\mathrm{vir}}
    = \frac{2T}{|W|}
    = \frac{M\,\sigma_{\mathrm{3D}}^{2}}{|W|}
    = \frac{3\,M\,\sigma^{2}}{|W|}.
\end{equation}
The gravitational energy of any spherical mass distribution can be written
\begin{equation} \label{eq:W_general_app}
  W = -a\,\frac{G\,M^{2}}{r_{\mathrm{eff}}},
\end{equation}
where $a$ is the dimensionless structure constant that describes how
centrally concentrated the profile is. Comparing Eqs.~(\ref{eq:alpha_def_app}) and (\ref{eq:W_general_app}) with Eq.~(\ref{eq:virial_param_app}) gives $\eta = 3/a$. So to estimate the prefactor,  we require $a$ for the deprojected Gaussian model.

\paragraph{Deprojection.}
For a circular 2D Gaussian surface density, $\Sigma(R) \propto \exp(-R^{2}/2s^{2})$, the Abel deprojection returns a 3D Gaussian with the \emph{same} scale length $s$,
\begin{equation} \label{eq:rho_app}
  \rho(r) = \rho_{0}\,\exp\!\left(-\frac{r^{2}}{2s^{2}}\right),
  \qquad
  \rho_{0} = \frac{M}{(2\pi)^{3/2}\,s^{3}},
\end{equation}
where $\rho_{0}$ is fixed by normalising to the total mass $M$. Integrating $\rho$ out to radius $r$ gives the enclosed mass,
\begin{equation} \label{eq:Menc_app}
  M(r) = M\left[\,\mathrm{erf}\!\left(\frac{r}{\sqrt{2}\,s}\right)
        - \sqrt{\frac{2}{\pi}}\,\frac{r}{s}\,
          \exp\!\left(-\frac{r^{2}}{2s^{2}}\right)\right],
\end{equation}
which we will be using to evaluate the gravitational energy.

\paragraph{Gravitational energy.}
The gravitational energy is assembled shell by shell,
\begin{equation} \label{eq:W_integral_app}
  W = -\int_{0}^{\infty} \frac{G\,M(r)}{r}\,\mathrm{d}M(r),
\end{equation}
where $G\,M(r)/r$ is the potential due to the mass interior to $r$ and
$\mathrm{d}M(r)$ is the mass in the shell at $r$. Substituting
Eqs.~(\ref{eq:rho_app}) and (\ref{eq:Menc_app}) and carrying out the
integral gives the closed-form result
\begin{equation} \label{eq:W_gauss_app}
  W = -\frac{1}{2\sqrt{\pi}}\,\frac{G\,M^{2}}{s}.
\end{equation}
Expressed in terms of the Gaussian scale length $s$, the structure
constant is therefore $a_{s} = 1/(2\sqrt{\pi}) \approx 0.282$, giving
$\eta_{s} = 3/a_{s} = 6\sqrt{\pi} \approx 10.6$.

\paragraph{Conversion to the half-light radius.}
We do not measure $s$ directly; the observable size is the projected
half-light radius. For a 2D Gaussian the two differ only by a fixed
factor,
\begin{equation} \label{eq:reff_app}
  r_{\mathrm{eff}} = s\,\sqrt{2\ln 2} \approx 1.177\,s.
\end{equation}
Substituting Eq.~(\ref{eq:reff_app}) into Eq.~(\ref{eq:W_gauss_app}),
\begin{equation} \label{eq:W_reff_app}
  W = -\frac{\sqrt{2\ln 2}}{2\sqrt{\pi}}\,\frac{G\,M^{2}}{r_{\mathrm{eff}}},
\end{equation}
so the structure constant referred to the half-light radius is
$a_{\mathrm{eff}} = \sqrt{2\ln 2}/(2\sqrt{\pi}) \approx 0.332$, and
therefore
\begin{equation} \label{eq:eta_result_app}
  \eta = \frac{3}{a_{\mathrm{eff}}}
       = \frac{6\sqrt{\pi}}{\sqrt{2\ln 2}}
       \approx 9.0,
\end{equation}
which is the value adopted in the main text.

\bibliography{main}{}
\bibliographystyle{aasjournalv7}



\end{document}